\documentclass[%
reprint,
superscriptaddress,
showkeys,
amsmath,amssymb,
aps,
twocolumns]{revtex4-2}

\usepackage{amssymb}
\usepackage{amsmath}
\usepackage{amsthm}
\usepackage{natbib}
\usepackage{eucal}
\usepackage{mathrsfs}
\usepackage{graphicx}
\usepackage{dcolumn}
\usepackage{color}
\usepackage{bm}
\usepackage{hyperref}
\usepackage[usenames,dvipsnames,x11names,table]{xcolor}
\usepackage{balance}
\usepackage{dsfont}
\usepackage{tikz}
\usetikzlibrary{shapes}
\usepackage{multirow}
\usepackage{tabularx}
\usepackage{enumitem}
\usepackage{cancel,soul}

\setlist[itemize]{align=parleft,left=0pt..1em}
\newcommand{\Ra}{\operatorname{Ra}}
\newcommand{\Nu}{\operatorname{Nu}}
\newcommand{\We}{\operatorname{We}}

\makeatletter
\let\@internalcite\cite
\def\cite{\def\citeauthoryear##1##2{##1, ##2}\@internalcite}
\def\shortcite{\def\citeauthoryear##1##2{##2}\@internalcite}
\def\@biblabel#1{\def\citeauthoryear##1##2{##1, ##2}[#1]\hfill}
\makeatother
\begin{document}
\preprint{APS/123-QED}
\title{Dynamical regimes of thermally convective emulsions}
\author{Francesca Pelusi}
\email{francesca.pelusi@cnr.it}
\affiliation{Istituto per le Applicazioni del Calcolo, CNR - Via Pietro Castellino 111, 80131 Naples, Italy}
\author{Andrea Scagliarini}
\affiliation{Istituto per le Applicazioni del Calcolo, CNR - Via dei Taurini 19, 00185 Rome, Italy}
\affiliation{INFN, Sezione Roma ``Tor Vergata", Via della Ricerca Scientifica 1, 00133 Rome, Italy}
\author{Mauro Sbragaglia}
\affiliation{Department of Physics \& INFN, Tor Vergata University of Rome, Via della Ricerca Scientifica 1, 00133 Rome, Italy}
\author{Massimo Bernaschi}
\affiliation{Istituto per le Applicazioni del Calcolo, CNR - Via dei Taurini 19, 00185 Rome, Italy}
\author{Roberto Benzi}
\affiliation{Sino-Europe Complex Science Center, School of Mathematics \\North University
of China, Shanxi, Taiyuan 030051, China}
\affiliation{Department of Physics \& INFN, Tor Vergata University of Rome, Via della Ricerca Scientifica 1, 00133 Rome, Italy}
\date{\today}
\begin{abstract}
\noindent
Emulsions are paramount in various interdisciplinary topical areas, yet a satisfactory understanding of their behavior in buoyancy-driven thermal flows has not been established. In the present work, we unravel the dynamical regimes of thermal convection in emulsions by leveraging a large set of mesoscale numerical simulations. Emulsions are prepared with a given volume fraction of the initially dispersed phase, $\phi$, ranging from dilute (low values of $\phi$) to jammed emulsions (high values of $\phi$), resulting in different rheological responses of the emulsion, i.e., from Newtonian to non-Newtonian yield-stress behaviors, respectively. We then characterize the dynamics of the emulsions in the paradigmatic setup of the Rayleigh-B\'enard convection, i.e., when confined between two parallel walls at different temperatures under the effect of buoyancy forces, the latter encoded in the dimensionless Rayleigh number $\Ra$. We thoroughly investigated the dynamics of the emulsion in the changing of $\phi$ and $\Ra$. For a given $\phi$, at increasing $\Ra$, we observe that the emulsion exhibits convection states, where structural changes may appear (i.e., droplet breakup, coalescence or phase inversion), which inevitably impact the emulsion rheology. For sufficiently high values of $\Ra$, two states of convection are observed: for low/moderate values of $\phi$ (Newtonian emulsions), we observe breakup-dominated dynamics, whereas, for high values of $\phi$ (non-Newtonian emulsions), we observe phase-inverted states. For both scenarios, the droplet size distribution depends on $\Ra$, and scaling laws for the average droplet size are analyzed and quantified. Our results offer new insights into the rich dynamics of emulsions under thermal convection, offering the first detailed characterization of the various dynamical regimes to be expected and their relation with structural changes occurring in such complex fluids.
\end{abstract}
\maketitle
\section{Introduction}\label{sec:intro}
Emulsions are heterogeneous systems formed by a dispersion of droplets of a liquid phase (e.g., oil) in another immiscible liquid (e.g., water). Adding an emulsifier (e.g., surfactants) stabilizes the dispersion by hindering the coalescence of droplets~\cite{Barnes94, Mason99,Derkach09,Tadros13}. Owing to the presence of interfaces that can store part of the injected energy via elastic deformations~\cite{FerreroMartensBarrat14,Bonn17,BarratReview17,Nicolas18,Divoux24}, an emulsion exhibits a variety of rheological responses when stimulated with an external force. An important parameter -- discriminating different rheological responses of the emulsion -- is the volume fraction, $\phi$, i.e., the ratio between the volume of the dispersed phase and the total volume, which is related to the total interface area for a given number of droplets. For low values of $\phi$, the emulsion behaves as a Newtonian system presenting an augmented effective viscosity that depends on $\phi$ and the viscosity ratio of the two phases~\cite{Taylor32,Zinchenko84,Ghigliottietal10}. However, for higher values of $\phi$, the rheology of the emulsion is non-Newtonian in that it exhibits a shear-thinning rheology~\cite{Pal2000}, which changes to yield stress behavior when $\phi$ is high enough~\cite{Barnes94,Balmforthetal14,Bonn17,BarratReview17}. If the emulsion is solicited with a sufficiently large external force, droplets can break up and coalesce~\cite{Stone94,Cristini2003,Vankova07a,Vankova07b,Vankova07c,Perlekar2012,Scarbolo15,Roccon17,Soligo19,Liu21,Girotto22,Girotto24,CrialesiEsposito2024small}, thus inducing fluctuations of the total interface area and, hence, modifying the rheological response of the system. The application of flowing emulsions in different manufacturing contexts, such as pharmaceutical~\cite{Khan11}, food~\cite{Mcclements15} or energy industries~\cite{Goodarzi19}, made these materials the subject of intense scrutiny for decades since a precise control of their stability and rheological properties under specific external forcing and boundary conditions is required in such contexts~\cite{Gallegos99,Mcclements07}. Flowing emulsions exhibit highly complex features. Hence, numerical simulations are desirable tools of investigation to address regimes and processes particularly difficult to reproduce in experiments, thus unraveling multiple relevant questions. 
This aspect is witnessed in the extensive literature on computational studies on droplet/emulsion dynamics, with various methods developed~\cite{Benzi09,Roudsari12,Goodarzi19,Rosti19,wang2020molecular,Lohse20}. However, if, on the one hand, considerable effort has been made to investigate the behavior of emulsions under shear, pressure-driven, or even more complex flows (e.g., turbulence), on the other hand, a satisfactory understanding of the behavior of emulsions in buoyancy-driven thermal flows remains incomplete.
Thermally driven emulsions have a significant impact in industrial applications, particularly in petrochemical engineering~\cite{abdulredha2020overview}, where they influence oil recovery, refining, and transportation processes; but also in food science, where emulsion-based heat transfer is crucial for cooking and processing techniques~\cite{Jeffreys57}, affecting texture, stability, and nutrient retention. More generally, emulsions may be regarded as ``soft materials", whose behavior in thermal flows is particularly important: in geophysics, for example, these materials are essential for modeling Earth's mantle convection~\cite{Orowan65,Morgan71,Montelli06,French15,Davaille2018} -- where buoyancy-driven currents transport heat from the planet’s interior to its surface, shaping plate tectonics and volcanic activity -- and for understanding both explosive and nonexplosive lava flows~\cite{Griffiths2000,Lavallee15} -- where the interaction between molten rock, gas bubbles, and crystals determines magma rheology.
Effective analog materials for geodynamic modeling are worth being mentioned~\cite{SchellartStrak,DiGiuseppe2014,Reber2020}.

Some computational studies considered droplets (or also bubbles) in convective flows, but they pertain to the case of convective {\it multiphase flows} with point-like dispersed objects~\cite{Orestaetal09,Lakkaraju13}, or multiphase flows where finite-sized droplets/bubbles undergo a dynamics without any stabilization mechanism against their coalescence~\cite{Biferale12,Garoosi21,santos21,Liu21,Liu21a,Liu22,Brandt24,Mangani24}, thus resulting in a system with overall Newtonian rheology. On the other hand, some theoretical/numerical works investigated the role played by the non-Newtonian rheology -- and in particular the yield-stress rheology -- on the onset of thermal convection using {\it effective single fluid models} ~\cite{Zhang06,BalmforthRust09,Vikhansky09,Vikhansky10,AlbaalbakiKhayat11,Turanetal12,Massmeyer13,Hassanetal15,Karimfazli16}. These models rely on the use of the equations of continuum mechanics, supplemented with some non-trivial constitutive relations between the stress and the shear rate to account for non-Newtonian rheology. From the analysis of these effective single fluid models with yield-stress rheology, it has been found that conductive states become linearly stable in the presence of a finite yield stress~\cite{Zhang06,BalmforthRust09}. Moreover, the system can enter a convective state only after applying a perturbation above a given intensity threshold. The latter increases upon approaching the critical Rayleigh number, marking the transition from conduction to convection~\cite{Zhang06}. Effective single fluid models, however, neglect the size of microscopic constituents (droplets/bubbles); hence, they cannot account for plasticity and structural changes at mesoscales~\cite{Goyon08,Goyon10}; the latter, in turn, may have a non-trivial correlation with thermal plumes, as evidenced by some experiments on yield-stress fluids under thermal convection~\cite{Davailleetal13}. Also, recent numerical studies based on {\it lattice Boltzmann models} (LBMs) underscored the importance of finite-sized droplets~\cite{PelusiSM21,PelusiSM23,Pelusi24intermittent}, showing that their presence is crucial in enhancing heat flux fluctuations as the volume fraction of the dispersed phase increases, especially by approaching from above the transition from conduction to convection. In addition, unlike Newtonian convection, thermal plumes in non-Newtonian systems can stop before the emergence of other plumes~\cite{Davailleetal13}. This arrest can take place after long periods of chaotic oscillations~\cite{Vikhansky09}, and it may be intimately connected with the finite size of the droplets, as recently suggested by experiments~\cite{Jadhav21} and numerical studies by the authors~\cite{Pelusi24intermittent}. Indeed, in Ref.~\cite{Pelusi24intermittent}, we considered a model yield-stress emulsion and pinpointed a transient intermittent regime in the heat flux. In the latter regime, prolonged conductive periods alternate with quick and intense convective heat bursts triggered by microscopic plasticity and the large spatial correlation of the emulsion; during the heat bursts, coalescence events occur and make the emulsion prone to a {\it phase inversion}, whereby the majority phase becomes continuous and the minority phase forms droplets  (e.g., an oil-in-water emulsion becomes a water-in-oil emulsion). Such sudden morphological change comes together with a dramatic reduction of the total interface area, which, in turn, determines a drastic variation of the rheological properties~\cite{Groeneweg98,Kumar15,Bouchama03,Perazzo15} and a consequent transition to thermal convection of the phase-inverted emulsion. In general, a phase inversion can be achieved following various cues, such as chemical treatments~\cite{Fernandez04,Borrin16}, or mechanical forcing (stirring speed)~\cite{Yeo2002,Perazzo15}, or via the presence of homogeneous isotropic turbulent conditions~\cite{Bakhuis21,Girotto24,Yi24}. To the best of the authors' knowledge, except for the study presented in Ref.~\cite{Pelusi24intermittent}, phase inversion of concentrated emulsions in thermally buoyant flows has not been reported elsewhere.\\

The scenario portrayed above is rather intriguing and rich. Still, a deep understanding of the behavior of emulsions under thermally driven flows remains elusive. In this work, we significantly advance our understanding of thermally driven emulsions using LBM numerical simulations in the paradigmatic setup of Rayleigh-Bénard thermal convection, i.e., with the emulsion placed between two parallel walls at different temperatures under the effects of buoyancy forces, the latter encoded in the dimensionless Rayleigh number $\Ra$. We systematically change both $\phi$ and $\Ra$ and offer a characterization of the emulsion dynamics and a detailed view of the structural changes that the emulsion can experience. As $\Ra$ increases for a given $\phi$, the emulsion enters convective states where structural changes -- such as droplet breakup, coalescence, or phase inversion -- significantly alter its rheology. Furthermore, we characterize the scaling properties of the droplet size distribution at sufficiently high values of $\Ra$, where two regimes emerge: for low-to-moderate values of $\phi$, the dynamics is dominated by droplet breakup, whereas at higher values of $\phi$, phase inversion becomes predominant.\\

The paper is organized as follows: in Sec.~\ref{sec:method}, we describe the system set-up and provide a review of the LBM for emulsion modeling; in Sec.~\ref{sec:dynamic-regimes}, we analyze the dynamical regimes experienced by the emulsions; in Sec.~\ref{sec:droplet_size_statistics}, we provide statistical analysis of droplet size distribution at high values of $\Ra$. Conclusions will be drawn in Sec.~\ref{sec:conclusions}.
\section{System set-up and numerical modelling}\label{sec:method}
In this section, we provide details on the system set-up and the numerical modeling of emulsions. In Sec.~\ref{subsec:setup}, we describe the Rayleigh-Bénard configuration we used, the protocol for the emulsion preparation, and the relevant observables. In Sec.~\ref{subsec:lbm}, a brief review of the employed LBM for the simulations of emulsions is provided.
\subsection{System set-up}\label{subsec:setup}
\begin{figure*}[t!]
    \centering    \includegraphics[width=1.\textwidth]{
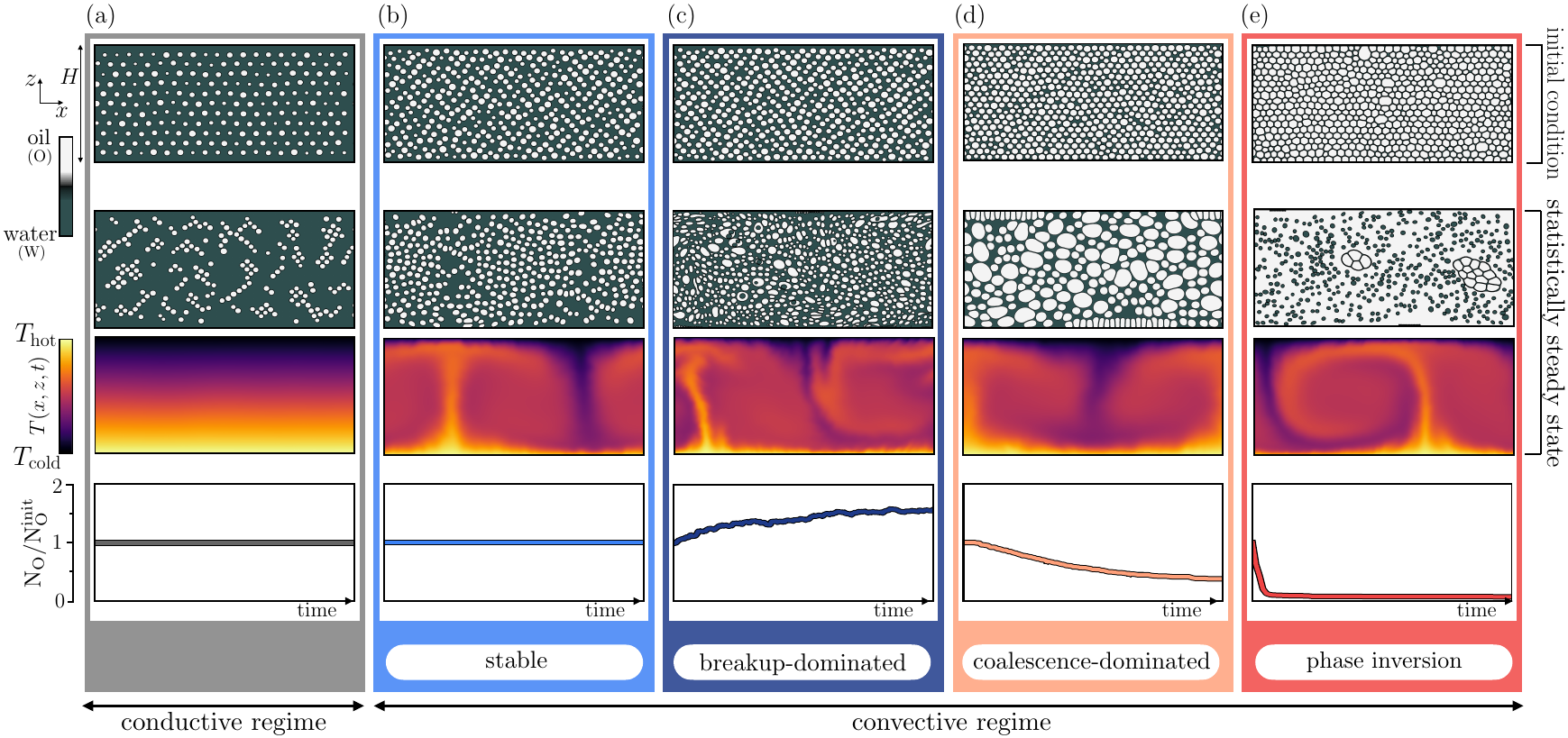}
    \caption{Characterization of dynamical regimes of thermally convective oil-in-water (O/W) emulsions in the setup of Rayleigh-Bénard thermal convection. Different regimes result from different combinations of the volume fraction of the initially dispersed oil phase $\phi$ and the Rayleigh number $\Ra$ (see text for more details). The characterization is done by considering the presence/absence of thermal convection and the variation in the number of oil droplets $\mathrm{N}_{\mathrm{O}}$ with respect to its initial value $\mathrm{N}^{\mathrm{init}}_{\mathrm{O}}$. Panel (a) (grey box): conductive regime, with no variation of $\mathrm{N}_{\mathrm{O}}$. Panel (b) (light-blue box): {\it stable} convective regime, with convective plumes and no variation of $\mathrm{N}_{\mathrm{O}}$. Panel (c) (blue box): {\it breakup-dominated} convective regime, with convective plumes and an increase of $\mathrm{N}_{\mathrm{O}}$ due to breakup events. Panel (d) (pink box): {\it coalescence-dominated} convective regime, with convective plumes and a decrease of $\mathrm{N}_{\mathrm{O}}$ caused by coalescence events. Panel (e) (red box): {\it phase-inverted} convective regime, with an initial O/W emulsion that enters a W/O steady convective state. Different colors are used to discriminate between the different regimes. For each pair $(\phi, \Ra)$, the corresponding regime in the statistically steady state is shown in Fig.~\ref{fig:phase_diagram}. Movies corresponding to panels (b)-(e) can be found in the Supplementary Material.}
    \label{fig:sketch}
\end{figure*}

We consider two fluid components with equal kinematic viscosity $\nu$ and equal thermal diffusivity $\kappa$. Without loss of generality, hereafter, we refer to ``oil (O) phase'', ``water (W) phase'', ``oil-in-water (O/W) emulsion'' and ``water-in-oil (W/O) emulsion''. We consider O/W emulsions in a 2D Rayleigh-B\'enard setup~\cite{Grossmann01,Ahlers09,Chilla12}. Using a 2D set-up is mainly driven by computational reasons, in that we need to properly resolve the emulsion droplets and achieve sufficient statistics in a reasonable amount of time. 2D convection is different from 3D convection, but it still captures many of its relevant features~\cite{van2013comparison}. We use a domain of size $L \times H$, with $L \approx 2H$ and wall-to-wall distance $H \approx 20d$, where $d$ is the initial average droplet diameter, which is fixed for all simulations. Two no-slip walls are set at $z=\pm H/2$, and periodic boundary conditions are applied along the $x$ direction. The walls are kept at constant temperatures, $T_{\mathrm{hot}}$ (bottom wall) and $T_{\mathrm{cold}}$ (top wall) (see Fig.~\ref{fig:sketch}), corresponding to a temperature jump $\Delta T = T_{\mathrm{hot}} - T_{\mathrm{cold}}$. Gravity acceleration ${\bm g}$ is acting in the negative $z$ direction, i.e., ${\bm g} = -g \hat{\bm z}$. The system is analyzed in terms of continuous fields depending on position ${\bm x} = (x,z)$ and time $t$: the hydrodynamical velocity ${\bm u}={\bm u}({\bm x},t)$, the density of the oil (water) phases, $\rho_{\text{O}}({\bm x},t)$ ($\rho_{\text{W}}({\bm x},t)$), and the temperature field $T({\bm x},t)$. In the bulk phases, the dynamics for ${\bm u}$ is set by the Navier-Stokes equation with viscosity $\nu$; furthermore, non-ideal interfaces with a non-zero surface tension $\Sigma$ separate regions with majority of water phase  ($\rho_{\text{O}} \ll \rho_{\text{W}}$) from regions with majority of the oil phase ($\rho_{\text{W}} \ll \rho_{\text{O}}$). The dynamics for the temperature field $T({\bm x},t)$ is governed by an advection-diffusion equation, with the advection set by the hydrodynamical velocity ${\bm u}$ and the diffusion regulated by the thermal diffusivity $\kappa$. Further details on the numerical modeling are provided in Sec~\ref{subsec:lbm}.\\
To create O/W emulsions, we follow a dedicated preparation protocol: the oil and water densities are initialized in such a way that an initial number of circular oil droplets, $\mathrm{N}^{\mathrm{init}}_{\mathrm{O}}$, corresponding to the desired volume fraction of the dispersed phase, $\phi$, are arranged in a honeycomb-like configuration. The number of initial droplets ranges from $\mathrm{N}^{\mathrm{init}}_{\mathrm{O}}=220$ for the most dilute emulsion ($\phi=0.16$) to $\mathrm{N}^{\mathrm{init}}_{\mathrm{O}}=700$ for the most packed emulsion ($\phi=0.84$). To create a slightly polydisperse emulsion, we take the honeycomb-like configuration, we add a small random perturbation to the initial centers-of-mass positions of the droplets and the density field of the continuous phase, and then we leave the emulsion free to relax, with no applied buoyancy force, towards its more energetically favorable configuration~\cite{TLBfind22} (see Fig.~\ref{fig:sketch}). We compute the volume fraction of the initially dispersed phase, $\phi$, as the fraction of the domain size initially occupied by the oil phase
\begin{equation}
\phi = \frac{A_{\mathrm{O}}}{A_{\mathrm{tot}}},
\end{equation}
where $A_{\mathrm{O}}$ is the area occupied by the oil phase, computed as in Refs.~\cite{PelusiSM21,PelusiSM23}, and $A_{\mathrm{tot}}=L \times H$. After this preparation step, buoyancy forces are added. In a homogeneous fluid, an important parameter that controls the various dynamical regimes of thermal convection is the Rayleigh number~\cite{Chandrasekhar61,Bodenschatz2000,Ahlers09,Lohse10,Chilla12}
\begin{equation}\label{eq:Ra}
\Ra = \frac{\alpha g \Delta T H^3}{\nu \kappa} ,
\end{equation}
where $\alpha$ is the thermal expansion coefficient. Other parameter are the Prandtl number $\Pr = \nu/\kappa$, and the aspect-ratio $\Gamma = L/H$. In particular, the conduction-to-convection transition occurs as $\Ra$ exceeds a critical threshold~\cite{Chandrasekhar61}. In the case of emulsions, however, as mentioned earlier, interfaces affect the viscous/elastic-visco-plastic characteristics of the system response to the forcing. Therefore, $\Ra$, as written in Eq.~\eqref{eq:Ra}, does not carry a uniquely determined dynamical information as $\phi$ increases since the effective viscosity of the emulsion acquires a dependency on the local shear rate and droplet plasticity at mesoscales is enhanced. In other words, since the emulsion is a structured fluid made of droplets at mesoscales, it is not correct to think that the thermal behavior of the emulsion can be obtained from that of a homogeneous system with a different viscosity. For these reasons, in the following, we consider the kinematic viscosity of the corresponding homogeneous system in the definition of both $\Ra$ and $\Pr$, and we keep $\Ra$ as a dimensionless measure of the imposed forcing. Another relevant parameter is the Weber number, defined as $\We = \rho U^2 \ell/\Sigma$ (where $\rho$ is the emulsion density, $U$ is a characteristic flow velocity and $\ell$ a typical droplet-scale length, e.g., the droplet mean radius). However, since we keep the surface tension $\Sigma$ constant, the $\We$ values are related to the $\Ra$ values (we will provide more details about it in the next section).
Similarly to previous studies~\cite{Liu21,Liu21a,Liu22,Brandt24,Mangani24}, we do not consider here the temperature dependence of the surface tension. This is indeed justified as in actual expertimental conditions the Marangoni number can be kept small, and then thermocapillary effects are negligible. The modelling of a temperature dependent surface tension and the study of phenomena where this aspect is relevant, are indeed an interesting matter for future works. \\
Taking all together, the choice of having unitary viscosity and thermal diffusivity ratios, together with imposing no-slip boundary conditions, reduces to four the number of parameters that define the behavior of the emulsion: $\Ra$, $\Pr$, $\Gamma$, and $\phi$. To simplify the problem, we fix $\Pr=1$ and $\Gamma=2$ in all the simulations and explore the various emerging dynamical regimes in the parameter space spanned by $\Ra$ and $\phi$.\\ 
To access the heat flux properties, we measure the Nusselt number, $\Nu$, that is the dimensionless ratio between the total and the conductive heat fluxes~\cite{Shraiman90,Ahlers09,Verzicco10,Chilla12}:
\begin{equation}\label{eq:Nusselt}
\Nu = 1 + \frac{\langle u_z T \rangle_{x,z}}{\kappa \frac{\Delta T}{H}}\ ,
\end{equation}
where $u_z$ is the vertical component of the hydrodynamical velocity and $\langle \dots \rangle_{x,z}$ denotes the space average. Notice that when emulsions reach their statistically steady states, the Nusselt number in Eq.~\eqref{eq:Nusselt} exhibits fluctuations in time around a mean value~\cite{PelusiSM21,PelusiSM23},  hereafter indicated as $\overline{\Nu}$ (see also Sec.~\ref{subsec:phase_diagram}):
\begin{equation*}
    \overline{\Nu}=\langle \Nu \rangle_t 
\end{equation*}
where $\langle \dots \rangle_{t}$ denotes the time average.
\subsection{Lattice Boltzmann modelling of emulsions}\label{subsec:lbm}
Simulations of emulsions are performed by employing the open source TLBfind code~\cite{TLBfind22} based on LBMs~\cite{Benzi92,Kruger17,Succi18}. To the best of our knowledge, the LBM approach is so far the only one allowing the simulation of realistic emulsion systems with finite-size droplets and non-trivial rheology~\cite{Benzietal14,Dollet15,LulliBenziSbragaglia18,Negro23,Girotto24,PelusiSM21,PelusiSM23,Pelusi24rheology,Pelusi24intermittent,Tiribocchi25review}. The essential features of the numerical methodology implemented in the TLBfind code are here illustrated; more details on the numerical model can be found in~\cite{TLBfind22}.\\ 
We consider a thermal non-ideal fluid with two components in the framework of the Shan-Chen interaction included in a LBM~\cite{ShanChen93,Sbragaglia07,Benzi09,Sbragagliaetal12}. The emulsion dynamics is numerically studied on a two-dimensional lattice by integrating the lattice Boltzmann equation, describing the evolution of the probability distribution function $f_{\beta, i} ({\bm x},t)$ of finding a fluid particle of the $\beta=\mathrm{O,W}$ phase with kinetic velocity ${\bm c}_i$ at time $t$ in the position ${\bm x} = (x,z)$. We assume unitary lattice spacing $\Delta x$ and time-lapse $\Delta t$. Kinetic velocities are discretized, and the index $i$ runs over a discrete set of 9 lattice velocities ${\bm c}_i$ ($i= 0, 1, \dots, 8)$ with constant Cartesian components (D2Q9 LBM scheme): ${\bm c}_0=(0,0)$, ${\bm c}_{1,3}=(\pm 1,0)$, ${\bm c}_{2,4}=(0,\pm 1)$, ${\bm c}_{5,8}=(+1,\pm 1)$, ${\bm c}_{6,7}=(-1,\pm 1)$. We consider the following lattice Boltzmann equation: 
\begin{equation}\label{eq:LBM}
f_{\beta,i}({\bm x}+{\bm c}_i,t+1) -  f_{\beta,i}({\bm x},t) = \Omega_{\beta,i}({\bm x},t),
\end{equation}
embedding exact streaming dynamics on the lattice supplemented with a relaxation process towards a local equilibrium distribution function $f_{\beta,i}^{(\mathrm{eq})}$ in the Bhatnagar–Gross–Krook (BGK) approximation~\cite{BGK54}:
\begin{equation}
\Omega_{\beta,i}({\bm x},t)=-\frac{1}{\tau}\left[f_{\beta,i}({\bm x},t)-f_{\beta,i}^{(\mathrm{eq})}({\bm x},t)  \right]
\end{equation}
where $\tau$ is a characteristic time of the relaxation process. The local equilibrium distribution function $f_{\beta,i}^{(\mathrm{eq})}$ depends on $({\bm x},t)$ via the coarse-grained
density $\rho_{\beta}=\rho_{\beta}({\bm x},t)$ and equilibrium velocity $\bar{{\bf u}}_{\beta}=\bar{{\bf u}}_{\beta}({\bm x},t)$ fields as
\begin{equation}\label{eq:feq}
\small
f_{\beta,i}^{(\mathrm{eq})}(\rho_{\beta},\bar{{\bf u}}_{\beta})= w_i \rho_{\beta} \left[1+\frac{\bar{{\bf u}}_{\beta} \cdot {\bm c}_{i}}{c_s^2}-\frac{\bar{{\bf u}}_{\beta} \cdot \bar{{\bf u}}_{\beta}}{2 c_s^2} + \frac{(\bar{{\bf u}}_{\beta} \cdot {\bm c_{i}})^2}{2 c_s^4} \right].
\end{equation}
In Eq.~\eqref{eq:feq}, $c_s = 1/\sqrt{3}$ is the characteristic speed of sound (a constant in the model), and $w_i$ represent the lattice weights~\cite{Kruger17,Succi18,TLBfind22}. Coarse-grained density and velocity fields can be related to the zeroth and first-order momenta of the distribution $f_{\beta,i}$ as
\begin{equation}
\begin{split}
    \rho_{\beta}({\bm x},t) = & \sum_{i=0}^{8} f_{\beta,i}({\bm x},t) \ , \\ {\bf u}({\bm x},t) = \frac{1}{\rho({\bm x},t)} & \sum_{\beta} \sum_{i=0}^{8} {\bm c}_i f_{\beta,i}({\bm x},t) \ ,
\end{split}
\end{equation}
where $\rho = \sum_\beta \rho_\beta$ is the total density. Furthermore, the action of internal and external forces enters via the equilibrium velocity
$$
\bar{{\bf u}}_{\beta}={\bf u}+\frac{\tau {\bm F}_{\beta}}{\rho_{\beta}}
$$
where the force term ${\bm F}_{\beta}={\bm F}_{\beta}({\bm x},t)$ encodes both inter-component and intra-component interactions, together with external forces, ${\bm F}_{\beta}={\bm F}^{\mathrm{intra}}_{\beta}+{\bm F}^{\mathrm{inter}}_{\beta}+{\bm F}_{\beta}^{\mathrm{ext}}$. Intra-component forces introduce phase segregation and promote the emergence of interfaces with non-negligible surface tension $\Sigma$~\cite{ShanChen93,Benzi09}:
\begin{equation}\label{eq:F12}
{\bm F}^{\mathrm{intra}}_{\beta}({\bm x},t) = - G_{\text{OW}} \psi_{\beta}({\bm x},t) \sum_{i=0}^{8} w_i \psi_{\beta'}({\bm x}+{\bm c}_i,t) {\bm c}_i \ ,
\end{equation}
where $\beta' \neq \beta$, $\psi_{\beta} = \rho_\beta/\rho_0$ is the pseudo-potential function~\cite{ShanChen93} (with $\rho_0$ being a reference density value) and $G_{\text{OW}}$ is a positive coupling constant controlling the intensity of the surface tension, $\Sigma$. We also introduce a disjoining pressure between approaching interfaces, thus mimicking the presence of surfactants stabilizing the emulsion against droplets' coalescence~\cite{Benzi09,Sbragagliaetal12,Dollet15}. To this aim, we include additional attractive ($a$) and repulsive ($r$) inter-component interactions:
\begin{align}\label{eq:F_stabilization}
{\bm F}^{\mathrm{inter}}_{\beta}({\bm x},t)  = & - G^a_{\beta \beta} \psi_{\beta}({\bm x},t) \sum_{i=0}^{8} w_i \psi_{\beta}({\bm x}+{\bm c}_i,t) {\bm c}_i + \nonumber \\
& - G^r_{\beta \beta} \psi_{\beta}({\bm x},t) \sum_{i=0}^{24} p_i \psi_{\beta} ({\bm x}+{\bm c}_i,t) {\bm c}_i , 
\end{align}
where $G_{\beta \beta}^a < 0$, $G_{\beta \beta}^r > 0$ are coupling constants and the second summation is performed with dedicated weights $p_i$~\cite{Benzi09} on a set of links larger than the standard D2Q9 scheme, thus including also next-to-nearest neighbor lattice directions~\cite{Benzi09,TLBfind22}. The pseudo-potential used in Eq. ~\eqref{eq:F_stabilization} is set to the original form proposed by Shan \& Chen \cite{ShanChen93}, $\psi_{\beta}=\rho_0\,(1-\exp(-\rho_{\beta}/\rho_0))$. Inter-component interactions play a fundamental role since they support the simulation of stable emulsions, which differ from simple mixtures of two immiscible fluids. At the hydrodynamic level, LBMs reproduce the Navier-Stokes equations for the hydrodynamic velocity ${\bm u}={\bf u}+{\bm F}/2 \rho$ with kinematic viscosity, $\nu$, related to the relaxation time $\tau$ as $\nu = c_s^2 (\tau - 1/2)$~\cite{Kruger17,Succi18}. Notice that, although spurious currents are present in the employed LBM~\cite{silva2023effect,silva2024lattice}, they do not affect the observed phenomenology.\\
The Navier-Stokes equations are coupled with the advection-diffusion dynamics of a temperature field $T=T({\bm x},t)$. Specifically, in the momentum equation, a buoyancy term in the Boussinesq approximation~\cite{Spiegel60} is added as
\begin{equation}\label{eq:F_ext}
{\bm F}_{\beta}^{\mathrm{ext}}({\bm x},t) = - \rho_\beta({\bm x},t) \alpha {\bm g} T({\bm x},t) .
\end{equation}
To reproduce the advection-diffusion dynamics for the temperature field, we introduce an additional probability distribution function, $h_i({\bm x},t)$, which evolves following a dedicated lattice Boltzmann equation~\cite{Kruger17,Succi18}:
\begin{equation}\label{eq:LBM_thermal}
\small
h_{i}({\bm x}+{\bm c}_i,t+1) - h_{i}({\bm x},t) = -\frac{1}{\tau_h} \left[h_{i}({\bm x},t)-h_{i}^{(\mathrm{eq})}({\bm x},t) \right] .
\end{equation}
In Eq.~\eqref{eq:LBM_thermal}, $\tau_h$ defines the relaxation time for the temperature dynamics. Then, the temperature field can be computed as the zeroth order momentum of the distribution function
\begin{equation}
    T({\bm x},t)=\sum_{i=0}^{8} h_i({\bm x},t) ,
\end{equation}
whereas the equilibrium distribution $h_{i}^{(\mathrm{eq})}$ depends on $({\bm x},t)$ via the temperature $T=T({\bm x},t)$ and the hydrodynamic velocity ${\bm u}={\bm u}({\bm x},t)$ as
\begin{equation}\label{eq:teq}
h_{i}^{(\mathrm{eq})}(T,{\bm u})= w_i T \left[1+\frac{{\bm u} \cdot {\bm c}_{i}}{c_s^2}-\frac{{\bm u} \cdot {\bm u}}{2 c_s^2} + \frac{({\bm u} \cdot {\bm c_{i}})^2}{2 c_s^4} \right].
\end{equation}
At the hydrodynamic level, the LBM scheme for the distribution $h_i$ solves the advection-diffusion equation for the temperature $T$, where the advecting velocity is set by ${\bm u}$ and the thermal diffusivity, $\kappa$, is related to $\tau_h$ as $\kappa = c_s^2(\tau_h-1/2)$.\\
The system domain is resolved with $L \times H = 2048 \times 1024$ lattice points and we used $\tau=\tau_h=1$ in Eqs.~\eqref{eq:LBM} and~\eqref{eq:LBM_thermal}. In all simulations, multi-component model parameters are fixed to: $\rho^{\mathrm{max}}_\mathrm{O,W} = 1.18$, $\rho^{\mathrm{min}}_\mathrm{O,W} = 0.18$, $\rho_0 = 0.83$, $G_{\mathrm{OW}} = 0.405$, $G^a_{\mathrm{WW}} = -9.0$, $G^a_{\mathrm{OO}} = -8.0$, $G^r_{\mathrm{WW}} = 8.1$, $G^r_{\mathrm{OO}} = 7.1$. At the walls, we impose bounce-back boundary conditions for fluid components and Dirichlet boundary conditions for the temperature field. Numerical simulations have been performed on Nvidia V100 and A30 GPUs. We performed about 300 numerical simulations, each typically requiring $\sim48$ GPU-h of elapsed time. The duration of these simulations in terms of the characteristic free-fall time $t_{FF} \sim \sqrt{H/(\alpha g \Delta T)}$ depends on the amplitude of the applied buoyancy force. It lays in the range between $t_{\mathrm{FF}} \approx 100$ ($\Ra \approx 4 \times 10^4$) and $t_{\mathrm{FF}} \approx 4.5 \times 10^{3}$ ($\Ra \approx 8 \times 10^6$). Time-averaged statistics in the statistically steady state (see Sec.~\ref{subsec:phase_diagram}) is performed considering intervals of time in the range [$15 \div 2000] \,t_{\mathrm{FF}}$, depending on $\Ra$. To analyze droplet sizes (see Sec.~\ref{sec:droplet_size_statistics}), we collect data considering all droplets at any time once the statistically steady state is reached.\\
Throughout the paper, all dimensional quantities will be given in lattice Boltzmann simulation units.
\section{Dynamical regimes}\label{sec:dynamic-regimes}
In this section, we provide a detailed view of the dynamical regimes of the emulsions under thermal convection at varying both $\phi$ and $\Ra$. In Sec.~\ref{subsec:phase_diagram}, we report on the properties of statistically steady states by also characterizing the associated rheology and the heat flux properties, whereas in Sec.~\ref{subsec:transient}, we report on the transient dynamics and fluctuations in time in the heat flux.
\subsection{Statistically steady states}\label{subsec:phase_diagram}
Previous works~\cite{PelusiSM21,PelusiSM23} showed that the dynamical behavior of concentrated emulsions in a Rayleigh-B\'enard setup is strongly affected by both $\phi$ and $\Ra$. However, situations where the emulsion structure changes during its dynamical evolution via breakup or coalescence events have not been considered, thus focusing only on convective states preserving the number of droplets of the initial condition. In this work, we take a major step forward and complete that analysis by spanning over $\phi \in [0,0.84]$ and $\Ra \in [4 \times 10^3, 8 \times 10^6]$ (correspondingly the range 
of Weber number is, roughly, $\We \in [10^{-3},10]$) with no discrimination on the structure of the emulsion. For each pair $(\phi, \Ra)$, we classify the statistically steady state in terms of the heat flux characteristics (i.e., conduction or convection) and the structural changes in the emulsions, as outlined in Fig.~\ref{fig:sketch}. Thus, we can distinguish between states of i) conduction ($\Nu=1$), whereby the emulsion structure is preserved, ii) {\it stable} convection ($\Nu>1$), with preserved emulsion structure (i.e., neither droplet breakup nor coalescence is detected), iii) {\it breakup-dominated} convection, iv) {\it coalescence-dominated} convection, and v) convection in emulsions that undergo a {\it phase inversion}. In this scenario, situations explored in Refs.~\cite{PelusiSM21,PelusiSM23} lay in the stable region outlined by $0.1 \lesssim \phi \lesssim 0.7$ and $10^{4} \lesssim \Ra \lesssim 6 \times 10^{4}$. Indeed, the structural changes take place at relatively high values of $\Ra$, where strong convective flows trigger continuous breakup and coalescence of droplets, eventually leading to a dynamical equilibrium characterized by a mean droplet size $r_m$. Notice that, in multiphase turbulent flows, $r_m$ corresponds to the so-called Hinze's scale $r_H$~\cite{Hinze1955,Anderson2006}, which decreases with the energy dissipation rate, hence with $\Ra$. 
\begin{figure*}[t!]
    \centering
    \includegraphics[width=0.95\textwidth]{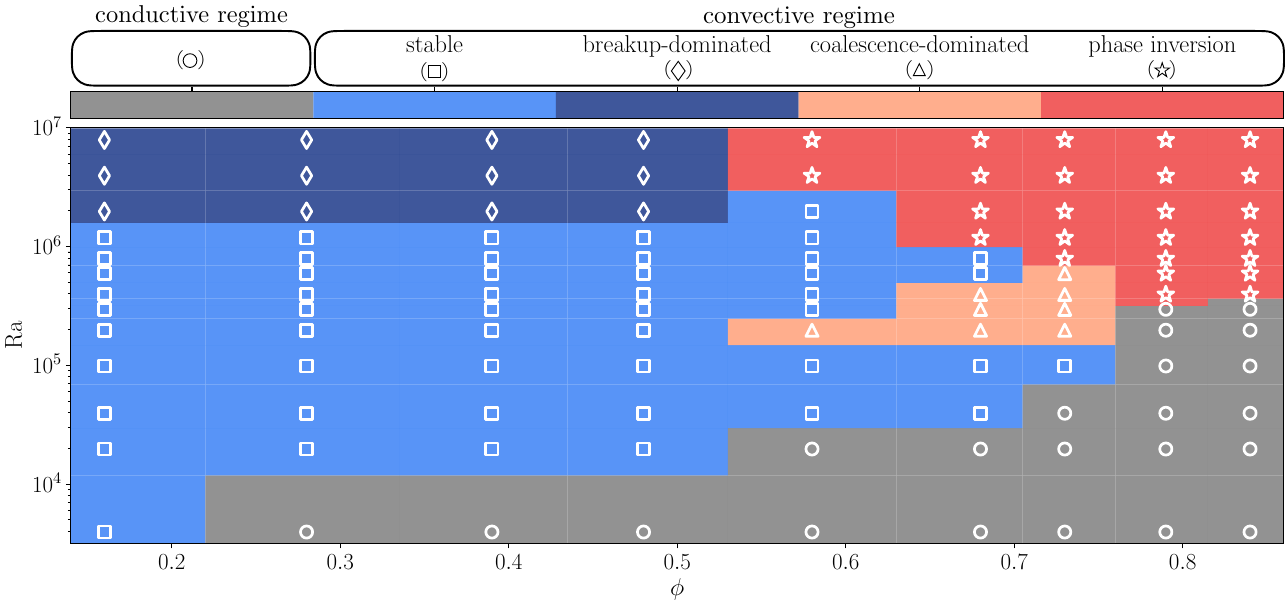}
    \caption{Phase diagram reporting a characterization of statistically steady states for different combinations of $\Ra$ and $\phi$ (same colors as Fig.~\ref{fig:sketch}).\label{fig:phase_diagram}}
\end{figure*}
For low-to-moderate values of $\phi$, if the initial mean droplet size is larger than $r_m$ then breakup overcomes coalescence in the initial transient, due to the low droplet-droplet collision rate, whence we call this regime ``breakup-dominated". Analogously, there exists a range in the pair $(\phi,\Ra)$ where the balance in the transient state is broken in favor of coalescence, thus producing ``coalescence-dominated" scenarios with fewer larger droplets compared to the initial condition. For emulsions with $\phi$ beyond the jamming point, a sustained convective state is only possible if $\Ra$ is high enough to trigger a phase inversion in the emulsion, as anticipated in Ref.~\cite{Pelusi24intermittent}.
In Fig.~\ref{fig:phase_diagram}, we report the phase diagram of the above-mentioned dynamical regimes. Although we conducted a large number of simulations, they were still not enough to discern whether the transition regions between different regimes are smooth or sharp; therefore, the transition lines drawn in Fig.~\ref{fig:phase_diagram} have just been used to facilitate the localization of the different regimes in the $(\phi,\Ra)$ phase diagram. Still, they have not to be understood as sharp transition lines. 
Note that at a fixed value of $\phi$ around the jamming point, the system's behavior as $\Ra$ increases is experiencing a ``non-monotonic" sequence: initially, stable convection is observed, which then goes through the coalescence-dominated regime before becoming stable again, eventually leading to phase inversion at higher values of $\Ra$. This non-monotonic behavior arises because of the appearance of the coalescence-dominated regime, which is a precursor of the phase-inverted states, as discussed further in Sec.~\ref{subsec:transient}.
\\\\
A fundamental aspect is the impact of thermal dynamics on the emulsion rheology. For this purpose, it is necessary to investigate the rheology of the emulsions that reach a statistically steady state compared to the rheology of the emulsions after the preparation step, i.e., before the buoyancy forces are switched on. Hence, we performed simulations with a standard Couette rheometer setup without buoyancy forces~\cite{TLBfind22}, where walls are moved with constant and opposing velocities along the $x$-axis, ${\bm u}_{\mathrm{wall}} = (u_{x} (x,z = \pm H/2,t) = \pm u_{w},0)$. Thus, we impose a shear rate $\dot{\gamma} = 2 u_w/H$ on the emulsion, and then we measure the shear stress $\sigma$ after the system has reached the statistically steady state. The corresponding rheological flow curves are reported in Fig.~\ref{fig:rheology_initial_cond}. Notice that data for $\phi=0.79$ coincide with the ones reported in the Supplementary Material of Ref.~\cite{Pelusi24intermittent}, and we stress again the good match with the Herschel–Bulkley (HB) law (dashed black line)
\begin{equation}\label{eq:HB_law}
\sigma = \sigma_0 + a \dot{\gamma}^n \ ,
\end{equation}
with parameters $\sigma_0  = 2.89 \times 10^{-5}$ (i.e., the yield stress), $a = 2.2 \times 10^{-2}$ (i.e., the consistency index), and $n = 0.577$ (i.e., the flux index), thus marking the emergence of a yield stress for high values of $\phi$. We then performed a rheological characterization of the configurations of the statistically steady states. Specifically, we captured the latest emulsion configuration of some representative pairs $(\phi, \Ra)$ pinched from the phase diagram, switched off the buoyancy forces, and performed a rheological experiment with the Couette rheometer setup described above.
\begin{figure*}[t!]
    \centering    
    \includegraphics[width=0.95\textwidth]{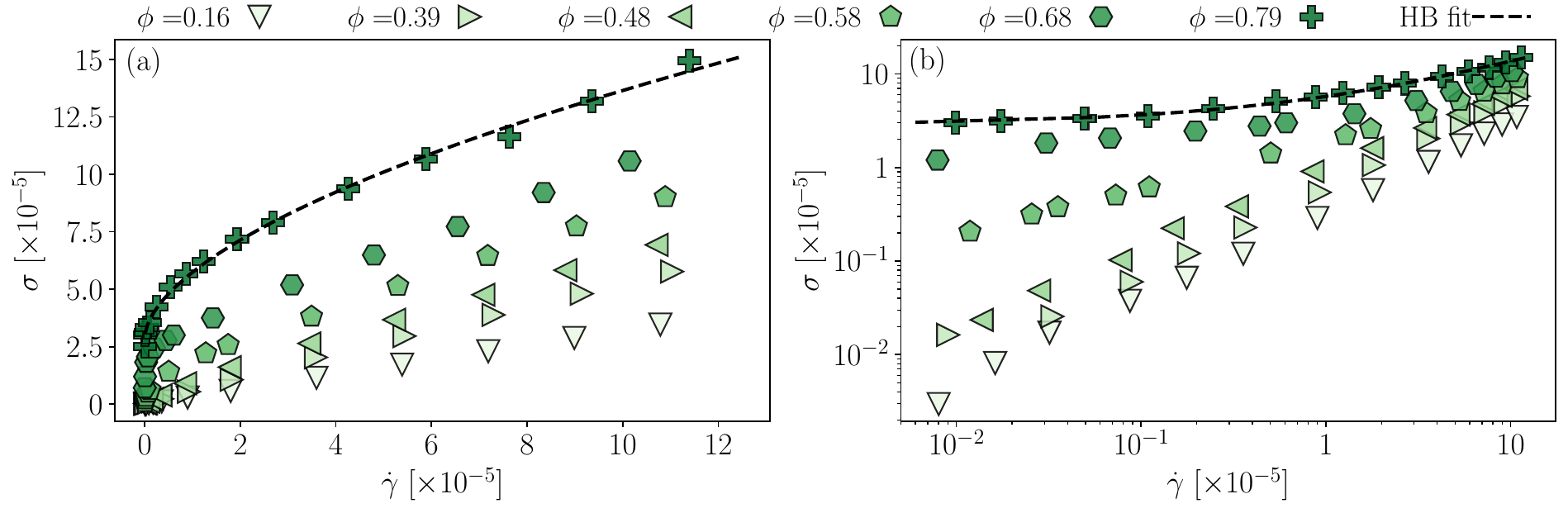}
    \caption{Rheological characterization of emulsions after the preparation step, before buoyancy forces are switched on. We report the shear stress $\sigma$ as a function of the shear rate $\dot{\gamma}$. Panel (a) refers to the lin-lin representation, whereas panel (b) refers to the log-log one. Different symbols/colors correspond to different values of $\phi$ (increasing values from lighter to darker colors). A fit with the Herschel-Bulkley (HB) law (cfr. Eq.~\eqref{eq:HB_law}) is drawn for $\phi=0.79$ (dashed black line).}\label{fig:rheology_initial_cond}
\end{figure*}
\begin{figure*}[t!]
    \centering    \includegraphics[width=1.\textwidth]{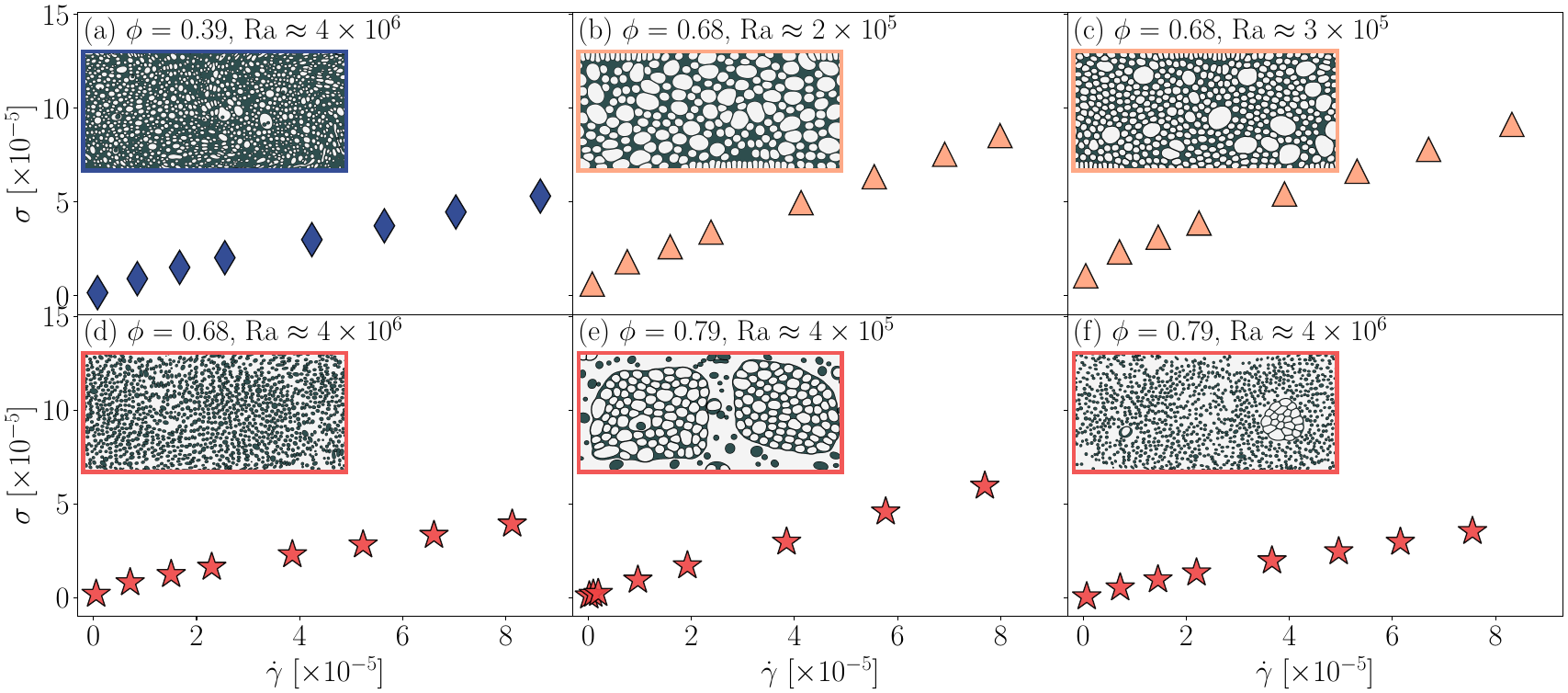}
    \caption{Rheological characterization of emulsions in the statistically steady state for different combinations of $\phi$ and $\Ra$ (same symbols/colors as Fig.~\ref{fig:phase_diagram}). Density maps of the corresponding emulsions are also reported.}
    \label{fig:rheology}
\end{figure*}
\begin{figure*}[t!]
    \centering    \includegraphics[width=.8\textwidth]{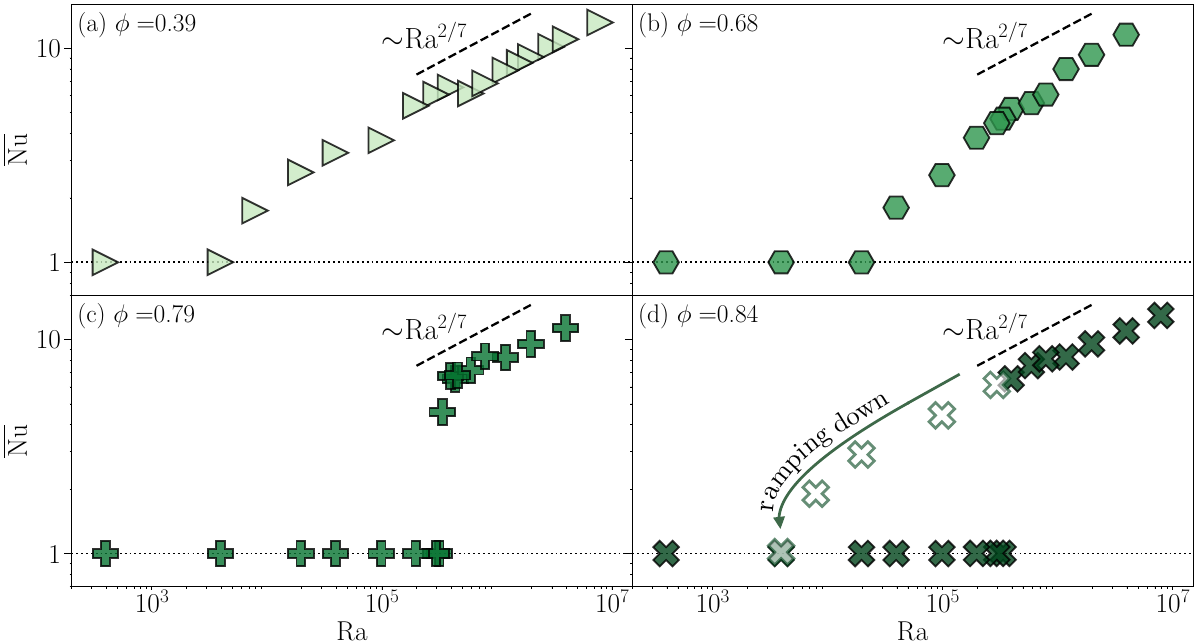}
    \caption{Characterization of the heat flux in the statistically steady state in terms of the Nusselt number averaged in time over the statistically steady state, $\overline{\Nu}$, as a function of $\Ra$, for different values of $\phi$ (same symbols/colors as Fig.~\ref{fig:rheology_initial_cond}). All filled symbols refer to statistically steady states obtained starting from the emulsion at rest and then applying the buoyancy forces. Open crosses in panel (d) refer to statistically steady states obtained starting from an emulsion configuration resulting from a convective statistically steady state and ramping down the intensity of buoyancy forces.}
    \label{fig:Nu_vs_Ra}
\end{figure*}
We report the corresponding rheological flow curves in Fig.~\ref{fig:rheology}. The flow curve for the breakup-dominated regime (panel (a)) presents an increase of about $20\%$ of the stress at fixed shear rate $\dot{\gamma}$ due to the increase in the number of droplets and, in turn, of the total interfacial energy. Since the breakup-dominated regime comes up only for dilute emulsions, rheological Newtonianity persists. For all other dynamical regimes, i.e., the coalescence-dominated regimes (panels (b) and (c)) and the phase-inverted emulsions (panels (d)-(f)), the activation of the buoyancy force causes a transition from a non-Newtonian to a Newtonian behavior. In the coalescence-dominated regime, the Newtonian behavior is triggered by a decrease in the number of droplets due to coalescence; in the phase inversion regime, phase inversion generates a Newtonian dilute W/O emulsion in marked contrast with the initial non-Newtonian O/W emulsion. To summarize, the resulting emulsions are always Newtonian despite the structural variations induced by the convective states. It is important to remark that this Newtonian behavior cannot be predicted {\it a priori} since each pair $(\phi, \Ra)$ leads to different emulsion structures regarding the number and size of droplets.\\\\ 
As a further step, we investigated how the aforementioned non-Newtonian-to-Newtonian transition at increasing values of $\Ra$ is reflected in the heat flux properties. We considered the Nusselt number averaged in time over the statistically steady state, $\overline{\Nu}$.  
Figure ~\ref{fig:Nu_vs_Ra} shows its relation with $\Ra$ for different values of $\phi$. In all cases (filled symbols), the statistically steady state is reached starting from the emulsion prepared in a static case (with no buoyancy forces) and then applying a given buoyancy amplitude corresponding to the desired $\Ra$. We observe that at increasing $\phi$, the transition to a convective state ($\overline{\Nu} >1$) occurs at higher
values of $\Ra$, as expected since the emulsion becomes more viscous at increasing $\phi$. We recall that $\Ra$, which we introduced as an appropriate dimensionless buoyancy force, is defined in terms of the ``bare" kinematic viscosity, whereas the ``effective" (dynamically relevant) viscosity grows with $\phi$. Indeed, for what concerns the convective heat flux,
emulsions with $\phi$ below jamming behave essentially as a single-phase fluid (but for the shift in the nominal value of the critical $\Ra$), with $\overline{\Nu}$ growing continuously with $\Ra$ above transition. For higher values of $\phi$ (i.e., when the emulsion exhibits yield stress), then the transition to convection is discontinuous, 
with a jump in $\overline{\Nu}$ witnessing the sudden structural transition from a jammed (conductive) O/W emulsion to a (convective) phase-inverted diluted W/O emulsion whose effective viscosity is reduced to the one dictated by the volume fraction of the water phase. 
Furthermore, the transition to a phase-inverted state is irreversible, and once the phase inversion is achieved, the system persists in such a Newtonian state. This fact is naturally accompanied by hysteresis in the $\overline{\Nu}$ {\it vs.}\,$\Ra$ curve: starting from an emulsion configuration resulting from a convective state with $\overline{\Nu}>1$ and then ramping down the intensity of buoyancy forces, one observes that $\overline{\Nu}$ goes to zero in a relatively smooth way (panel (d)), echoing the behavior of dilute emulsions. Finally, we note that for $\Ra > 5 \times 10^{5}$ the curves exhibit a scaling law $\overline{\Nu} \sim \Ra^\lambda$, with an exponent close to $\lambda \approx 2/7$, which aligns with findings from studies of thermal convection in simple (single phase) Newtonian fluids~\cite{Cioni97,Castaing89,Shraiman90,Ciliberto96,Benzi98,Grossmann99,Grossmann01,Ahlers09,Chilla12,Stevens13}. 
\subsection{Transient dynamics and heat flux fluctuations}\label{subsec:transient}
\begin{figure*}[t!]
    \centering
    \includegraphics[width=1.0\linewidth]{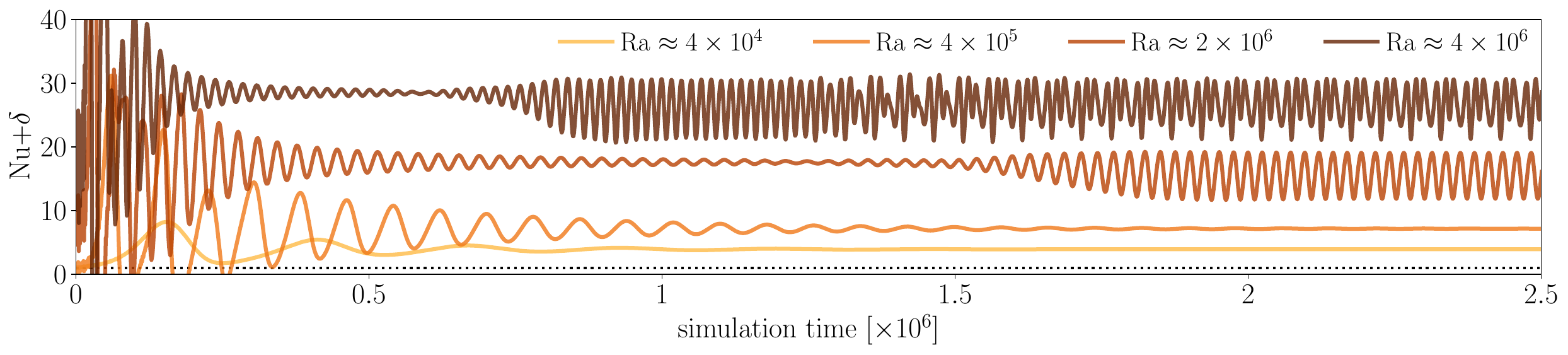}
    \caption{Transient dynamics and heat flux fluctuations in a Newtonian homogeneous fluid. We report the time evolution of the Nusselt number $\Nu$ (cfr. Eq.~\eqref{eq:Nusselt}) for different values of $\Ra$ (increasing values from lighter to darker colors). The dotted black line indicates the conductive regime ($\Nu = 1$). To facilitate readability, data are vertically shifted by a quantity $\delta$ which depends on $\Ra$: $\delta = 0$ for $\Ra \approx 4 \times 10^{4}$ and $\Ra \approx 4 \times 10^{5}$, while $\delta = 7$ for $\Ra \approx 2 \times 10^{6}$ and $\delta = 16$ for $\Ra \approx 4 \times 10^{6}$.}\label{fig:homogeneous}
\end{figure*}

\begin{figure*}[th!]
    \centering
\includegraphics[width=1.0\textwidth]{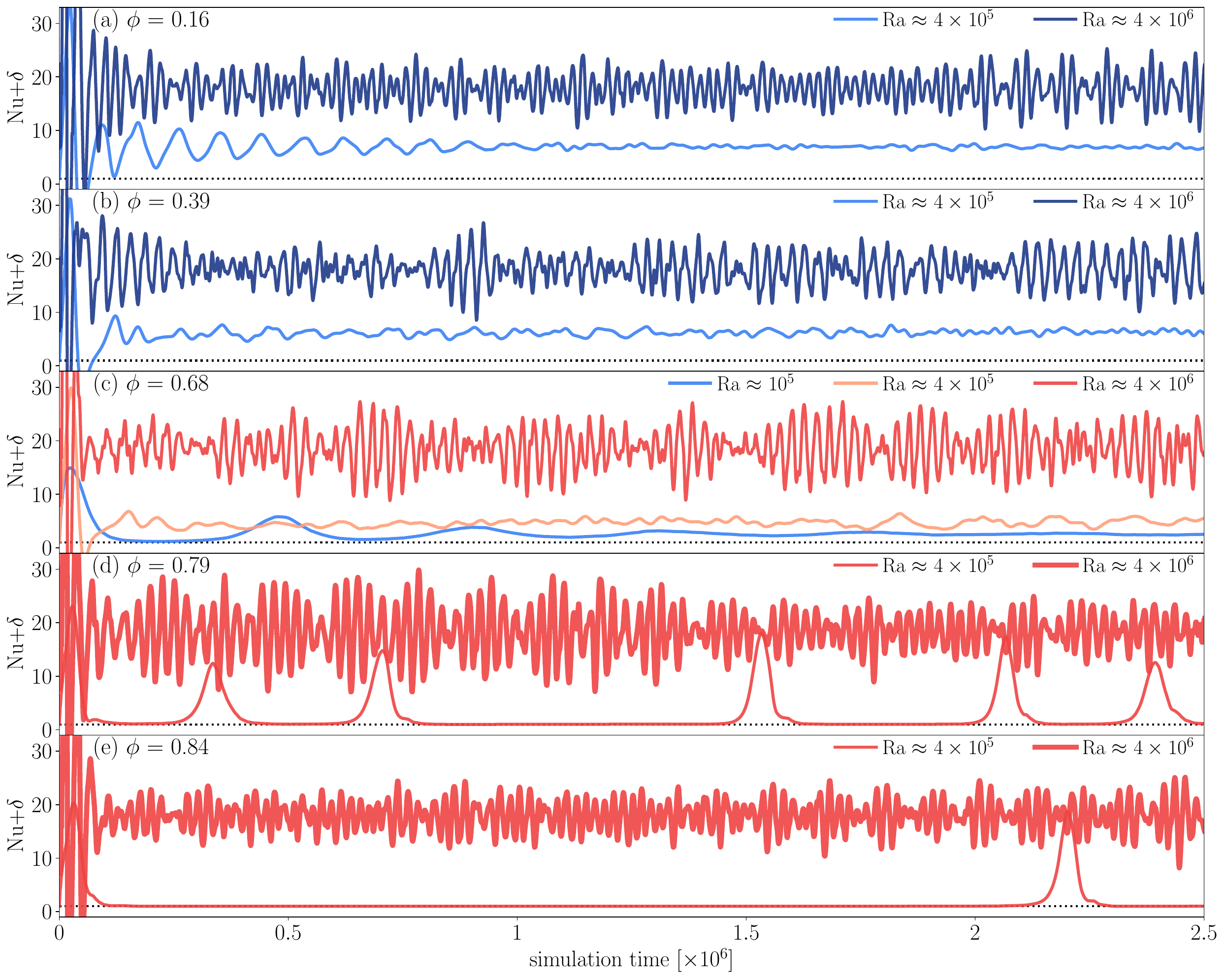}
    \caption{Transient dynamics and heat flux fluctuations in emulsions. We report the time evolution of the Nusselt number $\Nu$ (cfr. Eq.~\eqref{eq:Nusselt}) for different values of $\phi$ (different panels) and $\Ra$ (same colors as Fig.~\ref{fig:phase_diagram}; for the same color, higher values of $\Ra$ are represented with thicker lines). The dotted black line indicates the conductive regime ($\Nu = 1$). To facilitate readability, data are vertically shifted by a quantity $\delta$ which depends on $\Ra$: $\delta = 0$ for $\Ra \approx  10^{5}$ and $\Ra \approx 4 \times 10^{5}$, while $\delta = 7$ for $\Ra \approx 4 \times 10^{6}$.}\label{fig:Nu_time}
\end{figure*}

We complement the statistically steady state characterization of Sec.~\ref{subsec:phase_diagram} with insights into the transient dynamics and heat flux fluctuations. At first, to have a reference case to compare with, we simulated the dynamics of a Newtonian homogeneous fluid under thermal convection in the Rayleigh-B\'enard setup. In Fig.~\ref{fig:homogeneous}, we report the time evolution of the Nusselt number, $\Nu$, for different values of $\Ra$. As expected~\cite{Chandrasekhar61}, for values of $\Ra$ above a critical value, we observe stable convection in a wide range of values of $\Ra$, roughly between $\Ra \approx 2 \times 10^{3}$ and $\Ra \approx 10^{6}$. In this range, the transient dynamics of $\Nu$ shows oscillations whose amplitude decays exponentially over time~\cite{Ecke86}. Above this upper value, oscillations during transient dynamics are not damped anymore but persist with a periodic pattern, a scenario that is symptomatic of the progressive approach to turbulence~\cite{Heslot87}. We then select some representative pairs $(\phi,\Ra)$ and report the corresponding time evolution of $\Nu$ in Fig.~\ref{fig:Nu_time}. In particular, we consider a dilute emulsion showing a breakup-dominated regime ($\phi = 0.16$, panel (a)); a semi-dilute emulsion still showing a breakup-dominated regime ($\phi = 0.39$, panel (b)); a concentrated emulsion, just below the jamming point ($\phi = 0.68$, panel (c)); a jammed emulsion, just above the jamming point ($\phi = 0.79$, panel (d)); a highly packed, jammed emulsion ($\phi = 0.84$, panel (e)). Regular oscillations, which are peculiar to the Newtonian homogeneous fluid (see Fig.~\ref{fig:homogeneous}), tend to acquire a stochastic component when $\phi > 0$, both in the transient and in the statistically steady state. This randomness is rooted in the presence of finite-sized droplets, which collide while transported by the flow. For fixed $\Ra$, fluctuations in $\Nu$ increase at increasing $\phi$: this is particularly evident when $\Ra \approx 4 \times 10^{6}$ and we approach conditions triggering the phase inversion (see panels (a)-(c)). This behavior is similar to what has been observed in a recent study on a Taylor-Couette isothermal system~\cite{Yi24}, where increasing drag fluctuations are observed while approaching the catastrophic phase inversion. Interestingly, signatures of a beat pattern emerge for $\Ra \approx 4 \times 10^{6}$ and are more pronounced for the intermediate values of $\phi$. The combination ($\phi=0.79$, $\Ra \approx 4 \times 10^{5}$) reported in panel (d) coincides with the regime explored in Ref.~\cite{Pelusi24intermittent}, with intermittency characterizing the transient dynamics, where long conductive periods with $\Nu \approx 1$ alternate with intermittent heat bursts where $\Nu \gg 1$, overall resulting in a very long transient dynamics (much longer than the time interval shown in Fig.~\ref{fig:Nu_time}). 

\begin{figure*}[t!]
\centering    \includegraphics[width=.95\textwidth]{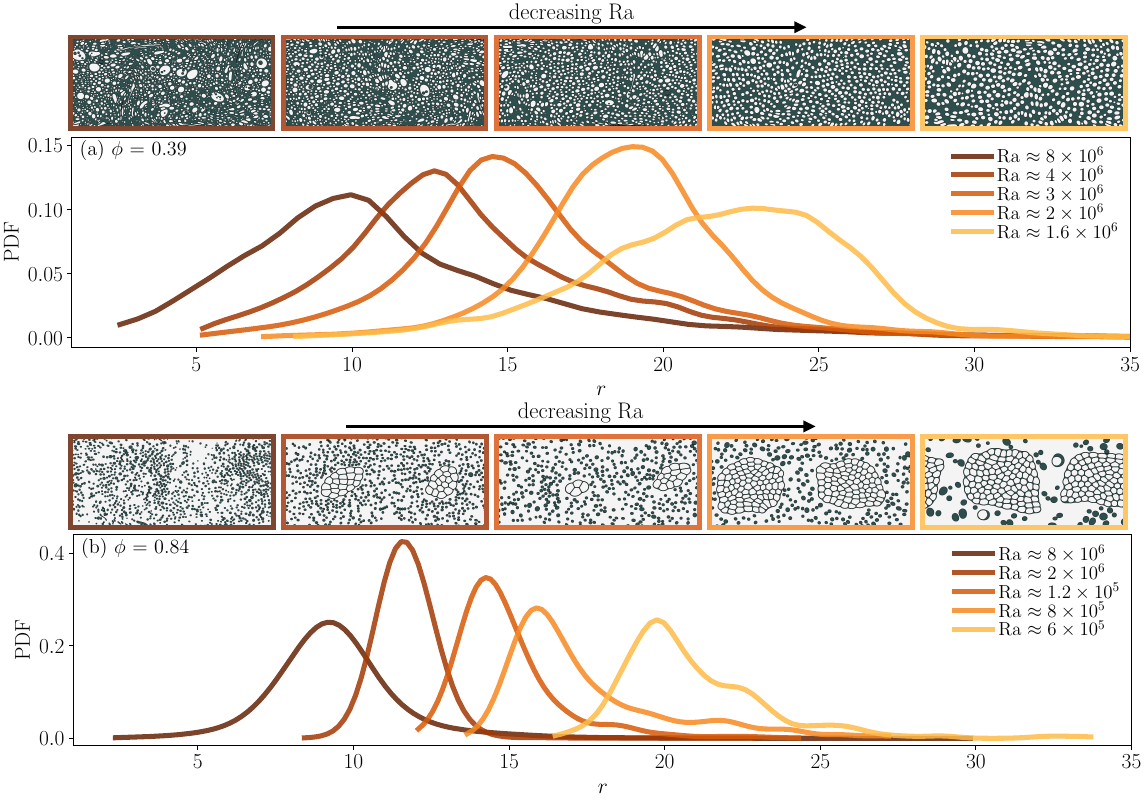}
\caption{Probability distribution functions (PDFs) for the radius of droplets, $r$, for different values of $\phi$ and $\Ra$ (increasing values from lighter to darker colors). Panel (a): PDFs for the radius of oil droplets in a breakup-dominated regime with $\phi=0.39$. Panel (b): PDFs for the radius of water droplets in a phase inversion regime with $\phi=0.84$.}\label{fig:PDF_radius}
\end{figure*}
\begin{figure}[t!]
    \centering
    \includegraphics[width=.475\textwidth]{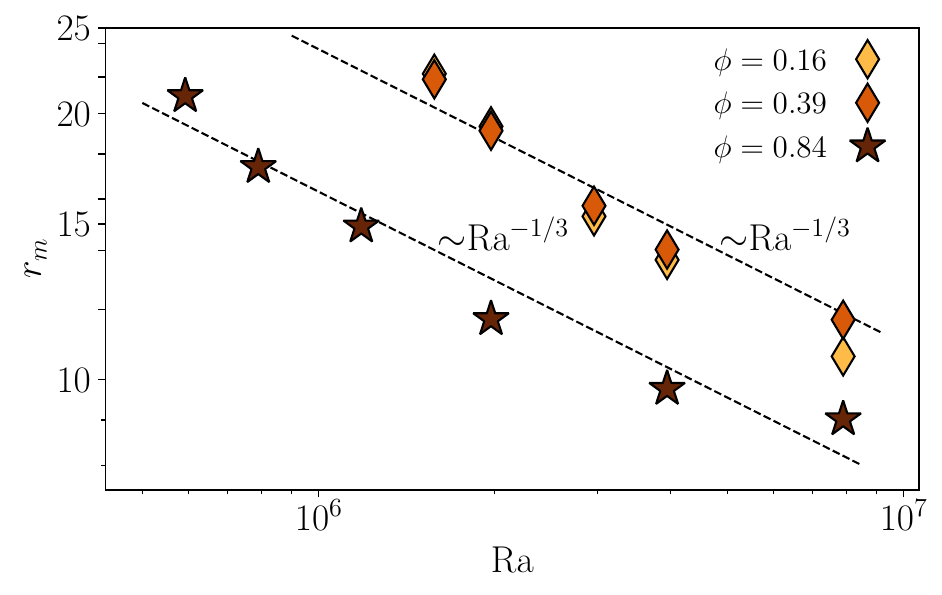}
    \caption{Average value of the droplet radius, $r_m$, as a function of $\Ra$ for different values of $\phi$ (increasing values from lighter to darker colors, same symbols as Fig.~\ref{fig:phase_diagram}). The black dashed lines show a fit to a power-law behavior with exponent $-1/3$.}
    \label{fig:dropsize_exponent}
\end{figure}
\begin{figure*}[t!]
\centering    
\includegraphics[width=.9\textwidth]{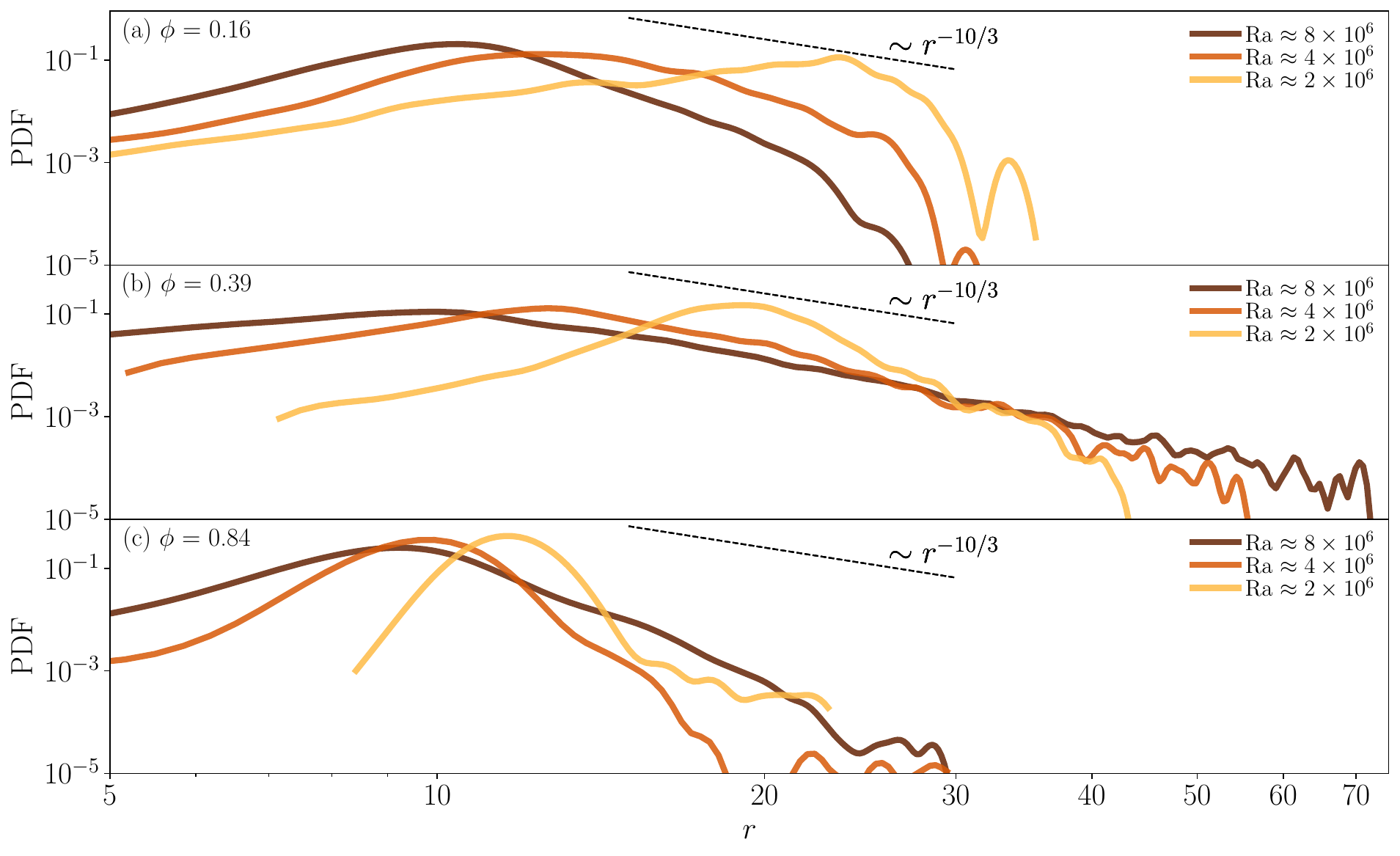}
\caption{Log-log PDFs for the radius of droplets, $r$, for different values of $\phi$ and $\Ra$ (increasing values from lighter to darker colors).  Panel (a): PDF for the radius of oil droplet in a break-up dominated regime with $\phi=0.16$. Panel (b): PDFs for the radius of oil droplet in a breakup-dominated regime with $\phi=0.39$. Panel (c): PDFs for the radius of water droplets in a phase inversion regime with $\phi=0.84$. The dashed black line marks the scaling law with exponent $-10/3$.\label{fig:PDF_log_comparison}}
\end{figure*}
Furthermore, the transient dynamics observed for $\Ra \approx 4 \times 10^5$ at changing $\phi$ provides information on how the intermittent regime starts from dumped oscillations (panel (a)), which become more irregular at increasing $\phi$ (panels (b)-(c)), before entering the intermittent regime (panels (d)-(e)). Overall, the intermittent transient dynamics emerges in a narrow range of values of $\Ra$ and $\phi$, corresponding to highly packed emulsions just above the transition from conduction to convection. As discussed in Ref.~\cite{Pelusi24intermittent}, the intermittency prompts a change in the O/W emulsion structure, whereby heat bursts trigger coalescence events of oil droplets and lead to the phase-inverted W/O emulsion undergoing steady convection. It is noteworthy to observe (see panels (c)-(e) and also Fig.~\ref{fig:phase_diagram}) that the coalescence-dominated regime emerges at values of $\Ra$ comparable to those for which the transient intermittency is observed, but at lower values of $\phi$. Following the phase diagram at constant $\Ra \approx 4 \times 10^{5}$, starting from high values of $\phi$, the intermittent dynamics is progressively lost as $\phi$ decreases, yet coalescence events are likely to occur. However, since the droplets are less tightly packed, their mobility is enhanced, and the emulsion is more prone to enter statistically steady convective states in shorter times without inducing phase inversion (see panel (c)). In other words, for $\Ra \approx 4 \times 10^{5}$ at increasing $\phi$, the coalescence-dominated regime may be seen as a sort of precursor signature of the phase-inverted scenario.\\
Overall, one can conclude that the particular transient path that the dynamical evolution selects for a given volume fraction and buoyancy amplitude depends on the initial condition and, in turn, determines the morphology of the statistically steady state. These results are evident, especially in concentrated regimes ($\phi>0.5$) undergoing phase inversion. Such complex phase-inverted states display distinctive morphological and dynamical characteristics compared to an emulsion prepared at the ``symmetric" volume fraction $1-\phi$. This ``asymmetry" is evident when looking at the transient dynamics at $\Ra \approx 4 \times 10^{5}$ for $\phi=0.16$ and $\phi=0.84$ (see panels (a) and (e)), where the dilute emulsion exhibits an almost regular transient state. In contrast, the emulsion with $\phi=0.84$ exhibits the intermittency described in~\cite{Pelusi24intermittent}. Nevertheless, there are signatures that the asymmetry is progressively lost as $\Ra$ increases: looking at the data for $\Ra \approx 4 \times 10^{6}$, we observe, at least qualitatively, that the behaviors of $\Nu$ as a function of time of these two emulsions are much more similar than the behaviors at lower values of $\Ra$, a fact that we will discuss in more detail when studying droplet size statistics in Sec.~\ref{sec:droplet_size_statistics}.\\
\section{Droplet size statistics}\label{sec:droplet_size_statistics}
The phase diagram shown in Fig.~\ref{fig:phase_diagram} reveals that, for sufficiently high values of $\Ra$, only two statistically steady states of convection are observed: a breakup-dominated regime for low-to-moderate values of $\phi$, and a phase inversion regime for higher values of $\phi$. For a given $\phi$, the system settles into a convective state corresponding to one of these two regimes. However, merely analyzing the heat flux properties in these scenarios does not fully capture the structural changes occurring within the emulsion as $\Ra$ varies. Hence, we performed a detailed investigation by examining the droplet size distribution at the statistically steady state in three representative cases: a semi-dilute emulsion in the breakup-dominated regime ($\phi = 0.39$), a yield-stress emulsion undergoing phase inversion ($\phi = 0.84$) and a dilute emulsion ($\phi=0.16$), which 
represents the symmetrical counterpart of the phase-inverted emulsion with $\phi=0.84$. We analyzed the probability distribution functions (PDFs) for the radius, $r$, of the droplets: for this statistical analysis, we consider the oil droplets for the O/W emulsion with $\phi=0.16, 0.39$ and the water droplets for the W/O phase-inverted emulsion with $\phi=0.84$. In Fig.~\ref{fig:PDF_radius}, we report the PDFs for $\phi=0.39$ and 
$\phi=0.84$.  We also report density maps of the emulsions as $\Ra$ decreases. For 
$\phi=0.39$, droplets of the oil phase are homogeneously distributed in space, and the breakup events become more frequent as $\Ra$ increases, resulting in a distribution of, mostly, many smaller droplets. For the phase-inverted emulsion, we observe that the phase inversion becomes progressively more and more homogeneous as $\Ra$ increases, with a reduction in the size of the regions occupied by the jammed O/W emulsion. Moreover, the size of the newborn droplets of the water phase changes, producing droplets of smaller size at increasing $\Ra$. Overall, both PDFs share the qualitative feature of a decreasing average value of the droplet size, $r_m$, at increasing $\Ra$. This behavior can be understood from a force balance perspective, whereby the turbulent stress imparted by the convective flow to a droplet 
competes with the surface tension resisting breakup. Put in equations, this means that the velocity 
fluctuation (squared) at the droplet scale, $(\delta_{r_m} u)^2$, must balance the Laplace pressure,
$\Sigma/r_m$, i.e., 
\begin{equation} \label{eq:balance}
(\delta_{r_m} u)^2 \sim \Sigma/r_m.
\end{equation}
Since the Reynolds number is not exceptionally high to justify a fully developed turbulent cascade, the velocity fluctuation can be estimated from the large-scale shear as $\delta_{r_m} u \sim (U_{FF}/H)r_m$, where $U_{FF} = H/t_{FF} \sim \sqrt{\alpha g \Delta T H}$ is the free-fall velocity~\cite{Chilla12}.
Inserting such expression in Eq.~\eqref{eq:balance}, we get $(\alpha g \Delta T/H) r_m^3 \sim \Sigma$, 
whence
\begin{equation}\label{eq:scaling}
r_m \sim (\Sigma \, H)^{1/3} \, (\alpha g \Delta T)^{-1/3} \propto \Ra^{-1/3}.
\end{equation}
This prediction is borne out by analyzing the data on the average droplet size extracted from the PDFs. Results are reported in Fig.~\ref{fig:dropsize_exponent}, where the data for the dilute emulsion at $\phi=0.16$ are also shown.
For all three cases, the scaling law in Eq.~\eqref{eq:scaling} is verified for almost a decade in $\Ra$. 
We remark that, in the Hinze's framework~\cite{Hinze1955}, the prefactor in Eq.~\eqref{eq:balance} is related to the critical Weber number for the breakup, $\We_{\text{crit}}$. Such critical value generally depends on the flow pattern around the breaking drops. The intrinsically inhomogeneous and anisotropic nature of flows in the Rayleigh-B\'enard setup and the different paths followed to reach the statistically steady state by the emulsions
at $\phi=0.16,0.39$ and by the emulsion at $\phi=0.84$ may then lead to different values of $\We_{\text{crit}}$ and account for the different prefactors observed in Fig.~\ref{fig:dropsize_exponent}. In Fig.~\ref{fig:PDF_log_comparison}, we show a log-log plot of the PDFs for the same volume fractions considered in Fig.~\ref{fig:dropsize_exponent}.
For the emulsion with $\phi=0.39$, we observe a power-law behavior $\sim r^{-10/3}$ for high values of $\Ra$. This agrees with dimensional arguments~\cite{Deane02} and with previous studies of drop-laden turbulence~\cite{Deane02,Mukherjee19,Soligo19,CrialesiEsposito23interaction,Roccon23} and multi-component flows in a Rayleigh-B\'enard convection~\cite{Liu21}. The same scaling law, however, is not observed for the more dilute emulsion with $\phi=0.16$. To understand why this happens, one must recall that the argument justifying the $-10/3$ scaling, originally devised for the bubbles size distribution in the superficial layers of the ocean, assumes a constant injection of relatively large bubbles being acted upon by turbulent fluctuations, thus breaking up into smaller bubbles which give rise to the PDF tail~\cite{Garrett2000}. 
To sustain this breakup-driven process, some mechanism that refills the distribution at large sizes is required. In the case of emulsions, where a net influx entraining the dispersed phase from the environment is absent, the supply of large droplets is provided by coalescence events of smaller ones. The system eventually reaches a dynamical equilibrium with a droplet size distribution exhibiting the tail with the $-10/3$ power law. In emulsions, though, because of the effect of the emulsifier, the dispersed phase must not be too dilute to get a significant coalescence rate (which grows with $\phi$) such as to appreciate the emergence of the $-10/3$ scaling. Notice that, this is at odds with unstabilized immiscible liquid
mixtures, where the coalescence rate is comparatively high, even at low volume fractions~\cite{Roccon23}. Finally, echoing the discussion done for Fig.~\ref{fig:Nu_time}, we note that for the cases $\phi=0.16$ and $\phi=0.84$, the PDF tails become progressively similar at increasing $\Ra$, whereas at lower values of $\Ra$ they differentiate.
\section{Concluding Remarks}\label{sec:conclusions}
On the basis of mesoscale numerical simulations, we have provided the first detailed and comprehensive characterization of the dynamics of O/W emulsions under thermal convection. Emulsions are characterized by finite-sized oil droplets with a positive disjoining pressure at the droplets' interfaces, stabilizing them against coalescence. The main focus of the study is on the interplay between the volume fraction of the initially dispersed phase, $\phi$, and the strength of the buoyancy forces encoded in the Rayleigh number, $\Ra$. Ranging from dilute Newtonian emulsions (low values of $\phi$) to yield-stress packed emulsions (high values of $\phi$), our analysis reveals a broad spectrum of dynamical regimes for different combinations of $\phi$ and $\Ra$, concerning the nature of the convective state and the associated structural changes within the emulsions. The latter feature droplet breakup, coalescence, or phase inversion processes.\\ 
The transition to convective states takes place in a quite asymmetric way: for low-to-moderate values of $\phi$ ($\phi \lesssim 0.5$), which are peculiar of Newtonian emulsions, convection is entered without any structural change, before exhibiting a breakup-dominated dynamics at higher values of $\Ra$. In marked contrast, at high values of $\phi$ ($\phi \gtrsim 0.5$), which are peculiar to non-Newtonian emulsions, a convective state is entered by a phase inversion process featuring intermittent transient states that favor droplet coalescence and the transition from a non-Newtonian O/W emulsion to a Newtonian W/O emulsion. This asymmetry is a distinctive feature of actual emulsions and is not expected to be observed in liquid mixtures without emulsifiers~\cite{Brandt24}. We also remark that the asymmetry is progressively lost at higher values of $\Ra$, where droplet size statistics, either for the oil phase in the O/W emulsions or for the water phase in the W/O emulsions, appears similar, although referred to complementary phases.\\
We emphasize that all emulsions studied in this work are prepared in a stable state, with a given surface tension at the interface of the droplets and a non-negligible disjoining pressure, effectively hindering droplet coalescence. As discussed in the text, both the surface tension and the disjoining pressure are considered as independent of the temperature. Hence, the presented scenarios of dynamical regimes under thermal convection have to be considered valid for the employed intensity of disjoining pressure and surface tension: we expect that the boundaries of each region in the phase diagram $\Ra$ {\it vs.} $\phi$ will be modified by a variation of these intensities and/or a possible dependence of these quantities on the temperature.\\ 
Actually, the stability induced by the disjoining pressure is a crucial distinction between a simple mixture of two immiscible fluids and a proper emulsion: we conducted preliminary numerical simulations by removing the disjoining pressure in our simulations while leaving the same surface tension at the droplet interface: for very dilute cases ($\phi \rightarrow 0$) the disjoining pressure has no effect at all, as expected, because of the reduced number of droplet interactions; for semi-dilute cases ($\phi \approx 0.4,0.5$) we preliminarily observed that the system structure, as well as the statistical features on droplet size, are manifestly affected. This surely deserves a dedicated study in the future. Another interesting aspect to investigate as a future perspective is the possibility to study convective dynamics of emulsions where droplets plasticity at mesoscales is introduced via fluidity-models~\cite{Goyon08,benzi2021continuum,benzi2021stress}, i.e., effective single fluid models with a non-local constitutive relation between stress and shear rate.\\
We also remark that a peculiar aspect of our study is a thorough characterization of the phase-inverted states; other recent studies focused on the phase inversion process triggered by homogeneous isotropic turbulence with imposed steady forcing~\cite{Girotto24,Yi24}. Our analysis in thermally convective flows revealed new properties of the phase inversion close to the thermal instability point, marking the transition from conduction to convection. In particular, phase inversion manifests via an irreversible discontinuous jump in the average heat flux $\overline{\Nu}$ as a function of $\Ra$. This fact further supports the view that phase inversion is a complex out-of-equilibrium phenomenon bearing similarity with critical-like systems~\cite{Yi24}.
\\ 
From a more general perspective, our study offers inspiring ideas for future research. Indeed, we remark that our simulation suite~\cite{TLBfind22} is equipped with a Lagrangian tracking tool for droplets. Hence, one could envisage future research studies where simulations are used to generate ground-truth data for AI studies on the dynamics of deformable objects in complex convective flows. Finally, we argue that our simulations offer inspiring insights to envisage novel experiments on thermally convective emulsions and push the field of thermally driven multiphase flows into novel and exciting directions.
\section*{Acknowledgements}
FP, MS, and MB acknowledge the support of the National Center for HPC, Big Data and Quantum Computing, Project CN\_00000013 – CUP E83C22003230001 and CUP B93C22000620006, Mission 4 Component 2 Investment 1.4, funded by the European Union – NextGenerationEU. This work received funding from the European Research Council (ERC) under the European Union's Horizon 2020 research and innovation programme (grant agreement No 882340). \\

\bibliographystyle{apsrev4-2}
\bibliography{francesca}

\begin{thebibliography}{126}%
\makeatletter
\providecommand \@ifxundefined [1]{%
 \@ifx{#1\undefined}
}%
\providecommand \@ifnum [1]{%
 \ifnum #1\expandafter \@firstoftwo
 \else \expandafter \@secondoftwo
 \fi
}%
\providecommand \@ifx [1]{%
 \ifx #1\expandafter \@firstoftwo
 \else \expandafter \@secondoftwo
 \fi
}%
\providecommand \natexlab [1]{#1}%
\providecommand \enquote  [1]{``#1''}%
\providecommand \bibnamefont  [1]{#1}%
\providecommand \bibfnamefont [1]{#1}%
\providecommand \citenamefont [1]{#1}%
\providecommand \href@noop [0]{\@secondoftwo}%
\providecommand \href [0]{\begingroup \@sanitize@url \@href}%
\providecommand \@href[1]{\@@startlink{#1}\@@href}%
\providecommand \@@href[1]{\endgroup#1\@@endlink}%
\providecommand \@sanitize@url [0]{\catcode `\\12\catcode `\$12\catcode
  `\&12\catcode `\#12\catcode `\^12\catcode `\_12\catcode `\%12\relax}%
\providecommand \@@startlink[1]{}%
\providecommand \@@endlink[0]{}%
\providecommand \url  [0]{\begingroup\@sanitize@url \@url }%
\providecommand \@url [1]{\endgroup\@href {#1}{\urlprefix }}%
\providecommand \urlprefix  [0]{URL }%
\providecommand \Eprint [0]{\href }%
\providecommand \doibase [0]{https://doi.org/}%
\providecommand \selectlanguage [0]{\@gobble}%
\providecommand \bibinfo  [0]{\@secondoftwo}%
\providecommand \bibfield  [0]{\@secondoftwo}%
\providecommand \translation [1]{[#1]}%
\providecommand \BibitemOpen [0]{}%
\providecommand \bibitemStop [0]{}%
\providecommand \bibitemNoStop [0]{.\EOS\space}%
\providecommand \EOS [0]{\spacefactor3000\relax}%
\providecommand \BibitemShut  [1]{\csname bibitem#1\endcsname}%
\let\auto@bib@innerbib\@empty
\bibitem [{\citenamefont {Barnes}(1994)}]{Barnes94}%
  \BibitemOpen
  \bibfield  {author} {\bibinfo {author} {\bibfnamefont {H.~A.}\ \bibnamefont
  {Barnes}},\ }\href
  {https://www.sciencedirect.com/science/article/pii/092777579302719U}
  {\bibfield  {journal} {\bibinfo  {journal} {Colloids and Surfaces A:
  Physicochemical and Engineering Aspects}\ }\textbf {\bibinfo {volume} {91}},\
  \bibinfo {pages} {89} (\bibinfo {year} {1994})}\BibitemShut {NoStop}%
\bibitem [{\citenamefont {Mason}(1999)}]{Mason99}%
  \BibitemOpen
  \bibfield  {author} {\bibinfo {author} {\bibfnamefont {T.}~\bibnamefont
  {Mason}},\ }\href
  {https://www.sciencedirect.com/science/article/pii/S1359029499000357}
  {\bibfield  {journal} {\bibinfo  {journal} {Current Opinion in Colloid \&
  Interface Science}\ }\textbf {\bibinfo {volume} {4}},\ \bibinfo {pages} {231}
  (\bibinfo {year} {1999})}\BibitemShut {NoStop}%
\bibitem [{\citenamefont {Derkach}(2009)}]{Derkach09}%
  \BibitemOpen
  \bibfield  {author} {\bibinfo {author} {\bibfnamefont {S.~R.}\ \bibnamefont
  {Derkach}},\ }\href
  {https://www.sciencedirect.com/science/article/pii/S0001868609000517}
  {\bibfield  {journal} {\bibinfo  {journal} {Advances in colloid and interface
  science}\ }\textbf {\bibinfo {volume} {151}},\ \bibinfo {pages} {1} (\bibinfo
  {year} {2009})}\BibitemShut {NoStop}%
\bibitem [{\citenamefont {Tadros}(2013)}]{Tadros13}%
  \BibitemOpen
  \bibfield  {author} {\bibinfo {author} {\bibfnamefont {T.~F.}\ \bibnamefont
  {Tadros}},\ }\href
  {http://kinampark.com/PLGARef/files/Tadros%202013%2C%20Emulsion%20formation%2C%20stability%2C%20and%20rheology.pdf}
  {\bibfield  {journal} {\bibinfo  {journal} {Emulsion formation and
  stability}\ }\textbf {\bibinfo {volume} {1}},\ \bibinfo {pages} {1} (\bibinfo
  {year} {2013})}\BibitemShut {NoStop}%
\bibitem [{\citenamefont {Ferrero}\ \emph {et~al.}(2014)\citenamefont
  {Ferrero}, \citenamefont {Martens},\ and\ \citenamefont
  {Barrat}}]{FerreroMartensBarrat14}%
  \BibitemOpen
  \bibfield  {author} {\bibinfo {author} {\bibfnamefont {E.~E.}\ \bibnamefont
  {Ferrero}}, \bibinfo {author} {\bibfnamefont {K.}~\bibnamefont {Martens}},\
  and\ \bibinfo {author} {\bibfnamefont {J.-L.}\ \bibnamefont {Barrat}},\
  }\href {https://doi.org/10.1103/PhysRevLett.113.248301} {\bibfield  {journal}
  {\bibinfo  {journal} {Physical Review Letters}\ }\textbf {\bibinfo {volume}
  {113}},\ \bibinfo {pages} {1} (\bibinfo {year} {2014})}\BibitemShut {NoStop}%
\bibitem [{\citenamefont {Bonn}\ \emph {et~al.}(2017)\citenamefont {Bonn},
  \citenamefont {Denn}, \citenamefont {Berthier}, \citenamefont {Divoux},\ and\
  \citenamefont {Manneville}}]{Bonn17}%
  \BibitemOpen
  \bibfield  {author} {\bibinfo {author} {\bibfnamefont {D.}~\bibnamefont
  {Bonn}}, \bibinfo {author} {\bibfnamefont {M.~M.}\ \bibnamefont {Denn}},
  \bibinfo {author} {\bibfnamefont {L.}~\bibnamefont {Berthier}}, \bibinfo
  {author} {\bibfnamefont {T.}~\bibnamefont {Divoux}},\ and\ \bibinfo {author}
  {\bibfnamefont {S.}~\bibnamefont {Manneville}},\ }\href
  {https://doi.org/10.1103/RevModPhys.89.035005} {\bibfield  {journal}
  {\bibinfo  {journal} {Reviews of Modern Physics}\ }\textbf {\bibinfo {volume}
  {89}},\ \bibinfo {pages} {035005} (\bibinfo {year} {2017})}\BibitemShut
  {NoStop}%
\bibitem [{\citenamefont {Barrat}(2017)}]{BarratReview17}%
  \BibitemOpen
  \bibfield  {author} {\bibinfo {author} {\bibfnamefont {J.-L.}\ \bibnamefont
  {Barrat}},\ }\href {https://doi.org/10.1016/j.physa.2017.11.146} {\bibfield
  {journal} {\bibinfo  {journal} {Physica A: Statistical Mechanics and its
  Applications}\ }\textbf {\bibinfo {volume} {504}},\ \bibinfo {pages} {20}
  (\bibinfo {year} {2017})}\BibitemShut {NoStop}%
\bibitem [{\citenamefont {Nicolas}\ \emph {et~al.}(2018)\citenamefont
  {Nicolas}, \citenamefont {Ferrero}, \citenamefont {Martens},\ and\
  \citenamefont {Barrat}}]{Nicolas18}%
  \BibitemOpen
  \bibfield  {author} {\bibinfo {author} {\bibfnamefont {A.}~\bibnamefont
  {Nicolas}}, \bibinfo {author} {\bibfnamefont {E.~E.}\ \bibnamefont
  {Ferrero}}, \bibinfo {author} {\bibfnamefont {K.}~\bibnamefont {Martens}},\
  and\ \bibinfo {author} {\bibfnamefont {J.-L.}\ \bibnamefont {Barrat}},\
  }\href {https://doi.org/10.1103/RevModPhys.90.045006} {\bibfield  {journal}
  {\bibinfo  {journal} {Reviews of Modern Physics}\ }\textbf {\bibinfo {volume}
  {90}},\ \bibinfo {pages} {045006} (\bibinfo {year} {2018})}\BibitemShut
  {NoStop}%
\bibitem [{\citenamefont {Divoux}\ \emph {et~al.}(2024)\citenamefont {Divoux},
  \citenamefont {Agoritsas}, \citenamefont {Aime}, \citenamefont {Barentin},
  \citenamefont {Barrat}, \citenamefont {Benzi}, \citenamefont {Berthier},
  \citenamefont {Bi}, \citenamefont {Biroli}, \citenamefont {Bonn} \emph
  {et~al.}}]{Divoux24}%
  \BibitemOpen
  \bibfield  {author} {\bibinfo {author} {\bibfnamefont {T.}~\bibnamefont
  {Divoux}}, \bibinfo {author} {\bibfnamefont {E.}~\bibnamefont {Agoritsas}},
  \bibinfo {author} {\bibfnamefont {S.}~\bibnamefont {Aime}}, \bibinfo {author}
  {\bibfnamefont {C.}~\bibnamefont {Barentin}}, \bibinfo {author}
  {\bibfnamefont {J.-L.}\ \bibnamefont {Barrat}}, \bibinfo {author}
  {\bibfnamefont {R.}~\bibnamefont {Benzi}}, \bibinfo {author} {\bibfnamefont
  {L.}~\bibnamefont {Berthier}}, \bibinfo {author} {\bibfnamefont
  {D.}~\bibnamefont {Bi}}, \bibinfo {author} {\bibfnamefont {G.}~\bibnamefont
  {Biroli}}, \bibinfo {author} {\bibfnamefont {D.}~\bibnamefont {Bonn}}, \emph
  {et~al.},\ }\href {https://doi.org/10.1039/D3SM01740K} {\bibfield  {journal}
  {\bibinfo  {journal} {Soft Matter}\ }\textbf {\bibinfo {volume} {20}},\
  \bibinfo {pages} {6868} (\bibinfo {year} {2024})}\BibitemShut {NoStop}%
\bibitem [{\citenamefont {Taylor}(1932)}]{Taylor32}%
  \BibitemOpen
  \bibfield  {author} {\bibinfo {author} {\bibfnamefont {G.~I.}\ \bibnamefont
  {Taylor}},\ }\href
  {https://royalsocietypublishing.org/doi/abs/10.1098/rspa.1932.0169}
  {\bibfield  {journal} {\bibinfo  {journal} {Proceedings of the Royal Society
  of London. Series A, Containing Papers of a Mathematical and Physical
  Character}\ }\textbf {\bibinfo {volume} {138}},\ \bibinfo {pages} {41}
  (\bibinfo {year} {1932})}\BibitemShut {NoStop}%
\bibitem [{\citenamefont {Zinchenko}(1984)}]{Zinchenko84}%
  \BibitemOpen
  \bibfield  {author} {\bibinfo {author} {\bibfnamefont {A.}~\bibnamefont
  {Zinchenko}},\ }\href
  {https://www.sciencedirect.com/science/article/pii/0021892884900893}
  {\bibfield  {journal} {\bibinfo  {journal} {Journal of Applied Mathematics
  and Mechanics}\ }\textbf {\bibinfo {volume} {48}},\ \bibinfo {pages} {198}
  (\bibinfo {year} {1984})}\BibitemShut {NoStop}%
\bibitem [{\citenamefont {Ghigliotti}\ \emph {et~al.}(2010)\citenamefont
  {Ghigliotti}, \citenamefont {Biben},\ and\ \citenamefont
  {Misbah}}]{Ghigliottietal10}%
  \BibitemOpen
  \bibfield  {author} {\bibinfo {author} {\bibfnamefont {G.}~\bibnamefont
  {Ghigliotti}}, \bibinfo {author} {\bibfnamefont {T.}~\bibnamefont {Biben}},\
  and\ \bibinfo {author} {\bibfnamefont {C.}~\bibnamefont {Misbah}},\ }\href
  {https://search.proquest.com/openview/c8b2f07135b02789a5493203c4d3aaf8/1?pq-origsite=gscholar&cbl=34769}
  {\bibfield  {journal} {\bibinfo  {journal} {Journal of Fluid Mechanics}\
  }\textbf {\bibinfo {volume} {653}},\ \bibinfo {pages} {489} (\bibinfo {year}
  {2010})}\BibitemShut {NoStop}%
\bibitem [{\citenamefont {Pal}(2000)}]{Pal2000}%
  \BibitemOpen
  \bibfield  {author} {\bibinfo {author} {\bibfnamefont {R.}~\bibnamefont
  {Pal}},\ }\href
  {https://www.sciencedirect.com/science/article/pii/S0021979700967766}
  {\bibfield  {journal} {\bibinfo  {journal} {Journal of colloid and interface
  science}\ }\textbf {\bibinfo {volume} {225}},\ \bibinfo {pages} {359}
  (\bibinfo {year} {2000})}\BibitemShut {NoStop}%
\bibitem [{\citenamefont {Balmforth}\ \emph {et~al.}(2014)\citenamefont
  {Balmforth}, \citenamefont {Frigaard},\ and\ \citenamefont
  {Ovarlez}}]{Balmforthetal14}%
  \BibitemOpen
  \bibfield  {author} {\bibinfo {author} {\bibfnamefont {N.~J.}\ \bibnamefont
  {Balmforth}}, \bibinfo {author} {\bibfnamefont {I.~A.}\ \bibnamefont
  {Frigaard}},\ and\ \bibinfo {author} {\bibfnamefont {G.}~\bibnamefont
  {Ovarlez}},\ }\href {https://doi.org/10.1146/annurev-fluid-010313-141424}
  {\bibfield  {journal} {\bibinfo  {journal} {Annual Review of Fluid
  Mechanics}\ }\textbf {\bibinfo {volume} {46}},\ \bibinfo {pages} {121}
  (\bibinfo {year} {2014})}\BibitemShut {NoStop}%
\bibitem [{\citenamefont {Stone}(1994)}]{Stone94}%
  \BibitemOpen
  \bibfield  {author} {\bibinfo {author} {\bibfnamefont {H.~A.}\ \bibnamefont
  {Stone}},\ }\href {https://doi.org/10.1146/annurev.fl.26.010194.000433}
  {\bibfield  {journal} {\bibinfo  {journal} {Annual Review of Fluid
  Mechanics}\ }\textbf {\bibinfo {volume} {26}},\ \bibinfo {pages} {65}
  (\bibinfo {year} {1994})}\BibitemShut {NoStop}%
\bibitem [{\citenamefont {Cristini}\ \emph {et~al.}(2003)\citenamefont
  {Cristini}, \citenamefont {Guido}, \citenamefont {Alfani}, \citenamefont
  {B\l~awzdziewicz},\ and\ \citenamefont {Loewenberg}}]{Cristini2003}%
  \BibitemOpen
  \bibfield  {author} {\bibinfo {author} {\bibfnamefont {V.}~\bibnamefont
  {Cristini}}, \bibinfo {author} {\bibfnamefont {S.}~\bibnamefont {Guido}},
  \bibinfo {author} {\bibfnamefont {A.}~\bibnamefont {Alfani}}, \bibinfo
  {author} {\bibfnamefont {J.}~\bibnamefont {B\l~awzdziewicz}},\ and\ \bibinfo
  {author} {\bibfnamefont {M.}~\bibnamefont {Loewenberg}},\ }\href
  {https://doi.org/10.1122/1.1603240} {\bibfield  {journal} {\bibinfo
  {journal} {Journal of Rheology}\ }\textbf {\bibinfo {volume} {47}},\ \bibinfo
  {pages} {1283} (\bibinfo {year} {2003})}\BibitemShut {NoStop}%
\bibitem [{\citenamefont {Vankova}\ \emph
  {et~al.}(2007{\natexlab{a}})\citenamefont {Vankova}, \citenamefont
  {Tcholakova}, \citenamefont {Denkov}, \citenamefont {Ivanov}, \citenamefont
  {Vulchev},\ and\ \citenamefont {Danner}}]{Vankova07a}%
  \BibitemOpen
  \bibfield  {author} {\bibinfo {author} {\bibfnamefont {N.}~\bibnamefont
  {Vankova}}, \bibinfo {author} {\bibfnamefont {S.}~\bibnamefont {Tcholakova}},
  \bibinfo {author} {\bibfnamefont {N.~D.}\ \bibnamefont {Denkov}}, \bibinfo
  {author} {\bibfnamefont {I.~B.}\ \bibnamefont {Ivanov}}, \bibinfo {author}
  {\bibfnamefont {V.~D.}\ \bibnamefont {Vulchev}},\ and\ \bibinfo {author}
  {\bibfnamefont {T.}~\bibnamefont {Danner}},\ }\href
  {https://doi.org/10.1016/j.jcis.2007.03.059} {\bibfield  {journal} {\bibinfo
  {journal} {Journal of Colloid and Interface Science}\ }\textbf {\bibinfo
  {volume} {312}},\ \bibinfo {pages} {363} (\bibinfo {year}
  {2007}{\natexlab{a}})}\BibitemShut {NoStop}%
\bibitem [{\citenamefont {Vankova}\ \emph
  {et~al.}(2007{\natexlab{b}})\citenamefont {Vankova}, \citenamefont
  {Tcholakova}, \citenamefont {Denkov}, \citenamefont {Vulchev},\ and\
  \citenamefont {Danner}}]{Vankova07b}%
  \BibitemOpen
  \bibfield  {author} {\bibinfo {author} {\bibfnamefont {N.}~\bibnamefont
  {Vankova}}, \bibinfo {author} {\bibfnamefont {S.}~\bibnamefont {Tcholakova}},
  \bibinfo {author} {\bibfnamefont {N.}~\bibnamefont {Denkov}}, \bibinfo
  {author} {\bibfnamefont {V.}~\bibnamefont {Vulchev}},\ and\ \bibinfo {author}
  {\bibfnamefont {T.}~\bibnamefont {Danner}},\ }\href
  {https://doi.org/10.1016/j.jcis.2007.04.064} {\bibfield  {journal} {\bibinfo
  {journal} {Journal of Colloid and Interface Science}\ }\textbf {\bibinfo
  {volume} {313}},\ \bibinfo {pages} {612} (\bibinfo {year}
  {2007}{\natexlab{b}})}\BibitemShut {NoStop}%
\bibitem [{\citenamefont {Tcholakova}\ \emph {et~al.}(2007)\citenamefont
  {Tcholakova}, \citenamefont {Vankova}, \citenamefont {Denkov},\ and\
  \citenamefont {Danner}}]{Vankova07c}%
  \BibitemOpen
  \bibfield  {author} {\bibinfo {author} {\bibfnamefont {S.}~\bibnamefont
  {Tcholakova}}, \bibinfo {author} {\bibfnamefont {N.}~\bibnamefont {Vankova}},
  \bibinfo {author} {\bibfnamefont {N.}~\bibnamefont {Denkov}},\ and\ \bibinfo
  {author} {\bibfnamefont {T.}~\bibnamefont {Danner}},\ }\href
  {https://doi.org/10.1016/j.jcis.2007.01.097} {\bibfield  {journal} {\bibinfo
  {journal} {Journal of Colloid and Interface Science}\ }\textbf {\bibinfo
  {volume} {310}},\ \bibinfo {pages} {570} (\bibinfo {year}
  {2007})}\BibitemShut {NoStop}%
\bibitem [{\citenamefont {Perlekar}\ \emph {et~al.}(2012)\citenamefont
  {Perlekar}, \citenamefont {Biferale}, \citenamefont {Sbragaglia},
  \citenamefont {Srivastava},\ and\ \citenamefont {Toschi}}]{Perlekar2012}%
  \BibitemOpen
  \bibfield  {author} {\bibinfo {author} {\bibfnamefont {P.}~\bibnamefont
  {Perlekar}}, \bibinfo {author} {\bibfnamefont {L.}~\bibnamefont {Biferale}},
  \bibinfo {author} {\bibfnamefont {M.}~\bibnamefont {Sbragaglia}}, \bibinfo
  {author} {\bibfnamefont {S.}~\bibnamefont {Srivastava}},\ and\ \bibinfo
  {author} {\bibfnamefont {F.}~\bibnamefont {Toschi}},\ }\href
  {https://doi.org/10.1063/1.4719144} {\bibfield  {journal} {\bibinfo
  {journal} {Physics of Fluids}\ }\textbf {\bibinfo {volume} {24}},\ \bibinfo
  {pages} {065101} (\bibinfo {year} {2012})}\BibitemShut {NoStop}%
\bibitem [{\citenamefont {Scarbolo}\ \emph {et~al.}(2015)\citenamefont
  {Scarbolo}, \citenamefont {Bianco},\ and\ \citenamefont
  {Soldati}}]{Scarbolo15}%
  \BibitemOpen
  \bibfield  {author} {\bibinfo {author} {\bibfnamefont {L.}~\bibnamefont
  {Scarbolo}}, \bibinfo {author} {\bibfnamefont {F.}~\bibnamefont {Bianco}},\
  and\ \bibinfo {author} {\bibfnamefont {A.}~\bibnamefont {Soldati}},\ }\href
  {https://doi.org/10.1063/1.4923424} {\bibfield  {journal} {\bibinfo
  {journal} {Physics of Fluids}\ }\textbf {\bibinfo {volume} {27}} (\bibinfo
  {year} {2015})}\BibitemShut {NoStop}%
\bibitem [{\citenamefont {Roccon}\ \emph {et~al.}(2017)\citenamefont {Roccon},
  \citenamefont {De~Paoli}, \citenamefont {Zonta},\ and\ \citenamefont
  {Soldati}}]{Roccon17}%
  \BibitemOpen
  \bibfield  {author} {\bibinfo {author} {\bibfnamefont {A.}~\bibnamefont
  {Roccon}}, \bibinfo {author} {\bibfnamefont {M.}~\bibnamefont {De~Paoli}},
  \bibinfo {author} {\bibfnamefont {F.}~\bibnamefont {Zonta}},\ and\ \bibinfo
  {author} {\bibfnamefont {A.}~\bibnamefont {Soldati}},\ }\href
  {https://doi.org/10.1103/PhysRevFluids.2.083603} {\bibfield  {journal}
  {\bibinfo  {journal} {Physical Review Fluids}\ }\textbf {\bibinfo {volume}
  {2}},\ \bibinfo {pages} {083603} (\bibinfo {year} {2017})}\BibitemShut
  {NoStop}%
\bibitem [{\citenamefont {Soligo}\ \emph {et~al.}(2019)\citenamefont {Soligo},
  \citenamefont {Roccon},\ and\ \citenamefont {Soldati}}]{Soligo19}%
  \BibitemOpen
  \bibfield  {author} {\bibinfo {author} {\bibfnamefont {G.}~\bibnamefont
  {Soligo}}, \bibinfo {author} {\bibfnamefont {A.}~\bibnamefont {Roccon}},\
  and\ \bibinfo {author} {\bibfnamefont {A.}~\bibnamefont {Soldati}},\ }\href
  {https://doi.org/10.1017/jfm.2019.772} {\bibfield  {journal} {\bibinfo
  {journal} {Journal of Fluid Mechanics}\ }\textbf {\bibinfo {volume} {881}},\
  \bibinfo {pages} {244} (\bibinfo {year} {2019})}\BibitemShut {NoStop}%
\bibitem [{\citenamefont {Liu}\ \emph {et~al.}(2021)\citenamefont {Liu},
  \citenamefont {Chong}, \citenamefont {Wang}, \citenamefont {Ng},
  \citenamefont {Verzicco},\ and\ \citenamefont {Lohse}}]{Liu21}%
  \BibitemOpen
  \bibfield  {author} {\bibinfo {author} {\bibfnamefont {H.-R.}\ \bibnamefont
  {Liu}}, \bibinfo {author} {\bibfnamefont {K.~L.}\ \bibnamefont {Chong}},
  \bibinfo {author} {\bibfnamefont {Q.}~\bibnamefont {Wang}}, \bibinfo {author}
  {\bibfnamefont {C.~S.}\ \bibnamefont {Ng}}, \bibinfo {author} {\bibfnamefont
  {R.}~\bibnamefont {Verzicco}},\ and\ \bibinfo {author} {\bibfnamefont
  {D.}~\bibnamefont {Lohse}},\ }\href
  {https://doi.org/https://doi.org/10.1017/jfm.2021.14} {\bibfield  {journal}
  {\bibinfo  {journal} {Journal of Fluid Mechanics}\ }\textbf {\bibinfo
  {volume} {913}},\ \bibinfo {pages} {A9} (\bibinfo {year} {2021})}\BibitemShut
  {NoStop}%
\bibitem [{\citenamefont {Girotto}\ \emph {et~al.}(2022)\citenamefont
  {Girotto}, \citenamefont {Benzi}, \citenamefont {Di~Staso}, \citenamefont
  {Scagliarini}, \citenamefont {Schifano},\ and\ \citenamefont
  {Toschi}}]{Girotto22}%
  \BibitemOpen
  \bibfield  {author} {\bibinfo {author} {\bibfnamefont {I.}~\bibnamefont
  {Girotto}}, \bibinfo {author} {\bibfnamefont {R.}~\bibnamefont {Benzi}},
  \bibinfo {author} {\bibfnamefont {G.}~\bibnamefont {Di~Staso}}, \bibinfo
  {author} {\bibfnamefont {A.}~\bibnamefont {Scagliarini}}, \bibinfo {author}
  {\bibfnamefont {S.~F.}\ \bibnamefont {Schifano}},\ and\ \bibinfo {author}
  {\bibfnamefont {F.}~\bibnamefont {Toschi}},\ }\href
  {https://doi.org/10.1080/14685248.2022.2067333} {\bibfield  {journal}
  {\bibinfo  {journal} {Journal of Turbulence}\ }\textbf {\bibinfo {volume}
  {23}},\ \bibinfo {pages} {265} (\bibinfo {year} {2022})}\BibitemShut
  {NoStop}%
\bibitem [{\citenamefont {Girotto}\ \emph {et~al.}(2024)\citenamefont
  {Girotto}, \citenamefont {Scagliarini}, \citenamefont {Benzi},\ and\
  \citenamefont {Toschi}}]{Girotto24}%
  \BibitemOpen
  \bibfield  {author} {\bibinfo {author} {\bibfnamefont {I.}~\bibnamefont
  {Girotto}}, \bibinfo {author} {\bibfnamefont {A.}~\bibnamefont
  {Scagliarini}}, \bibinfo {author} {\bibfnamefont {R.}~\bibnamefont {Benzi}},\
  and\ \bibinfo {author} {\bibfnamefont {F.}~\bibnamefont {Toschi}},\ }\href
  {https://doi.org/10.1017/jfm.2024.364} {\bibfield  {journal} {\bibinfo
  {journal} {Journal of Fluid Mechanics}\ }\textbf {\bibinfo {volume} {986}},\
  \bibinfo {pages} {A33} (\bibinfo {year} {2024})}\BibitemShut {NoStop}%
\bibitem [{\citenamefont {Crialesi-Esposito}\ \emph {et~al.}(2024)\citenamefont
  {Crialesi-Esposito}, \citenamefont {Boffetta}, \citenamefont {Brandt},
  \citenamefont {Chibbaro},\ and\ \citenamefont
  {Musacchio}}]{CrialesiEsposito2024small}%
  \BibitemOpen
  \bibfield  {author} {\bibinfo {author} {\bibfnamefont {M.}~\bibnamefont
  {Crialesi-Esposito}}, \bibinfo {author} {\bibfnamefont {G.}~\bibnamefont
  {Boffetta}}, \bibinfo {author} {\bibfnamefont {L.}~\bibnamefont {Brandt}},
  \bibinfo {author} {\bibfnamefont {S.}~\bibnamefont {Chibbaro}},\ and\
  \bibinfo {author} {\bibfnamefont {S.}~\bibnamefont {Musacchio}},\ }\href
  {https://doi.org/10.1103/PhysRevFluids.9.L072301} {\bibfield  {journal}
  {\bibinfo  {journal} {Physical Review Fluids}\ }\textbf {\bibinfo {volume}
  {9}},\ \bibinfo {pages} {L072301} (\bibinfo {year} {2024})}\BibitemShut
  {NoStop}%
\bibitem [{\citenamefont {Khan}\ \emph {et~al.}(2011)\citenamefont {Khan},
  \citenamefont {Akhtar}, \citenamefont {Khan}, \citenamefont {Waseem},
  \citenamefont {Mahmood}, \citenamefont {Rasul}, \citenamefont {Iqbal},\ and\
  \citenamefont {Khan}}]{Khan11}%
  \BibitemOpen
  \bibfield  {author} {\bibinfo {author} {\bibfnamefont {B.~A.}\ \bibnamefont
  {Khan}}, \bibinfo {author} {\bibfnamefont {N.}~\bibnamefont {Akhtar}},
  \bibinfo {author} {\bibfnamefont {H.~M.~S.}\ \bibnamefont {Khan}}, \bibinfo
  {author} {\bibfnamefont {K.}~\bibnamefont {Waseem}}, \bibinfo {author}
  {\bibfnamefont {T.}~\bibnamefont {Mahmood}}, \bibinfo {author} {\bibfnamefont
  {A.}~\bibnamefont {Rasul}}, \bibinfo {author} {\bibfnamefont
  {M.}~\bibnamefont {Iqbal}},\ and\ \bibinfo {author} {\bibfnamefont
  {H.}~\bibnamefont {Khan}},\ }\href {https://doi.org/10.5897/AJPP11.698}
  {\bibfield  {journal} {\bibinfo  {journal} {African Journal of Pharmacy and
  Pharmacology}\ }\textbf {\bibinfo {volume} {5}},\ \bibinfo {pages} {2715}
  (\bibinfo {year} {2011})}\BibitemShut {NoStop}%
\bibitem [{\citenamefont {McClements}(2015)}]{Mcclements15}%
  \BibitemOpen
  \bibfield  {author} {\bibinfo {author} {\bibfnamefont {D.~J.}\ \bibnamefont
  {McClements}},\ }\href {https://doi.org/10.1201/9781420039436} {\emph
  {\bibinfo {title} {Food emulsions: principles, practices, and techniques}}}\
  (\bibinfo  {publisher} {CRC press},\ \bibinfo {year} {2015})\ p.\ \bibinfo
  {pages} {632}\BibitemShut {NoStop}%
\bibitem [{\citenamefont {Goodarzi}\ and\ \citenamefont
  {Zendehboudi}(2019)}]{Goodarzi19}%
  \BibitemOpen
  \bibfield  {author} {\bibinfo {author} {\bibfnamefont {F.}~\bibnamefont
  {Goodarzi}}\ and\ \bibinfo {author} {\bibfnamefont {S.}~\bibnamefont
  {Zendehboudi}},\ }\href {https://doi.org/10.1002/cjce.23336} {\bibfield
  {journal} {\bibinfo  {journal} {The Canadian Journal of Chemical
  Engineering}\ }\textbf {\bibinfo {volume} {97}},\ \bibinfo {pages} {281}
  (\bibinfo {year} {2019})}\BibitemShut {NoStop}%
\bibitem [{\citenamefont {Gallegos}\ and\ \citenamefont
  {Franco}(1999)}]{Gallegos99}%
  \BibitemOpen
  \bibfield  {author} {\bibinfo {author} {\bibfnamefont {C.}~\bibnamefont
  {Gallegos}}\ and\ \bibinfo {author} {\bibfnamefont {J.}~\bibnamefont
  {Franco}},\ }\href {https://doi.org/10.1016/S1359-0294(99)00003} {\bibfield
  {journal} {\bibinfo  {journal} {Current Opinion in Colloid \& Interface
  Science}\ }\textbf {\bibinfo {volume} {4}},\ \bibinfo {pages} {288} (\bibinfo
  {year} {1999})}\BibitemShut {NoStop}%
\bibitem [{\citenamefont {McClements}(2007)}]{Mcclements07}%
  \BibitemOpen
  \bibfield  {author} {\bibinfo {author} {\bibfnamefont {D.~J.}\ \bibnamefont
  {McClements}},\ }\href {https://doi.org/10.1080/10408390701289292} {\bibfield
   {journal} {\bibinfo  {journal} {Critical Reviews in Food Science and
  Nutrition}\ }\textbf {\bibinfo {volume} {47}},\ \bibinfo {pages} {611}
  (\bibinfo {year} {2007})}\BibitemShut {NoStop}%
\bibitem [{\citenamefont {Benzi}\ \emph {et~al.}(2009)\citenamefont {Benzi},
  \citenamefont {Sbragaglia}, \citenamefont {Succi}, \citenamefont
  {Bernaschi},\ and\ \citenamefont {Chibbaro}}]{Benzi09}%
  \BibitemOpen
  \bibfield  {author} {\bibinfo {author} {\bibfnamefont {R.}~\bibnamefont
  {Benzi}}, \bibinfo {author} {\bibfnamefont {M.}~\bibnamefont {Sbragaglia}},
  \bibinfo {author} {\bibfnamefont {S.}~\bibnamefont {Succi}}, \bibinfo
  {author} {\bibfnamefont {M.}~\bibnamefont {Bernaschi}},\ and\ \bibinfo
  {author} {\bibfnamefont {S.}~\bibnamefont {Chibbaro}},\ }\href
  {https://doi.org/10.1063/1.3216105} {\bibfield  {journal} {\bibinfo
  {journal} {Journal of Chemical Physics}\ }\textbf {\bibinfo {volume} {131}},\
  \bibinfo {eid} {104903} (\bibinfo {year} {2009})}\BibitemShut {NoStop}%
\bibitem [{\citenamefont {Roudsari}\ \emph {et~al.}(2012)\citenamefont
  {Roudsari}, \citenamefont {Turcotte}, \citenamefont {Dhib},\ and\
  \citenamefont {Ein-Mozaffari}}]{Roudsari12}%
  \BibitemOpen
  \bibfield  {author} {\bibinfo {author} {\bibfnamefont {S.~F.}\ \bibnamefont
  {Roudsari}}, \bibinfo {author} {\bibfnamefont {G.}~\bibnamefont {Turcotte}},
  \bibinfo {author} {\bibfnamefont {R.}~\bibnamefont {Dhib}},\ and\ \bibinfo
  {author} {\bibfnamefont {F.}~\bibnamefont {Ein-Mozaffari}},\ }\href
  {https://doi.org/10.1016/j.compchemeng.2012.06.013} {\bibfield  {journal}
  {\bibinfo  {journal} {Computers \& Chemical Engineering}\ }\textbf {\bibinfo
  {volume} {45}},\ \bibinfo {pages} {124} (\bibinfo {year} {2012})}\BibitemShut
  {NoStop}%
\bibitem [{\citenamefont {Rosti}\ \emph {et~al.}(2019)\citenamefont {Rosti},
  \citenamefont {Ge}, \citenamefont {Jain}, \citenamefont {Dodd},\ and\
  \citenamefont {Brandt}}]{Rosti19}%
  \BibitemOpen
  \bibfield  {author} {\bibinfo {author} {\bibfnamefont {M.~E.}\ \bibnamefont
  {Rosti}}, \bibinfo {author} {\bibfnamefont {Z.}~\bibnamefont {Ge}}, \bibinfo
  {author} {\bibfnamefont {S.~S.}\ \bibnamefont {Jain}}, \bibinfo {author}
  {\bibfnamefont {M.~S.}\ \bibnamefont {Dodd}},\ and\ \bibinfo {author}
  {\bibfnamefont {L.}~\bibnamefont {Brandt}},\ }\href
  {https://doi.org/10.1017/jfm.2019.581} {\bibfield  {journal} {\bibinfo
  {journal} {Journal of Fluid Mechanics}\ }\textbf {\bibinfo {volume} {876}},\
  \bibinfo {pages} {962} (\bibinfo {year} {2019})}\BibitemShut {NoStop}%
\bibitem [{\citenamefont {Wang}\ \emph {et~al.}(2020)\citenamefont {Wang},
  \citenamefont {Xu}, \citenamefont {Liu}, \citenamefont {Liu},\ and\
  \citenamefont {Rui}}]{wang2020molecular}%
  \BibitemOpen
  \bibfield  {author} {\bibinfo {author} {\bibfnamefont {Z.}~\bibnamefont
  {Wang}}, \bibinfo {author} {\bibfnamefont {Y.}~\bibnamefont {Xu}}, \bibinfo
  {author} {\bibfnamefont {Y.}~\bibnamefont {Liu}}, \bibinfo {author}
  {\bibfnamefont {X.}~\bibnamefont {Liu}},\ and\ \bibinfo {author}
  {\bibfnamefont {Z.}~\bibnamefont {Rui}},\ }\href
  {https://doi.org/10.1007/s13369-020-04840-9} {\bibfield  {journal} {\bibinfo
  {journal} {Arabian Journal for Science and Engineering}\ }\textbf {\bibinfo
  {volume} {45}},\ \bibinfo {pages} {7161} (\bibinfo {year}
  {2020})}\BibitemShut {NoStop}%
\bibitem [{\citenamefont {Lohse}\ and\ \citenamefont {Zhang}(2020)}]{Lohse20}%
  \BibitemOpen
  \bibfield  {author} {\bibinfo {author} {\bibfnamefont {D.}~\bibnamefont
  {Lohse}}\ and\ \bibinfo {author} {\bibfnamefont {X.}~\bibnamefont {Zhang}},\
  }\href {https://doi.org/10.1038/s42254-020-0199-z} {\bibfield  {journal}
  {\bibinfo  {journal} {Nature Reviews Physics}\ }\textbf {\bibinfo {volume}
  {2}},\ \bibinfo {pages} {426} (\bibinfo {year} {2020})}\BibitemShut {NoStop}%
\bibitem [{\citenamefont {Abdulredha}\ \emph {et~al.}(2020)\citenamefont
  {Abdulredha}, \citenamefont {Aslina},\ and\ \citenamefont
  {Luqman}}]{abdulredha2020overview}%
  \BibitemOpen
  \bibfield  {author} {\bibinfo {author} {\bibfnamefont {M.~M.}\ \bibnamefont
  {Abdulredha}}, \bibinfo {author} {\bibfnamefont {H.~S.}\ \bibnamefont
  {Aslina}},\ and\ \bibinfo {author} {\bibfnamefont {C.~A.}\ \bibnamefont
  {Luqman}},\ }\href {https://doi.org/10.1016/j.arabjc.2018.11.014} {\bibfield
  {journal} {\bibinfo  {journal} {Arabian Journal of Chemistry}\ }\textbf
  {\bibinfo {volume} {13}},\ \bibinfo {pages} {3403} (\bibinfo {year}
  {2020})}\BibitemShut {NoStop}%
\bibitem [{\citenamefont {Jeffreys}(1957)}]{Jeffreys57}%
  \BibitemOpen
  \bibfield  {author} {\bibinfo {author} {\bibfnamefont {H.}~\bibnamefont
  {Jeffreys}},\ }\href
  {http://janus.lib.cam.ac.uk/db/node.xsp?id=EAD%2FGBR%2F0275%2FJeffreys%2FC7-C29}
  {\bibfield  {journal} {\bibinfo  {journal} {St. John’s Coll. Camb. Sir
  Harold Jeffreys Pap.}\ }\textbf {\bibinfo {volume} {C17}} (\bibinfo {year}
  {1957})}\BibitemShut {NoStop}%
\bibitem [{\citenamefont {Orowan}(1965)}]{Orowan65}%
  \BibitemOpen
  \bibfield  {author} {\bibinfo {author} {\bibfnamefont {E.}~\bibnamefont
  {Orowan}},\ }\href {https://doi.org/10.1098/rsta.1965.0041} {\bibfield
  {journal} {\bibinfo  {journal} {Philosophical Transactions of the Royal
  Society A}\ }\textbf {\bibinfo {volume} {258}},\ \bibinfo {pages} {284}
  (\bibinfo {year} {1965})}\BibitemShut {NoStop}%
\bibitem [{\citenamefont {Morgan}(1971)}]{Morgan71}%
  \BibitemOpen
  \bibfield  {author} {\bibinfo {author} {\bibfnamefont {W.~J.}\ \bibnamefont
  {Morgan}},\ }\href {https://doi.org/10.1038/230042a0} {\bibfield  {journal}
  {\bibinfo  {journal} {Nature}\ }\textbf {\bibinfo {volume} {230}},\ \bibinfo
  {pages} {42} (\bibinfo {year} {1971})}\BibitemShut {NoStop}%
\bibitem [{\citenamefont {Montelli}\ \emph {et~al.}(2006)\citenamefont
  {Montelli}, \citenamefont {Nolet}, \citenamefont {Dahlen},\ and\
  \citenamefont {Masters}}]{Montelli06}%
  \BibitemOpen
  \bibfield  {author} {\bibinfo {author} {\bibfnamefont {R.}~\bibnamefont
  {Montelli}}, \bibinfo {author} {\bibfnamefont {G.}~\bibnamefont {Nolet}},
  \bibinfo {author} {\bibfnamefont {F.}~\bibnamefont {Dahlen}},\ and\ \bibinfo
  {author} {\bibfnamefont {G.}~\bibnamefont {Masters}},\ }\href
  {https://doi.org/10.1029/2006GC001248} {\bibfield  {journal} {\bibinfo
  {journal} {Geochemistry, Geophysics, Geosystems}\ }\textbf {\bibinfo {volume}
  {7}} (\bibinfo {year} {2006})}\BibitemShut {NoStop}%
\bibitem [{\citenamefont {French}\ and\ \citenamefont
  {Romanowicz}(2015)}]{French15}%
  \BibitemOpen
  \bibfield  {author} {\bibinfo {author} {\bibfnamefont {S.}~\bibnamefont
  {French}}\ and\ \bibinfo {author} {\bibfnamefont {B.}~\bibnamefont
  {Romanowicz}},\ }\href {https://doi.org/10.1038/nature14876} {\bibfield
  {journal} {\bibinfo  {journal} {Nature}\ }\textbf {\bibinfo {volume} {525}},\
  \bibinfo {pages} {95} (\bibinfo {year} {2015})}\BibitemShut {NoStop}%
\bibitem [{\citenamefont {Davaille}\ \emph {et~al.}(2018)\citenamefont
  {Davaille}, \citenamefont {Carrez},\ and\ \citenamefont
  {Cordier}}]{Davaille2018}%
  \BibitemOpen
  \bibfield  {author} {\bibinfo {author} {\bibfnamefont {A.}~\bibnamefont
  {Davaille}}, \bibinfo {author} {\bibfnamefont {P.}~\bibnamefont {Carrez}},\
  and\ \bibinfo {author} {\bibfnamefont {P.}~\bibnamefont {Cordier}},\ }\href
  {https://doi.org/doi.org/10.1002/2017GL076575} {\bibfield  {journal}
  {\bibinfo  {journal} {Geophysical Research Letters}\ }\textbf {\bibinfo
  {volume} {45}},\ \bibinfo {pages} {1349} (\bibinfo {year}
  {2018})}\BibitemShut {NoStop}%
\bibitem [{\citenamefont {Griffiths}(2000)}]{Griffiths2000}%
  \BibitemOpen
  \bibfield  {author} {\bibinfo {author} {\bibfnamefont {R.}~\bibnamefont
  {Griffiths}},\ }\href {https://doi.org/10.1146/annurev.fluid.32.1.477}
  {\bibfield  {journal} {\bibinfo  {journal} {Annual Review of Fluid
  Mechanics}\ }\textbf {\bibinfo {volume} {32}},\ \bibinfo {pages} {477}
  (\bibinfo {year} {2000})}\BibitemShut {NoStop}%
\bibitem [{\citenamefont {Lavall{\'e}e}\ \emph {et~al.}(2015)\citenamefont
  {Lavall{\'e}e}, \citenamefont {Dingwell}, \citenamefont {Johnson},
  \citenamefont {Cimarelli}, \citenamefont {Hornby}, \citenamefont {Kendrick},
  \citenamefont {Von~Aulock}, \citenamefont {Kennedy}, \citenamefont {Andrews},
  \citenamefont {Wadsworth} \emph {et~al.}}]{Lavallee15}%
  \BibitemOpen
  \bibfield  {author} {\bibinfo {author} {\bibfnamefont {Y.}~\bibnamefont
  {Lavall{\'e}e}}, \bibinfo {author} {\bibfnamefont {D.~B.}\ \bibnamefont
  {Dingwell}}, \bibinfo {author} {\bibfnamefont {J.~B.}\ \bibnamefont
  {Johnson}}, \bibinfo {author} {\bibfnamefont {C.}~\bibnamefont {Cimarelli}},
  \bibinfo {author} {\bibfnamefont {A.~J.}\ \bibnamefont {Hornby}}, \bibinfo
  {author} {\bibfnamefont {J.~E.}\ \bibnamefont {Kendrick}}, \bibinfo {author}
  {\bibfnamefont {F.~W.}\ \bibnamefont {Von~Aulock}}, \bibinfo {author}
  {\bibfnamefont {B.~M.}\ \bibnamefont {Kennedy}}, \bibinfo {author}
  {\bibfnamefont {B.~J.}\ \bibnamefont {Andrews}}, \bibinfo {author}
  {\bibfnamefont {F.~B.}\ \bibnamefont {Wadsworth}}, \emph {et~al.},\ }\href
  {https://doi.org/10.1038/nature16153} {\bibfield  {journal} {\bibinfo
  {journal} {Nature}\ }\textbf {\bibinfo {volume} {528}},\ \bibinfo {pages}
  {544} (\bibinfo {year} {2015})}\BibitemShut {NoStop}%
\bibitem [{\citenamefont {Schellart}\ and\ \citenamefont
  {Strak}(2016)}]{SchellartStrak}%
  \BibitemOpen
  \bibfield  {author} {\bibinfo {author} {\bibfnamefont {W.}~\bibnamefont
  {Schellart}}\ and\ \bibinfo {author} {\bibfnamefont {V.}~\bibnamefont
  {Strak}},\ }\href {https://doi.org/10.1016/j.jog.2016.03.009} {\bibfield
  {journal} {\bibinfo  {journal} {Journal of Geodynamics}\ }\textbf {\bibinfo
  {volume} {100}},\ \bibinfo {pages} {7} (\bibinfo {year} {2016})}\BibitemShut
  {NoStop}%
\bibitem [{\citenamefont {Di~Giuseppe}\ \emph {et~al.}(2014)\citenamefont
  {Di~Giuseppe}, \citenamefont {Corbi}, \citenamefont {Funiciello},
  \citenamefont {Massmeyer}, \citenamefont {Santimano}, \citenamefont
  {Rosenau},\ and\ \citenamefont {Davaille}}]{DiGiuseppe2014}%
  \BibitemOpen
  \bibfield  {author} {\bibinfo {author} {\bibfnamefont {E.}~\bibnamefont
  {Di~Giuseppe}}, \bibinfo {author} {\bibfnamefont {F.}~\bibnamefont {Corbi}},
  \bibinfo {author} {\bibfnamefont {F.}~\bibnamefont {Funiciello}}, \bibinfo
  {author} {\bibfnamefont {A.}~\bibnamefont {Massmeyer}}, \bibinfo {author}
  {\bibfnamefont {T.}~\bibnamefont {Santimano}}, \bibinfo {author}
  {\bibfnamefont {M.}~\bibnamefont {Rosenau}},\ and\ \bibinfo {author}
  {\bibfnamefont {A.}~\bibnamefont {Davaille}},\ }\href
  {https://doi.org/10.1016/j.tecto.2014.12.005} {\bibfield  {journal} {\bibinfo
   {journal} {Tectonophysics}\ }\textbf {\bibinfo {volume} {642}},\ \bibinfo
  {pages} {29} (\bibinfo {year} {2014})}\BibitemShut {NoStop}%
\bibitem [{\citenamefont {Reber}\ \emph {et~al.}(2020)\citenamefont {Reber},
  \citenamefont {Cooke},\ and\ \citenamefont {Dooley}}]{Reber2020}%
  \BibitemOpen
  \bibfield  {author} {\bibinfo {author} {\bibfnamefont {J.}~\bibnamefont
  {Reber}}, \bibinfo {author} {\bibfnamefont {M.}~\bibnamefont {Cooke}},\ and\
  \bibinfo {author} {\bibfnamefont {T.}~\bibnamefont {Dooley}},\ }\href
  {https://doi.org/10.1016/j.earscirev.2020.103107} {\bibfield  {journal}
  {\bibinfo  {journal} {Earth-Science Reviews}\ }\textbf {\bibinfo {volume}
  {202}},\ \bibinfo {pages} {103107} (\bibinfo {year} {2020})}\BibitemShut
  {NoStop}%
\bibitem [{\citenamefont {Oresta}\ \emph {et~al.}(2009)\citenamefont {Oresta},
  \citenamefont {Verzicco}, \citenamefont {Lohse},\ and\ \citenamefont
  {Prosperetti}}]{Orestaetal09}%
  \BibitemOpen
  \bibfield  {author} {\bibinfo {author} {\bibfnamefont {P.}~\bibnamefont
  {Oresta}}, \bibinfo {author} {\bibfnamefont {R.}~\bibnamefont {Verzicco}},
  \bibinfo {author} {\bibfnamefont {D.}~\bibnamefont {Lohse}},\ and\ \bibinfo
  {author} {\bibfnamefont {A.}~\bibnamefont {Prosperetti}},\ }\href
  {https://doi.org/10.1103/PhysRevE.80.026304} {\bibfield  {journal} {\bibinfo
  {journal} {Physical Review E}\ }\textbf {\bibinfo {volume} {80}},\ \bibinfo
  {pages} {026304} (\bibinfo {year} {2009})}\BibitemShut {NoStop}%
\bibitem [{\citenamefont {Lakkaraju}\ \emph {et~al.}(2013)\citenamefont
  {Lakkaraju}, \citenamefont {Stevens}, \citenamefont {Oresta}, \citenamefont
  {Verzicco}, \citenamefont {Lohse},\ and\ \citenamefont
  {Prosperetti}}]{Lakkaraju13}%
  \BibitemOpen
  \bibfield  {author} {\bibinfo {author} {\bibfnamefont {R.}~\bibnamefont
  {Lakkaraju}}, \bibinfo {author} {\bibfnamefont {R.~J. A.~M.}\ \bibnamefont
  {Stevens}}, \bibinfo {author} {\bibfnamefont {P.}~\bibnamefont {Oresta}},
  \bibinfo {author} {\bibfnamefont {R.}~\bibnamefont {Verzicco}}, \bibinfo
  {author} {\bibfnamefont {D.}~\bibnamefont {Lohse}},\ and\ \bibinfo {author}
  {\bibfnamefont {A.}~\bibnamefont {Prosperetti}},\ }\href
  {https://doi.org/10.1073/pnas.1217546110} {\bibfield  {journal} {\bibinfo
  {journal} {Proceedings of the National Academy of Sciences}\ }\textbf
  {\bibinfo {volume} {110}},\ \bibinfo {pages} {9237} (\bibinfo {year}
  {2013})}\BibitemShut {NoStop}%
\bibitem [{\citenamefont {Biferale}\ \emph {et~al.}(2012)\citenamefont
  {Biferale}, \citenamefont {Perlekar}, \citenamefont {Sbragaglia},\ and\
  \citenamefont {Toschi}}]{Biferale12}%
  \BibitemOpen
  \bibfield  {author} {\bibinfo {author} {\bibfnamefont {L.}~\bibnamefont
  {Biferale}}, \bibinfo {author} {\bibfnamefont {P.}~\bibnamefont {Perlekar}},
  \bibinfo {author} {\bibfnamefont {M.}~\bibnamefont {Sbragaglia}},\ and\
  \bibinfo {author} {\bibfnamefont {F.}~\bibnamefont {Toschi}},\ }\href
  {https://doi.org/10.1103/PhysRevLett.108.104502} {\bibfield  {journal}
  {\bibinfo  {journal} {Physical Review Letters}\ }\textbf {\bibinfo {volume}
  {108}},\ \bibinfo {pages} {104502} (\bibinfo {year} {2012})}\BibitemShut
  {NoStop}%
\bibitem [{\citenamefont {Garoosi}\ and\ \citenamefont
  {Mahdi}(2021)}]{Garoosi21}%
  \BibitemOpen
  \bibfield  {author} {\bibinfo {author} {\bibfnamefont {F.}~\bibnamefont
  {Garoosi}}\ and\ \bibinfo {author} {\bibfnamefont {T.-F.}\ \bibnamefont
  {Mahdi}},\ }\href {https://doi.org/10.1016/j.ijheatmasstransfer.2021.121163}
  {\bibfield  {journal} {\bibinfo  {journal} {International Journal of Heat and
  Mass Transfer}\ }\textbf {\bibinfo {volume} {172}},\ \bibinfo {pages}
  {121163} (\bibinfo {year} {2021})}\BibitemShut {NoStop}%
\bibitem [{\citenamefont {Santos}\ \emph {et~al.}(2021)\citenamefont {Santos},
  \citenamefont {Lugarini}, \citenamefont {Junqueira},\ and\ \citenamefont
  {Franco}}]{santos21}%
  \BibitemOpen
  \bibfield  {author} {\bibinfo {author} {\bibfnamefont {P.~R.}\ \bibnamefont
  {Santos}}, \bibinfo {author} {\bibfnamefont {A.}~\bibnamefont {Lugarini}},
  \bibinfo {author} {\bibfnamefont {S.~L.}\ \bibnamefont {Junqueira}},\ and\
  \bibinfo {author} {\bibfnamefont {A.~T.}\ \bibnamefont {Franco}},\ }\href
  {https://doi.org/10.1016/j.ijthermalsci.2021.106991} {\bibfield  {journal}
  {\bibinfo  {journal} {International Journal of Thermal Sciences}\ }\textbf
  {\bibinfo {volume} {166}},\ \bibinfo {pages} {106991} (\bibinfo {year}
  {2021})}\BibitemShut {NoStop}%
\bibitem [{\citenamefont {Liu}\ \emph {et~al.}(2022{\natexlab{a}})\citenamefont
  {Liu}, \citenamefont {Chong}, \citenamefont {Ng}, \citenamefont {Verzicco},\
  and\ \citenamefont {Lohse}}]{Liu21a}%
  \BibitemOpen
  \bibfield  {author} {\bibinfo {author} {\bibfnamefont {H.-R.}\ \bibnamefont
  {Liu}}, \bibinfo {author} {\bibfnamefont {K.~L.}\ \bibnamefont {Chong}},
  \bibinfo {author} {\bibfnamefont {C.~S.}\ \bibnamefont {Ng}}, \bibinfo
  {author} {\bibfnamefont {R.}~\bibnamefont {Verzicco}},\ and\ \bibinfo
  {author} {\bibfnamefont {D.}~\bibnamefont {Lohse}},\ }\href
  {https://doi.org/10.1017/jfm.2021.1068} {\bibfield  {journal} {\bibinfo
  {journal} {Journal of Fluid Mechanics}\ }\textbf {\bibinfo {volume} {933}},\
  \bibinfo {pages} {R1} (\bibinfo {year} {2022}{\natexlab{a}})}\BibitemShut
  {NoStop}%
\bibitem [{\citenamefont {Liu}\ \emph {et~al.}(2022{\natexlab{b}})\citenamefont
  {Liu}, \citenamefont {Chong}, \citenamefont {Yang}, \citenamefont
  {Verzicco},\ and\ \citenamefont {Lohse}}]{Liu22}%
  \BibitemOpen
  \bibfield  {author} {\bibinfo {author} {\bibfnamefont {H.-R.}\ \bibnamefont
  {Liu}}, \bibinfo {author} {\bibfnamefont {K.~L.}\ \bibnamefont {Chong}},
  \bibinfo {author} {\bibfnamefont {R.}~\bibnamefont {Yang}}, \bibinfo {author}
  {\bibfnamefont {R.}~\bibnamefont {Verzicco}},\ and\ \bibinfo {author}
  {\bibfnamefont {D.}~\bibnamefont {Lohse}},\ }\href
  {https://doi.org/10.1017/jfm.2022.181} {\bibfield  {journal} {\bibinfo
  {journal} {Journal of Fluid Mechanics}\ }\textbf {\bibinfo {volume} {938}},\
  \bibinfo {pages} {A31} (\bibinfo {year} {2022}{\natexlab{b}})}\BibitemShut
  {NoStop}%
\bibitem [{\citenamefont {Bilondi}\ \emph {et~al.}(2024)\citenamefont
  {Bilondi}, \citenamefont {Scapin}, \citenamefont {Brandt},\ and\
  \citenamefont {Mirbod}}]{Brandt24}%
  \BibitemOpen
  \bibfield  {author} {\bibinfo {author} {\bibfnamefont {A.~M.}\ \bibnamefont
  {Bilondi}}, \bibinfo {author} {\bibfnamefont {N.}~\bibnamefont {Scapin}},
  \bibinfo {author} {\bibfnamefont {L.}~\bibnamefont {Brandt}},\ and\ \bibinfo
  {author} {\bibfnamefont {P.}~\bibnamefont {Mirbod}},\ }\href
  {https://doi.org/10.1017/jfm.2024.765} {\bibfield  {journal} {\bibinfo
  {journal} {Journal of Fluid Mechanics}\ }\textbf {\bibinfo {volume} {999}},\
  \bibinfo {pages} {A4} (\bibinfo {year} {2024})}\BibitemShut {NoStop}%
\bibitem [{\citenamefont {Mangani}\ \emph {et~al.}(2024)\citenamefont
  {Mangani}, \citenamefont {Roccon}, \citenamefont {Zonta},\ and\ \citenamefont
  {Soldati}}]{Mangani24}%
  \BibitemOpen
  \bibfield  {author} {\bibinfo {author} {\bibfnamefont {F.}~\bibnamefont
  {Mangani}}, \bibinfo {author} {\bibfnamefont {A.}~\bibnamefont {Roccon}},
  \bibinfo {author} {\bibfnamefont {F.}~\bibnamefont {Zonta}},\ and\ \bibinfo
  {author} {\bibfnamefont {A.}~\bibnamefont {Soldati}},\ }\href
  {https://doi.org/10.1017/jfm.2023.1002} {\bibfield  {journal} {\bibinfo
  {journal} {Journal of Fluid Mechanics}\ }\textbf {\bibinfo {volume} {978}},\
  \bibinfo {pages} {A12} (\bibinfo {year} {2024})}\BibitemShut {NoStop}%
\bibitem [{\citenamefont {Zhang}\ \emph {et~al.}(2006)\citenamefont {Zhang},
  \citenamefont {Vola},\ and\ \citenamefont {Frigaard}}]{Zhang06}%
  \BibitemOpen
  \bibfield  {author} {\bibinfo {author} {\bibfnamefont {J.}~\bibnamefont
  {Zhang}}, \bibinfo {author} {\bibfnamefont {D.}~\bibnamefont {Vola}},\ and\
  \bibinfo {author} {\bibfnamefont {I.~A.}\ \bibnamefont {Frigaard}},\ }\href
  {https://doi.org/10.1017/S002211200600200X} {\bibfield  {journal} {\bibinfo
  {journal} {Journal of Fluid Mechanics}\ }\textbf {\bibinfo {volume} {566}},\
  \bibinfo {pages} {389} (\bibinfo {year} {2006})}\BibitemShut {NoStop}%
\bibitem [{\citenamefont {Balmforth}\ and\ \citenamefont
  {Rust}(2009)}]{BalmforthRust09}%
  \BibitemOpen
  \bibfield  {author} {\bibinfo {author} {\bibfnamefont {N.~J.}\ \bibnamefont
  {Balmforth}}\ and\ \bibinfo {author} {\bibfnamefont {A.~C.}\ \bibnamefont
  {Rust}},\ }\href
  {https://doi.org/https://doi.org/10.1016/j.jnnfm.2008.07.012} {\bibfield
  {journal} {\bibinfo  {journal} {Journal of Non-Newtonian Fluid Mechanics}\
  }\textbf {\bibinfo {volume} {158}},\ \bibinfo {pages} {36 } (\bibinfo {year}
  {2009})}\BibitemShut {NoStop}%
\bibitem [{\citenamefont {Vikhansky}(2009)}]{Vikhansky09}%
  \BibitemOpen
  \bibfield  {author} {\bibinfo {author} {\bibfnamefont {A.}~\bibnamefont
  {Vikhansky}},\ }\href {https://doi.org/10.1063/1.3256166} {\bibfield
  {journal} {\bibinfo  {journal} {Physics of Fluids}\ }\textbf {\bibinfo
  {volume} {21}} (\bibinfo {year} {2009})}\BibitemShut {NoStop}%
\bibitem [{\citenamefont {Vikhansky}(2010)}]{Vikhansky10}%
  \BibitemOpen
  \bibfield  {author} {\bibinfo {author} {\bibfnamefont {A.}~\bibnamefont
  {Vikhansky}},\ }\href
  {https://doi.org/https://doi.org/10.1016/j.jnnfm.2010.09.003} {\bibfield
  {journal} {\bibinfo  {journal} {Journal of Non-Newtonian Fluid Mechanics}\
  }\textbf {\bibinfo {volume} {165}},\ \bibinfo {pages} {1713 } (\bibinfo
  {year} {2010})}\BibitemShut {NoStop}%
\bibitem [{\citenamefont {Albaalbaki}\ and\ \citenamefont
  {Khayat}(2011)}]{AlbaalbakiKhayat11}%
  \BibitemOpen
  \bibfield  {author} {\bibinfo {author} {\bibfnamefont {B.}~\bibnamefont
  {Albaalbaki}}\ and\ \bibinfo {author} {\bibfnamefont {R.~E.}\ \bibnamefont
  {Khayat}},\ }\href {https://doi.org/10.1017/S0022112010004775} {\bibfield
  {journal} {\bibinfo  {journal} {Journal of Fluid Mechanics}\ }\textbf
  {\bibinfo {volume} {668}},\ \bibinfo {pages} {500} (\bibinfo {year}
  {2011})}\BibitemShut {NoStop}%
\bibitem [{\citenamefont {Turan}\ \emph {et~al.}(2012)\citenamefont {Turan},
  \citenamefont {Chakraborty},\ and\ \citenamefont {Poole}}]{Turanetal12}%
  \BibitemOpen
  \bibfield  {author} {\bibinfo {author} {\bibfnamefont {O.}~\bibnamefont
  {Turan}}, \bibinfo {author} {\bibfnamefont {N.}~\bibnamefont {Chakraborty}},\
  and\ \bibinfo {author} {\bibfnamefont {R.~J.}\ \bibnamefont {Poole}},\ }\href
  {https://doi.org/https://doi.org/10.1016/j.jnnfm.2012.01.006} {\bibfield
  {journal} {\bibinfo  {journal} {Journal of Non-Newtonian Fluid Mechanics}\
  }\textbf {\bibinfo {volume} {171-172}},\ \bibinfo {pages} {83 } (\bibinfo
  {year} {2012})}\BibitemShut {NoStop}%
\bibitem [{\citenamefont {Massmeyer}\ \emph {et~al.}(2013)\citenamefont
  {Massmeyer}, \citenamefont {Di~Giuseppe}, \citenamefont {Davaille},
  \citenamefont {Rolf},\ and\ \citenamefont {Tackley}}]{Massmeyer13}%
  \BibitemOpen
  \bibfield  {author} {\bibinfo {author} {\bibfnamefont {A.}~\bibnamefont
  {Massmeyer}}, \bibinfo {author} {\bibfnamefont {E.}~\bibnamefont
  {Di~Giuseppe}}, \bibinfo {author} {\bibfnamefont {A.}~\bibnamefont
  {Davaille}}, \bibinfo {author} {\bibfnamefont {T.}~\bibnamefont {Rolf}},\
  and\ \bibinfo {author} {\bibfnamefont {P.~J.}\ \bibnamefont {Tackley}},\
  }\href
  {https://www.sciencedirect.com/science/article/pii/S0377025712002741?casa_token=eTzk51eEitQAAAAA:Se89x-3mKwiJaeWS-TWZxd0HgvJl48KcsXRx4kSnODwgqUy2MTPOVM5S9WXFLhZat-JkZkZ2cXs}
  {\bibfield  {journal} {\bibinfo  {journal} {Journal of Non-Newtonian Fluid
  Mechanics}\ }\textbf {\bibinfo {volume} {195}},\ \bibinfo {pages} {32}
  (\bibinfo {year} {2013})}\BibitemShut {NoStop}%
\bibitem [{\citenamefont {Hassan}\ \emph {et~al.}(2015)\citenamefont {Hassan},
  \citenamefont {Pathak},\ and\ \citenamefont {Khan}}]{Hassanetal15}%
  \BibitemOpen
  \bibfield  {author} {\bibinfo {author} {\bibfnamefont {M.}~\bibnamefont
  {Hassan}}, \bibinfo {author} {\bibfnamefont {M.}~\bibnamefont {Pathak}},\
  and\ \bibinfo {author} {\bibfnamefont {M.~K.}\ \bibnamefont {Khan}},\ }\href
  {https://doi.org/https://doi.org/10.1016/j.jnnfm.2015.10.003} {\bibfield
  {journal} {\bibinfo  {journal} {Journal of Non-Newtonian Fluid Mechanics}\
  }\textbf {\bibinfo {volume} {226}},\ \bibinfo {pages} {32 } (\bibinfo {year}
  {2015})}\BibitemShut {NoStop}%
\bibitem [{\citenamefont {Karimfazli}\ \emph {et~al.}(2016)\citenamefont
  {Karimfazli}, \citenamefont {Frigaard},\ and\ \citenamefont
  {Wachs}}]{Karimfazli16}%
  \BibitemOpen
  \bibfield  {author} {\bibinfo {author} {\bibfnamefont {I.}~\bibnamefont
  {Karimfazli}}, \bibinfo {author} {\bibfnamefont {I.}~\bibnamefont
  {Frigaard}},\ and\ \bibinfo {author} {\bibfnamefont {A.}~\bibnamefont
  {Wachs}},\ }\href
  {https://www.cambridge.org/core/journals/journal-of-fluid-mechanics/article/thermal-plumes-in-viscoplastic-fluids-flow-onset-and-development/24EE61DE6E13233CB75DAA35292B2E30}
  {\bibfield  {journal} {\bibinfo  {journal} {Journal of Fluid Mechanics}\
  }\textbf {\bibinfo {volume} {787}},\ \bibinfo {pages} {474} (\bibinfo {year}
  {2016})}\BibitemShut {NoStop}%
\bibitem [{\citenamefont {Goyon}\ \emph {et~al.}(2008)\citenamefont {Goyon},
  \citenamefont {Colin}, \citenamefont {Ovarlez}, \citenamefont {Ajdari},\ and\
  \citenamefont {Bocquet}}]{Goyon08}%
  \BibitemOpen
  \bibfield  {author} {\bibinfo {author} {\bibfnamefont {J.}~\bibnamefont
  {Goyon}}, \bibinfo {author} {\bibfnamefont {A.}~\bibnamefont {Colin}},
  \bibinfo {author} {\bibfnamefont {G.}~\bibnamefont {Ovarlez}}, \bibinfo
  {author} {\bibfnamefont {A.}~\bibnamefont {Ajdari}},\ and\ \bibinfo {author}
  {\bibfnamefont {L.}~\bibnamefont {Bocquet}},\ }\href
  {https://www.nature.com/articles/nature07026} {\bibfield  {journal} {\bibinfo
   {journal} {Nature}\ }\textbf {\bibinfo {volume} {454}},\ \bibinfo {pages}
  {84} (\bibinfo {year} {2008})}\BibitemShut {NoStop}%
\bibitem [{\citenamefont {Goyon}\ \emph {et~al.}(2010)\citenamefont {Goyon},
  \citenamefont {Colin}, \citenamefont {Ovarlez}, \citenamefont {Ajdari},\ and\
  \citenamefont {Bocquet}}]{Goyon10}%
  \BibitemOpen
  \bibfield  {author} {\bibinfo {author} {\bibfnamefont {J.}~\bibnamefont
  {Goyon}}, \bibinfo {author} {\bibfnamefont {A.}~\bibnamefont {Colin}},
  \bibinfo {author} {\bibfnamefont {G.}~\bibnamefont {Ovarlez}}, \bibinfo
  {author} {\bibfnamefont {A.}~\bibnamefont {Ajdari}},\ and\ \bibinfo {author}
  {\bibfnamefont {L.}~\bibnamefont {Bocquet}},\ }\href
  {https://pubs.rsc.org/en/content/articlelanding/2010/sm/c001930e/unauth#!divAbstract}
  {\bibfield  {journal} {\bibinfo  {journal} {Soft Matter}\ }\textbf {\bibinfo
  {volume} {6}},\ \bibinfo {pages} {2668} (\bibinfo {year} {2010})}\BibitemShut
  {NoStop}%
\bibitem [{\citenamefont {Davaille}\ \emph {et~al.}(2013)\citenamefont
  {Davaille}, \citenamefont {Gueslin}, \citenamefont {Massmeyer},\ and\
  \citenamefont {Di~Giuseppe}}]{Davailleetal13}%
  \BibitemOpen
  \bibfield  {author} {\bibinfo {author} {\bibfnamefont {A.}~\bibnamefont
  {Davaille}}, \bibinfo {author} {\bibfnamefont {B.}~\bibnamefont {Gueslin}},
  \bibinfo {author} {\bibfnamefont {A.}~\bibnamefont {Massmeyer}},\ and\
  \bibinfo {author} {\bibfnamefont {E.}~\bibnamefont {Di~Giuseppe}},\ }\href
  {https://doi.org/10.1016/j.jnnfm.2012.10.008} {\bibfield  {journal} {\bibinfo
   {journal} {Journal of Non-Newtonian Fluid Mechanics}\ }\textbf {\bibinfo
  {volume} {193}},\ \bibinfo {pages} {144} (\bibinfo {year}
  {2013})}\BibitemShut {NoStop}%
\bibitem [{\citenamefont {Pelusi}\ \emph {et~al.}(2021)\citenamefont {Pelusi},
  \citenamefont {Sbragaglia}, \citenamefont {Benzi}, \citenamefont
  {Scagliarini}, \citenamefont {Bernaschi},\ and\ \citenamefont
  {Succi}}]{PelusiSM21}%
  \BibitemOpen
  \bibfield  {author} {\bibinfo {author} {\bibfnamefont {F.}~\bibnamefont
  {Pelusi}}, \bibinfo {author} {\bibfnamefont {M.}~\bibnamefont {Sbragaglia}},
  \bibinfo {author} {\bibfnamefont {R.}~\bibnamefont {Benzi}}, \bibinfo
  {author} {\bibfnamefont {A.}~\bibnamefont {Scagliarini}}, \bibinfo {author}
  {\bibfnamefont {M.}~\bibnamefont {Bernaschi}},\ and\ \bibinfo {author}
  {\bibfnamefont {S.}~\bibnamefont {Succi}},\ }\href
  {https://doi.org/10.1039/D0SM01777A} {\bibfield  {journal} {\bibinfo
  {journal} {Soft Matter}\ }\textbf {\bibinfo {volume} {17}},\ \bibinfo {pages}
  {3709} (\bibinfo {year} {2021})}\BibitemShut {NoStop}%
\bibitem [{\citenamefont {Pelusi}\ \emph {et~al.}(2023)\citenamefont {Pelusi},
  \citenamefont {Ascione}, \citenamefont {Sbragaglia},\ and\ \citenamefont
  {Bernaschi}}]{PelusiSM23}%
  \BibitemOpen
  \bibfield  {author} {\bibinfo {author} {\bibfnamefont {F.}~\bibnamefont
  {Pelusi}}, \bibinfo {author} {\bibfnamefont {S.}~\bibnamefont {Ascione}},
  \bibinfo {author} {\bibfnamefont {M.}~\bibnamefont {Sbragaglia}},\ and\
  \bibinfo {author} {\bibfnamefont {M.}~\bibnamefont {Bernaschi}},\ }\href
  {https://doi.org/10.1039/D3SM00716B} {\bibfield  {journal} {\bibinfo
  {journal} {Soft Matter}\ }\textbf {\bibinfo {volume} {19}},\ \bibinfo {pages}
  {7192} (\bibinfo {year} {2023})}\BibitemShut {NoStop}%
\bibitem [{\citenamefont {Pelusi}\ \emph
  {et~al.}(2024{\natexlab{a}})\citenamefont {Pelusi}, \citenamefont
  {Scagliarini}, \citenamefont {Sbragaglia}, \citenamefont {Bernaschi},\ and\
  \citenamefont {Benzi}}]{Pelusi24intermittent}%
  \BibitemOpen
  \bibfield  {author} {\bibinfo {author} {\bibfnamefont {F.}~\bibnamefont
  {Pelusi}}, \bibinfo {author} {\bibfnamefont {A.}~\bibnamefont {Scagliarini}},
  \bibinfo {author} {\bibfnamefont {M.}~\bibnamefont {Sbragaglia}}, \bibinfo
  {author} {\bibfnamefont {M.}~\bibnamefont {Bernaschi}},\ and\ \bibinfo
  {author} {\bibfnamefont {R.}~\bibnamefont {Benzi}},\ }\href
  {https://doi.org/10.1103/PhysRevLett.133.244001} {\bibfield  {journal}
  {\bibinfo  {journal} {Physical Review Letters}\ }\textbf {\bibinfo {volume}
  {133}},\ \bibinfo {pages} {244001} (\bibinfo {year}
  {2024}{\natexlab{a}})}\BibitemShut {NoStop}%
\bibitem [{\citenamefont {Jadhav}\ \emph {et~al.}(2021)\citenamefont {Jadhav},
  \citenamefont {Rossi},\ and\ \citenamefont {Karimfazli}}]{Jadhav21}%
  \BibitemOpen
  \bibfield  {author} {\bibinfo {author} {\bibfnamefont {K.}~\bibnamefont
  {Jadhav}}, \bibinfo {author} {\bibfnamefont {P.}~\bibnamefont {Rossi}},\ and\
  \bibinfo {author} {\bibfnamefont {I.}~\bibnamefont {Karimfazli}},\ }\href
  {https://doi.org/10.1017/jfm.2020.1096} {\bibfield  {journal} {\bibinfo
  {journal} {Journal of Fluid Mechanics}\ }\textbf {\bibinfo {volume} {912}},\
  \bibinfo {pages} {A10} (\bibinfo {year} {2021})}\BibitemShut {NoStop}%
\bibitem [{\citenamefont {Groeneweg}\ \emph {et~al.}(1998)\citenamefont
  {Groeneweg}, \citenamefont {Agterof}, \citenamefont {Jaeger}, \citenamefont
  {Janssen}, \citenamefont {Wieringa},\ and\ \citenamefont
  {Klahn}}]{Groeneweg98}%
  \BibitemOpen
  \bibfield  {author} {\bibinfo {author} {\bibfnamefont {F.}~\bibnamefont
  {Groeneweg}}, \bibinfo {author} {\bibfnamefont {W.}~\bibnamefont {Agterof}},
  \bibinfo {author} {\bibfnamefont {P.}~\bibnamefont {Jaeger}}, \bibinfo
  {author} {\bibfnamefont {J.}~\bibnamefont {Janssen}}, \bibinfo {author}
  {\bibfnamefont {J.}~\bibnamefont {Wieringa}},\ and\ \bibinfo {author}
  {\bibfnamefont {J.}~\bibnamefont {Klahn}},\ }\href
  {https://doi.org/10.1205/026387698524596} {\bibfield  {journal} {\bibinfo
  {journal} {Chemical Engineering Research and Design}\ }\textbf {\bibinfo
  {volume} {76}},\ \bibinfo {pages} {55} (\bibinfo {year} {1998})}\BibitemShut
  {NoStop}%
\bibitem [{\citenamefont {Kumar}\ \emph {et~al.}(2015)\citenamefont {Kumar},
  \citenamefont {Li}, \citenamefont {Cheng},\ and\ \citenamefont
  {Lee}}]{Kumar15}%
  \BibitemOpen
  \bibfield  {author} {\bibinfo {author} {\bibfnamefont {A.}~\bibnamefont
  {Kumar}}, \bibinfo {author} {\bibfnamefont {S.}~\bibnamefont {Li}}, \bibinfo
  {author} {\bibfnamefont {C.-M.}\ \bibnamefont {Cheng}},\ and\ \bibinfo
  {author} {\bibfnamefont {D.}~\bibnamefont {Lee}},\ }\href
  {https://doi.org/10.1021/acs.iecr.5b01122} {\bibfield  {journal} {\bibinfo
  {journal} {Industrial \& Engineering Chemistry Research}\ }\textbf {\bibinfo
  {volume} {54}},\ \bibinfo {pages} {8375} (\bibinfo {year}
  {2015})}\BibitemShut {NoStop}%
\bibitem [{\citenamefont {Bouchama}\ \emph {et~al.}(2003)\citenamefont
  {Bouchama}, \citenamefont {Van~Aken}, \citenamefont {Autin},\ and\
  \citenamefont {Koper}}]{Bouchama03}%
  \BibitemOpen
  \bibfield  {author} {\bibinfo {author} {\bibfnamefont {F.}~\bibnamefont
  {Bouchama}}, \bibinfo {author} {\bibfnamefont {G.}~\bibnamefont {Van~Aken}},
  \bibinfo {author} {\bibfnamefont {A.}~\bibnamefont {Autin}},\ and\ \bibinfo
  {author} {\bibfnamefont {G.}~\bibnamefont {Koper}},\ }\href
  {https://doi.org/10.1016/j.colsurfa.2003.08.011} {\bibfield  {journal}
  {\bibinfo  {journal} {Colloids and Surfaces A: Physicochemical and
  engineering aspects}\ }\textbf {\bibinfo {volume} {231}},\ \bibinfo {pages}
  {11} (\bibinfo {year} {2003})}\BibitemShut {NoStop}%
\bibitem [{\citenamefont {Perazzo}\ \emph {et~al.}(2015)\citenamefont
  {Perazzo}, \citenamefont {Preziosi},\ and\ \citenamefont
  {Guido}}]{Perazzo15}%
  \BibitemOpen
  \bibfield  {author} {\bibinfo {author} {\bibfnamefont {A.}~\bibnamefont
  {Perazzo}}, \bibinfo {author} {\bibfnamefont {V.}~\bibnamefont {Preziosi}},\
  and\ \bibinfo {author} {\bibfnamefont {S.}~\bibnamefont {Guido}},\ }\href
  {https://doi.org/10.1016/j.cis.2015.01.001} {\bibfield  {journal} {\bibinfo
  {journal} {Advances in Colloid and Interface Science}\ }\textbf {\bibinfo
  {volume} {222}},\ \bibinfo {pages} {581} (\bibinfo {year}
  {2015})}\BibitemShut {NoStop}%
\bibitem [{\citenamefont {Fernandez}\ \emph {et~al.}(2004)\citenamefont
  {Fernandez}, \citenamefont {Andr{\'e}}, \citenamefont {Rieger},\ and\
  \citenamefont {K{\"u}hnle}}]{Fernandez04}%
  \BibitemOpen
  \bibfield  {author} {\bibinfo {author} {\bibfnamefont {P.}~\bibnamefont
  {Fernandez}}, \bibinfo {author} {\bibfnamefont {V.}~\bibnamefont
  {Andr{\'e}}}, \bibinfo {author} {\bibfnamefont {J.}~\bibnamefont {Rieger}},\
  and\ \bibinfo {author} {\bibfnamefont {A.}~\bibnamefont {K{\"u}hnle}},\
  }\href {https://doi.org/10.1016/j.colsurfa.2004.09.029} {\bibfield  {journal}
  {\bibinfo  {journal} {Colloids and Surfaces A: Physicochemical and
  Engineering Aspects}\ }\textbf {\bibinfo {volume} {251}},\ \bibinfo {pages}
  {53} (\bibinfo {year} {2004})}\BibitemShut {NoStop}%
\bibitem [{\citenamefont {Borrin}\ \emph {et~al.}(2016)\citenamefont {Borrin},
  \citenamefont {Georges}, \citenamefont {Moraes},\ and\ \citenamefont
  {Pinho}}]{Borrin16}%
  \BibitemOpen
  \bibfield  {author} {\bibinfo {author} {\bibfnamefont {T.~R.}\ \bibnamefont
  {Borrin}}, \bibinfo {author} {\bibfnamefont {E.~L.}\ \bibnamefont {Georges}},
  \bibinfo {author} {\bibfnamefont {I.~C.}\ \bibnamefont {Moraes}},\ and\
  \bibinfo {author} {\bibfnamefont {S.~C.}\ \bibnamefont {Pinho}},\ }\href
  {https://doi.org/10.1016/j.jfoodeng.2015.08.012} {\bibfield  {journal}
  {\bibinfo  {journal} {Journal of Food Engineering}\ }\textbf {\bibinfo
  {volume} {169}},\ \bibinfo {pages} {1} (\bibinfo {year} {2016})}\BibitemShut
  {NoStop}%
\bibitem [{\citenamefont {Yeo}\ \emph {et~al.}(2002)\citenamefont {Yeo},
  \citenamefont {Matar}, \citenamefont {de~Ortiz},\ and\ \citenamefont
  {Hewitt}}]{Yeo2002}%
  \BibitemOpen
  \bibfield  {author} {\bibinfo {author} {\bibfnamefont {L.}~\bibnamefont
  {Yeo}}, \bibinfo {author} {\bibfnamefont {O.}~\bibnamefont {Matar}}, \bibinfo
  {author} {\bibfnamefont {E.}~\bibnamefont {de~Ortiz}},\ and\ \bibinfo
  {author} {\bibfnamefont {G.}~\bibnamefont {Hewitt}},\ }\href
  {https://doi.org/10.1016/S0009-2509(02)00260-9} {\bibfield  {journal}
  {\bibinfo  {journal} {Chemical Engineering Science}\ }\textbf {\bibinfo
  {volume} {57}},\ \bibinfo {pages} {3505} (\bibinfo {year}
  {2002})}\BibitemShut {NoStop}%
\bibitem [{\citenamefont {Bakhuis}\ \emph {et~al.}(2021)\citenamefont
  {Bakhuis}, \citenamefont {Ezeta}, \citenamefont {Bullee}, \citenamefont
  {Marin}, \citenamefont {Lohse}, \citenamefont {Sun},\ and\ \citenamefont
  {Huisman}}]{Bakhuis21}%
  \BibitemOpen
  \bibfield  {author} {\bibinfo {author} {\bibfnamefont {D.}~\bibnamefont
  {Bakhuis}}, \bibinfo {author} {\bibfnamefont {R.}~\bibnamefont {Ezeta}},
  \bibinfo {author} {\bibfnamefont {P.~A.}\ \bibnamefont {Bullee}}, \bibinfo
  {author} {\bibfnamefont {A.}~\bibnamefont {Marin}}, \bibinfo {author}
  {\bibfnamefont {D.}~\bibnamefont {Lohse}}, \bibinfo {author} {\bibfnamefont
  {C.}~\bibnamefont {Sun}},\ and\ \bibinfo {author} {\bibfnamefont {S.~G.}\
  \bibnamefont {Huisman}},\ }\href
  {https://doi.org/10.1103/PhysRevLett.126.064501} {\bibfield  {journal}
  {\bibinfo  {journal} {Physical Review Letters}\ }\textbf {\bibinfo {volume}
  {126}},\ \bibinfo {pages} {064501} (\bibinfo {year} {2021})}\BibitemShut
  {NoStop}%
\bibitem [{\citenamefont {Yi}\ \emph {et~al.}(2024)\citenamefont {Yi},
  \citenamefont {Girotto}, \citenamefont {Toschi},\ and\ \citenamefont
  {Sun}}]{Yi24}%
  \BibitemOpen
  \bibfield  {author} {\bibinfo {author} {\bibfnamefont {L.}~\bibnamefont
  {Yi}}, \bibinfo {author} {\bibfnamefont {I.}~\bibnamefont {Girotto}},
  \bibinfo {author} {\bibfnamefont {F.}~\bibnamefont {Toschi}},\ and\ \bibinfo
  {author} {\bibfnamefont {C.}~\bibnamefont {Sun}},\ }\href
  {https://doi.org/10.1103/PhysRevLett.133.134001} {\bibfield  {journal}
  {\bibinfo  {journal} {Physical Review Letters}\ }\textbf {\bibinfo {volume}
  {133}},\ \bibinfo {pages} {134001} (\bibinfo {year} {2024})}\BibitemShut
  {NoStop}%
\bibitem [{\citenamefont {Grossmann}\ and\ \citenamefont
  {Lohse}(2001)}]{Grossmann01}%
  \BibitemOpen
  \bibfield  {author} {\bibinfo {author} {\bibfnamefont {S.}~\bibnamefont
  {Grossmann}}\ and\ \bibinfo {author} {\bibfnamefont {D.}~\bibnamefont
  {Lohse}},\ }\href {https://doi.org/10.1103/PhysRevLett.86.3316} {\bibfield
  {journal} {\bibinfo  {journal} {Physical Review Letters}\ }\textbf {\bibinfo
  {volume} {86}},\ \bibinfo {pages} {3316} (\bibinfo {year}
  {2001})}\BibitemShut {NoStop}%
\bibitem [{\citenamefont {Ahlers}\ \emph {et~al.}(2009)\citenamefont {Ahlers},
  \citenamefont {Grossmann},\ and\ \citenamefont {Lohse}}]{Ahlers09}%
  \BibitemOpen
  \bibfield  {author} {\bibinfo {author} {\bibfnamefont {G.}~\bibnamefont
  {Ahlers}}, \bibinfo {author} {\bibfnamefont {S.}~\bibnamefont {Grossmann}},\
  and\ \bibinfo {author} {\bibfnamefont {D.}~\bibnamefont {Lohse}},\ }\href
  {https://journals.aps.org/rmp/abstract/10.1103/RevModPhys.81.503} {\bibfield
  {journal} {\bibinfo  {journal} {Reviews of Modern Physics}\ }\textbf
  {\bibinfo {volume} {81}},\ \bibinfo {pages} {503} (\bibinfo {year}
  {2009})}\BibitemShut {NoStop}%
\bibitem [{\citenamefont {Chill{\`a}}\ and\ \citenamefont
  {Schumacher}(2012)}]{Chilla12}%
  \BibitemOpen
  \bibfield  {author} {\bibinfo {author} {\bibfnamefont {F.}~\bibnamefont
  {Chill{\`a}}}\ and\ \bibinfo {author} {\bibfnamefont {J.}~\bibnamefont
  {Schumacher}},\ }\href
  {https://link.springer.com/article/10.1140/epje/i2012-12058-1} {\bibfield
  {journal} {\bibinfo  {journal} {The European Physical Journal E}\ }\textbf
  {\bibinfo {volume} {35}},\ \bibinfo {pages} {1} (\bibinfo {year}
  {2012})}\BibitemShut {NoStop}%
\bibitem [{\citenamefont {Van Der~Poel}\ \emph {et~al.}(2013)\citenamefont {Van
  Der~Poel}, \citenamefont {Stevens},\ and\ \citenamefont
  {Lohse}}]{van2013comparison}%
  \BibitemOpen
  \bibfield  {author} {\bibinfo {author} {\bibfnamefont {E.~P.}\ \bibnamefont
  {Van Der~Poel}}, \bibinfo {author} {\bibfnamefont {R.~J.}\ \bibnamefont
  {Stevens}},\ and\ \bibinfo {author} {\bibfnamefont {D.}~\bibnamefont
  {Lohse}},\ }\href {https://doi.org/10.1017/jfm.2013.488} {\bibfield
  {journal} {\bibinfo  {journal} {Journal of Fluid Mechanics}\ }\textbf
  {\bibinfo {volume} {736}},\ \bibinfo {pages} {177} (\bibinfo {year}
  {2013})}\BibitemShut {NoStop}%
\bibitem [{\citenamefont {Pelusi}\ \emph {et~al.}(2022)\citenamefont {Pelusi},
  \citenamefont {Lulli}, \citenamefont {Sbragaglia},\ and\ \citenamefont
  {Bernaschi}}]{TLBfind22}%
  \BibitemOpen
  \bibfield  {author} {\bibinfo {author} {\bibfnamefont {F.}~\bibnamefont
  {Pelusi}}, \bibinfo {author} {\bibfnamefont {M.}~\bibnamefont {Lulli}},
  \bibinfo {author} {\bibfnamefont {M.}~\bibnamefont {Sbragaglia}},\ and\
  \bibinfo {author} {\bibfnamefont {M.}~\bibnamefont {Bernaschi}},\ }\href
  {https://doi.org/10.1016/J.CPC.2021.108259} {\bibfield  {journal} {\bibinfo
  {journal} {Computer Physics Communications}\ }\textbf {\bibinfo {volume}
  {273}},\ \bibinfo {pages} {108259} (\bibinfo {year} {2022})}\BibitemShut
  {NoStop}%
\bibitem [{\citenamefont {Chandrasekhar}(1961)}]{Chandrasekhar61}%
  \BibitemOpen
  \bibfield  {author} {\bibinfo {author} {\bibfnamefont {S.}~\bibnamefont
  {Chandrasekhar}},\ }\href
  {https://books.google.it/books?hl=en&lr=&id=Mg3CAgAAQBAJ&oi=fnd&pg=PP1&dq=Hydrodynamic+and+hydromagnetic+stability&ots=crkgkYPc2a&sig=o5sWcyCbL71myOImOEeU75I3wOg&redir_esc=y#v=onepage&q=Hydrodynamic%20and%20hydromagnetic%20stability&f=false}
  {\emph {\bibinfo {title} {Hydrodynamic and hydromagnetic stability}}}\
  (\bibinfo  {publisher} {Oxford University Press},\ \bibinfo {year}
  {1961})\BibitemShut {NoStop}%
\bibitem [{\citenamefont {Bodenschatz}\ \emph {et~al.}(2000)\citenamefont
  {Bodenschatz}, \citenamefont {Pesch},\ and\ \citenamefont
  {Ahlers}}]{Bodenschatz2000}%
  \BibitemOpen
  \bibfield  {author} {\bibinfo {author} {\bibfnamefont {E.}~\bibnamefont
  {Bodenschatz}}, \bibinfo {author} {\bibfnamefont {W.}~\bibnamefont {Pesch}},\
  and\ \bibinfo {author} {\bibfnamefont {G.}~\bibnamefont {Ahlers}},\ }\href
  {https://doi.org/10.1146/annurev.fluid.32.1.709} {\bibfield  {journal}
  {\bibinfo  {journal} {Annual Review of Fluid Mechanics}\ }\textbf {\bibinfo
  {volume} {32}},\ \bibinfo {pages} {709} (\bibinfo {year} {2000})}\BibitemShut
  {NoStop}%
\bibitem [{\citenamefont {Lohse}\ and\ \citenamefont {Xia}(2010)}]{Lohse10}%
  \BibitemOpen
  \bibfield  {author} {\bibinfo {author} {\bibfnamefont {D.}~\bibnamefont
  {Lohse}}\ and\ \bibinfo {author} {\bibfnamefont {K.-Q.}\ \bibnamefont
  {Xia}},\ }\href
  {https://www.annualreviews.org/doi/full/10.1146/annurev.fluid.010908.165152}
  {\bibfield  {journal} {\bibinfo  {journal} {Annual Review of Fluid
  Mechanics}\ }\textbf {\bibinfo {volume} {42}},\ \bibinfo {pages} {335}
  (\bibinfo {year} {2010})}\BibitemShut {NoStop}%
\bibitem [{\citenamefont {Shraiman}\ and\ \citenamefont
  {Siggia}(1990)}]{Shraiman90}%
  \BibitemOpen
  \bibfield  {author} {\bibinfo {author} {\bibfnamefont {B.~I.}\ \bibnamefont
  {Shraiman}}\ and\ \bibinfo {author} {\bibfnamefont {E.~D.}\ \bibnamefont
  {Siggia}},\ }\href
  {https://journals.aps.org/pra/abstract/10.1103/PhysRevA.42.3650} {\bibfield
  {journal} {\bibinfo  {journal} {Physical Review A}\ }\textbf {\bibinfo
  {volume} {42}},\ \bibinfo {pages} {3650} (\bibinfo {year}
  {1990})}\BibitemShut {NoStop}%
\bibitem [{\citenamefont {Stevens}\ \emph {et~al.}(2010)\citenamefont
  {Stevens}, \citenamefont {Verzicco},\ and\ \citenamefont
  {Lohse}}]{Verzicco10}%
  \BibitemOpen
  \bibfield  {author} {\bibinfo {author} {\bibfnamefont {R.}~\bibnamefont
  {Stevens}}, \bibinfo {author} {\bibfnamefont {R.}~\bibnamefont {Verzicco}},\
  and\ \bibinfo {author} {\bibfnamefont {D.}~\bibnamefont {Lohse}},\ }\href
  {https://doi.org/10.1017/S0022112009992461} {\bibfield  {journal} {\bibinfo
  {journal} {Journal of Fluid Mechanics}\ }\textbf {\bibinfo {volume} {643}},\
  \bibinfo {pages} {495} (\bibinfo {year} {2010})}\BibitemShut {NoStop}%
\bibitem [{\citenamefont {Benzi}\ \emph {et~al.}(1992)\citenamefont {Benzi},
  \citenamefont {Succi},\ and\ \citenamefont {Vergassola}}]{Benzi92}%
  \BibitemOpen
  \bibfield  {author} {\bibinfo {author} {\bibfnamefont {R.}~\bibnamefont
  {Benzi}}, \bibinfo {author} {\bibfnamefont {S.}~\bibnamefont {Succi}},\ and\
  \bibinfo {author} {\bibfnamefont {M.}~\bibnamefont {Vergassola}},\ }\href
  {https://doi.org/10.1016/0370-1573(92)90090-M} {\bibfield  {journal}
  {\bibinfo  {journal} {Physics Reports}\ }\textbf {\bibinfo {volume} {222}},\
  \bibinfo {pages} {145} (\bibinfo {year} {1992})}\BibitemShut {NoStop}%
\bibitem [{\citenamefont {Kr{\"u}ger}\ \emph {et~al.}(2017)\citenamefont
  {Kr{\"u}ger}, \citenamefont {Kusumaatmaja}, \citenamefont {Kuzmin},
  \citenamefont {Shardt}, \citenamefont {Silva},\ and\ \citenamefont
  {Viggen}}]{Kruger17}%
  \BibitemOpen
  \bibfield  {author} {\bibinfo {author} {\bibfnamefont {T.}~\bibnamefont
  {Kr{\"u}ger}}, \bibinfo {author} {\bibfnamefont {H.}~\bibnamefont
  {Kusumaatmaja}}, \bibinfo {author} {\bibfnamefont {A.}~\bibnamefont
  {Kuzmin}}, \bibinfo {author} {\bibfnamefont {O.}~\bibnamefont {Shardt}},
  \bibinfo {author} {\bibfnamefont {G.}~\bibnamefont {Silva}},\ and\ \bibinfo
  {author} {\bibfnamefont {E.~M.}\ \bibnamefont {Viggen}},\ }\href
  {https://link.springer.com/content/pdf/10.1007/978-3-319-44649-3.pdf}
  {\bibfield  {journal} {\bibinfo  {journal} {Springer International
  Publishing}\ }\textbf {\bibinfo {volume} {10}},\ \bibinfo {pages} {4}
  (\bibinfo {year} {2017})}\BibitemShut {NoStop}%
\bibitem [{\citenamefont {Succi}(2018)}]{Succi18}%
  \BibitemOpen
  \bibfield  {author} {\bibinfo {author} {\bibfnamefont {S.}~\bibnamefont
  {Succi}},\ }\href
  {https://books.google.it/books?hl=it&lr=&id=SHlUDwAAQBAJ&oi=fnd&pg=PP1&dq=succi+lattice+boltzmann+equation+2018&ots=RWlf5ZUevx&sig=aNvu-Q8CTTVLrzcM9uZCNMcPS0k&redir_esc=y#v=onepage&q=succi%20lattice%20boltzmann%20equation%202018&f=false}
  {\emph {\bibinfo {title} {The lattice Boltzmann Equation}}}\ (\bibinfo
  {publisher} {Oxford University Press},\ \bibinfo {year} {2018})\BibitemShut
  {NoStop}%
\bibitem [{\citenamefont {Benzi}\ \emph {et~al.}(2014)\citenamefont {Benzi},
  \citenamefont {Sbragaglia}, \citenamefont {Perlekar}, \citenamefont
  {Bernaschi}, \citenamefont {Succi},\ and\ \citenamefont
  {Toschi}}]{Benzietal14}%
  \BibitemOpen
  \bibfield  {author} {\bibinfo {author} {\bibfnamefont {R.}~\bibnamefont
  {Benzi}}, \bibinfo {author} {\bibfnamefont {M.}~\bibnamefont {Sbragaglia}},
  \bibinfo {author} {\bibfnamefont {P.}~\bibnamefont {Perlekar}}, \bibinfo
  {author} {\bibfnamefont {M.}~\bibnamefont {Bernaschi}}, \bibinfo {author}
  {\bibfnamefont {S.}~\bibnamefont {Succi}},\ and\ \bibinfo {author}
  {\bibfnamefont {F.}~\bibnamefont {Toschi}},\ }\href
  {https://doi.org/10.1039/C4SM00348A} {\bibfield  {journal} {\bibinfo
  {journal} {Soft Matter}\ }\textbf {\bibinfo {volume} {10}},\ \bibinfo {pages}
  {4615} (\bibinfo {year} {2014})}\BibitemShut {NoStop}%
\bibitem [{\citenamefont {Dollet}\ \emph {et~al.}(2015)\citenamefont {Dollet},
  \citenamefont {Scagliarini},\ and\ \citenamefont {Sbragaglia}}]{Dollet15}%
  \BibitemOpen
  \bibfield  {author} {\bibinfo {author} {\bibfnamefont {B.}~\bibnamefont
  {Dollet}}, \bibinfo {author} {\bibfnamefont {A.}~\bibnamefont
  {Scagliarini}},\ and\ \bibinfo {author} {\bibfnamefont {M.}~\bibnamefont
  {Sbragaglia}},\ }\href {https://doi.org/10.1017/jfm.2015.28} {\bibfield
  {journal} {\bibinfo  {journal} {Journal of Fluid Mechanics}\ }\textbf
  {\bibinfo {volume} {766}},\ \bibinfo {pages} {556} (\bibinfo {year}
  {2015})}\BibitemShut {NoStop}%
\bibitem [{\citenamefont {Lulli}\ \emph {et~al.}(2018)\citenamefont {Lulli},
  \citenamefont {Benzi},\ and\ \citenamefont
  {Sbragaglia}}]{LulliBenziSbragaglia18}%
  \BibitemOpen
  \bibfield  {author} {\bibinfo {author} {\bibfnamefont {M.}~\bibnamefont
  {Lulli}}, \bibinfo {author} {\bibfnamefont {R.}~\bibnamefont {Benzi}},\ and\
  \bibinfo {author} {\bibfnamefont {M.}~\bibnamefont {Sbragaglia}},\ }\href
  {https://doi.org/10.1103/PhysRevX.8.021031} {\bibfield  {journal} {\bibinfo
  {journal} {Physical Review X}\ }\textbf {\bibinfo {volume} {8}},\ \bibinfo
  {pages} {021031} (\bibinfo {year} {2018})}\BibitemShut {NoStop}%
\bibitem [{\citenamefont {Negro}\ \emph {et~al.}(2023)\citenamefont {Negro},
  \citenamefont {Carenza}, \citenamefont {Gonnella}, \citenamefont {Mackay},
  \citenamefont {Morozov},\ and\ \citenamefont {Marenduzzo}}]{Negro23}%
  \BibitemOpen
  \bibfield  {author} {\bibinfo {author} {\bibfnamefont {G.}~\bibnamefont
  {Negro}}, \bibinfo {author} {\bibfnamefont {L.~N.}\ \bibnamefont {Carenza}},
  \bibinfo {author} {\bibfnamefont {G.}~\bibnamefont {Gonnella}}, \bibinfo
  {author} {\bibfnamefont {F.}~\bibnamefont {Mackay}}, \bibinfo {author}
  {\bibfnamefont {A.}~\bibnamefont {Morozov}},\ and\ \bibinfo {author}
  {\bibfnamefont {D.}~\bibnamefont {Marenduzzo}},\ }\href
  {https://www.science.org/doi/full/10.1126/sciadv.adf8106} {\bibfield
  {journal} {\bibinfo  {journal} {Science Advances}\ }\textbf {\bibinfo
  {volume} {9}},\ \bibinfo {pages} {eadf8106} (\bibinfo {year}
  {2023})}\BibitemShut {NoStop}%
\bibitem [{\citenamefont {Pelusi}\ \emph
  {et~al.}(2024{\natexlab{b}})\citenamefont {Pelusi}, \citenamefont {Filippi},
  \citenamefont {Derzsi}, \citenamefont {Pierno},\ and\ \citenamefont
  {Sbragaglia}}]{Pelusi24rheology}%
  \BibitemOpen
  \bibfield  {author} {\bibinfo {author} {\bibfnamefont {F.}~\bibnamefont
  {Pelusi}}, \bibinfo {author} {\bibfnamefont {D.}~\bibnamefont {Filippi}},
  \bibinfo {author} {\bibfnamefont {L.}~\bibnamefont {Derzsi}}, \bibinfo
  {author} {\bibfnamefont {M.}~\bibnamefont {Pierno}},\ and\ \bibinfo {author}
  {\bibfnamefont {M.}~\bibnamefont {Sbragaglia}},\ }\href
  {https://pubs.rsc.org/en/content/articlehtml/2024/sm/d4sm00041b} {\bibfield
  {journal} {\bibinfo  {journal} {Soft Matter}\ } (\bibinfo {year}
  {2024}{\natexlab{b}})}\BibitemShut {NoStop}%
\bibitem [{\citenamefont {Tiribocchi}\ \emph {et~al.}(2025)\citenamefont
  {Tiribocchi}, \citenamefont {Durve}, \citenamefont {Lauricella},
  \citenamefont {Montessori}, \citenamefont {Tucny},\ and\ \citenamefont
  {Succi}}]{Tiribocchi25review}%
  \BibitemOpen
  \bibfield  {author} {\bibinfo {author} {\bibfnamefont {A.}~\bibnamefont
  {Tiribocchi}}, \bibinfo {author} {\bibfnamefont {M.}~\bibnamefont {Durve}},
  \bibinfo {author} {\bibfnamefont {M.}~\bibnamefont {Lauricella}}, \bibinfo
  {author} {\bibfnamefont {A.}~\bibnamefont {Montessori}}, \bibinfo {author}
  {\bibfnamefont {J.-M.}\ \bibnamefont {Tucny}},\ and\ \bibinfo {author}
  {\bibfnamefont {S.}~\bibnamefont {Succi}},\ }\href
  {https://doi.org/10.1016/j.physrep.2024.11.002} {\bibfield  {journal}
  {\bibinfo  {journal} {Physics Reports}\ }\textbf {\bibinfo {volume} {1105}},\
  \bibinfo {pages} {1} (\bibinfo {year} {2025})}\BibitemShut {NoStop}%
\bibitem [{\citenamefont {Shan}\ and\ \citenamefont {Chen}(1993)}]{ShanChen93}%
  \BibitemOpen
  \bibfield  {author} {\bibinfo {author} {\bibfnamefont {X.}~\bibnamefont
  {Shan}}\ and\ \bibinfo {author} {\bibfnamefont {H.}~\bibnamefont {Chen}},\
  }\href {https://doi.org/10.1103/PhysRevE.47.1815} {\bibfield  {journal}
  {\bibinfo  {journal} {Physical Review E}\ }\textbf {\bibinfo {volume} {47}},\
  \bibinfo {pages} {1815} (\bibinfo {year} {1993})}\BibitemShut {NoStop}%
\bibitem [{\citenamefont {Sbragaglia}\ \emph {et~al.}(2007)\citenamefont
  {Sbragaglia}, \citenamefont {Benzi}, \citenamefont {Biferale}, \citenamefont
  {Succi}, \citenamefont {Sugiyama},\ and\ \citenamefont
  {Toschi}}]{Sbragaglia07}%
  \BibitemOpen
  \bibfield  {author} {\bibinfo {author} {\bibfnamefont {M.}~\bibnamefont
  {Sbragaglia}}, \bibinfo {author} {\bibfnamefont {R.}~\bibnamefont {Benzi}},
  \bibinfo {author} {\bibfnamefont {L.}~\bibnamefont {Biferale}}, \bibinfo
  {author} {\bibfnamefont {S.}~\bibnamefont {Succi}}, \bibinfo {author}
  {\bibfnamefont {K.}~\bibnamefont {Sugiyama}},\ and\ \bibinfo {author}
  {\bibfnamefont {F.}~\bibnamefont {Toschi}},\ }\href
  {https://doi.org/10.1103/PhysRevE.75.026702} {\bibfield  {journal} {\bibinfo
  {journal} {Physical Review E}\ }\textbf {\bibinfo {volume} {75}} (\bibinfo
  {year} {2007})}\BibitemShut {NoStop}%
\bibitem [{\citenamefont {Sbragaglia}\ \emph {et~al.}(2012)\citenamefont
  {Sbragaglia}, \citenamefont {Benzi}, \citenamefont {Bernaschi},\ and\
  \citenamefont {Succi}}]{Sbragagliaetal12}%
  \BibitemOpen
  \bibfield  {author} {\bibinfo {author} {\bibfnamefont {M.}~\bibnamefont
  {Sbragaglia}}, \bibinfo {author} {\bibfnamefont {R.}~\bibnamefont {Benzi}},
  \bibinfo {author} {\bibfnamefont {M.}~\bibnamefont {Bernaschi}},\ and\
  \bibinfo {author} {\bibfnamefont {S.}~\bibnamefont {Succi}},\ }\href
  {http://dx.doi.org/10.1039/C2SM26167G} {\bibfield  {journal} {\bibinfo
  {journal} {Soft Matter}\ }\textbf {\bibinfo {volume} {8}},\ \bibinfo {pages}
  {10773} (\bibinfo {year} {2012})}\BibitemShut {NoStop}%
\bibitem [{\citenamefont {Bhatnagar}\ \emph {et~al.}(1954)\citenamefont
  {Bhatnagar}, \citenamefont {Gross},\ and\ \citenamefont {Krook}}]{BGK54}%
  \BibitemOpen
  \bibfield  {author} {\bibinfo {author} {\bibfnamefont {P.~L.}\ \bibnamefont
  {Bhatnagar}}, \bibinfo {author} {\bibfnamefont {E.~P.}\ \bibnamefont
  {Gross}},\ and\ \bibinfo {author} {\bibfnamefont {M.}~\bibnamefont {Krook}},\
  }\href {https://doi.org/10.1103/PhysRev.94.511} {\bibfield  {journal}
  {\bibinfo  {journal} {Physical Review}\ }\textbf {\bibinfo {volume} {94}},\
  \bibinfo {pages} {511} (\bibinfo {year} {1954})}\BibitemShut {NoStop}%
\bibitem [{\citenamefont {Silva}\ \emph {et~al.}(2023)\citenamefont {Silva},
  \citenamefont {Coelho}, \citenamefont {da~Gama},\ and\ \citenamefont
  {Ara{\'u}jo}}]{silva2023effect}%
  \BibitemOpen
  \bibfield  {author} {\bibinfo {author} {\bibfnamefont {D.~P.}\ \bibnamefont
  {Silva}}, \bibinfo {author} {\bibfnamefont {R.~C.}\ \bibnamefont {Coelho}},
  \bibinfo {author} {\bibfnamefont {M.~M.~T.}\ \bibnamefont {da~Gama}},\ and\
  \bibinfo {author} {\bibfnamefont {N.~A.}\ \bibnamefont {Ara{\'u}jo}},\ }\href
  {https://doi.org/10.1103/PhysRevE.107.035106} {\bibfield  {journal} {\bibinfo
   {journal} {Physical Review E}\ }\textbf {\bibinfo {volume} {107}},\ \bibinfo
  {pages} {035106} (\bibinfo {year} {2023})}\BibitemShut {NoStop}%
\bibitem [{\citenamefont {Silva}\ \emph {et~al.}(2024)\citenamefont {Silva},
  \citenamefont {Coelho}, \citenamefont {Pagonabarraga}, \citenamefont {Succi},
  \citenamefont {da~Gama},\ and\ \citenamefont
  {Ara{\'u}jo}}]{silva2024lattice}%
  \BibitemOpen
  \bibfield  {author} {\bibinfo {author} {\bibfnamefont {D.~P.}\ \bibnamefont
  {Silva}}, \bibinfo {author} {\bibfnamefont {R.~C.}\ \bibnamefont {Coelho}},
  \bibinfo {author} {\bibfnamefont {I.}~\bibnamefont {Pagonabarraga}}, \bibinfo
  {author} {\bibfnamefont {S.}~\bibnamefont {Succi}}, \bibinfo {author}
  {\bibfnamefont {M.~M.~T.}\ \bibnamefont {da~Gama}},\ and\ \bibinfo {author}
  {\bibfnamefont {N.~A.}\ \bibnamefont {Ara{\'u}jo}},\ }\href
  {https://doi.org/10.1039/D3SM01648J} {\bibfield  {journal} {\bibinfo
  {journal} {Soft Matter}\ } (\bibinfo {year} {2024})}\BibitemShut {NoStop}%
\bibitem [{\citenamefont {Spiegel}\ and\ \citenamefont
  {Veronis}(1960)}]{Spiegel60}%
  \BibitemOpen
  \bibfield  {author} {\bibinfo {author} {\bibfnamefont {E.~A.}\ \bibnamefont
  {Spiegel}}\ and\ \bibinfo {author} {\bibfnamefont {G.}~\bibnamefont
  {Veronis}},\ }\href {https://adsabs.harvard.edu/full/1960apj...131..442s}
  {\bibfield  {journal} {\bibinfo  {journal} {Astrophysical Journal}\ }\textbf
  {\bibinfo {volume} {131}},\ \bibinfo {pages} {442} (\bibinfo {year}
  {1960})}\BibitemShut {NoStop}%
\bibitem [{\citenamefont {Hinze}(1955)}]{Hinze1955}%
  \BibitemOpen
  \bibfield  {author} {\bibinfo {author} {\bibfnamefont {J.}~\bibnamefont
  {Hinze}},\ }\href {https://doi.org/10.1002/aic.690010303} {\bibfield
  {journal} {\bibinfo  {journal} {A.I.Ch.E. Journal}\ }\textbf {\bibinfo
  {volume} {1}},\ \bibinfo {pages} {289} (\bibinfo {year} {1955})}\BibitemShut
  {NoStop}%
\bibitem [{\citenamefont {Andersson}\ and\ \citenamefont
  {Andersson}(2006)}]{Anderson2006}%
  \BibitemOpen
  \bibfield  {author} {\bibinfo {author} {\bibfnamefont {R.}~\bibnamefont
  {Andersson}}\ and\ \bibinfo {author} {\bibfnamefont {B.}~\bibnamefont
  {Andersson}},\ }\href {https://doi.org/10.1002/aic.10831} {\bibfield
  {journal} {\bibinfo  {journal} {A.I.Ch.E. Journal}\ }\textbf {\bibinfo
  {volume} {52}},\ \bibinfo {pages} {2020} (\bibinfo {year}
  {2006})}\BibitemShut {NoStop}%
\bibitem [{\citenamefont {Cioni}\ \emph {et~al.}(1997)\citenamefont {Cioni},
  \citenamefont {Ciliberto},\ and\ \citenamefont {Sommeria}}]{Cioni97}%
  \BibitemOpen
  \bibfield  {author} {\bibinfo {author} {\bibfnamefont {S.}~\bibnamefont
  {Cioni}}, \bibinfo {author} {\bibfnamefont {S.}~\bibnamefont {Ciliberto}},\
  and\ \bibinfo {author} {\bibfnamefont {J.}~\bibnamefont {Sommeria}},\ }\href
  {https://doi.org/10.1017/S0022112096004491} {\bibfield  {journal} {\bibinfo
  {journal} {Journal of Fluid Mechanics}\ }\textbf {\bibinfo {volume} {335}},\
  \bibinfo {pages} {111} (\bibinfo {year} {1997})}\BibitemShut {NoStop}%
\bibitem [{\citenamefont {Castaing}\ \emph {et~al.}(1989)\citenamefont
  {Castaing}, \citenamefont {Gunaratne}, \citenamefont {Heslot}, \citenamefont
  {Kadanoff}, \citenamefont {Libchaber}, \citenamefont {Thomae}, \citenamefont
  {Wu}, \citenamefont {Zaleski},\ and\ \citenamefont {Zanetti}}]{Castaing89}%
  \BibitemOpen
  \bibfield  {author} {\bibinfo {author} {\bibfnamefont {B.}~\bibnamefont
  {Castaing}}, \bibinfo {author} {\bibfnamefont {G.}~\bibnamefont {Gunaratne}},
  \bibinfo {author} {\bibfnamefont {F.}~\bibnamefont {Heslot}}, \bibinfo
  {author} {\bibfnamefont {L.}~\bibnamefont {Kadanoff}}, \bibinfo {author}
  {\bibfnamefont {A.}~\bibnamefont {Libchaber}}, \bibinfo {author}
  {\bibfnamefont {S.}~\bibnamefont {Thomae}}, \bibinfo {author} {\bibfnamefont
  {X.-Z.}\ \bibnamefont {Wu}}, \bibinfo {author} {\bibfnamefont
  {S.}~\bibnamefont {Zaleski}},\ and\ \bibinfo {author} {\bibfnamefont
  {G.}~\bibnamefont {Zanetti}},\ }\href
  {https://doi.org/10.1017/S0022112089001643} {\bibfield  {journal} {\bibinfo
  {journal} {Journal of Fluid Mechanics}\ }\textbf {\bibinfo {volume} {204}},\
  \bibinfo {pages} {1} (\bibinfo {year} {1989})}\BibitemShut {NoStop}%
\bibitem [{\citenamefont {Ciliberto}\ \emph {et~al.}(1996)\citenamefont
  {Ciliberto}, \citenamefont {Cioni},\ and\ \citenamefont
  {Laroche}}]{Ciliberto96}%
  \BibitemOpen
  \bibfield  {author} {\bibinfo {author} {\bibfnamefont {S.}~\bibnamefont
  {Ciliberto}}, \bibinfo {author} {\bibfnamefont {S.}~\bibnamefont {Cioni}},\
  and\ \bibinfo {author} {\bibfnamefont {C.}~\bibnamefont {Laroche}},\ }\href
  {https://doi.org/10.1103/PhysRevE.54.R5901} {\bibfield  {journal} {\bibinfo
  {journal} {Physical Review E}\ }\textbf {\bibinfo {volume} {54}},\ \bibinfo
  {pages} {R5901} (\bibinfo {year} {1996})}\BibitemShut {NoStop}%
\bibitem [{\citenamefont {Benzi}\ \emph {et~al.}(1998)\citenamefont {Benzi},
  \citenamefont {Toschi},\ and\ \citenamefont {Tripiccione}}]{Benzi98}%
  \BibitemOpen
  \bibfield  {author} {\bibinfo {author} {\bibfnamefont {R.}~\bibnamefont
  {Benzi}}, \bibinfo {author} {\bibfnamefont {F.}~\bibnamefont {Toschi}},\ and\
  \bibinfo {author} {\bibfnamefont {R.}~\bibnamefont {Tripiccione}},\ }\href
  {https://doi.org/10.1023/B:JOSS.0000033168.36971.59} {\bibfield  {journal}
  {\bibinfo  {journal} {Journal of Statistical Physics}\ }\textbf {\bibinfo
  {volume} {93}},\ \bibinfo {pages} {901} (\bibinfo {year} {1998})}\BibitemShut
  {NoStop}%
\bibitem [{\citenamefont {Grossmann}\ and\ \citenamefont
  {Lohse}(2000)}]{Grossmann99}%
  \BibitemOpen
  \bibfield  {author} {\bibinfo {author} {\bibfnamefont {S.}~\bibnamefont
  {Grossmann}}\ and\ \bibinfo {author} {\bibfnamefont {D.}~\bibnamefont
  {Lohse}},\ }\href {https://doi.org/10.1017/S0022112099007545} {\bibfield
  {journal} {\bibinfo  {journal} {Journal of Fluid Mechanics}\ }\textbf
  {\bibinfo {volume} {407}},\ \bibinfo {pages} {27} (\bibinfo {year}
  {2000})}\BibitemShut {NoStop}%
\bibitem [{\citenamefont {Stevens}\ \emph {et~al.}(2013)\citenamefont
  {Stevens}, \citenamefont {van~der Poel}, \citenamefont {Grossmann},\ and\
  \citenamefont {Lohse}}]{Stevens13}%
  \BibitemOpen
  \bibfield  {author} {\bibinfo {author} {\bibfnamefont {R.}~\bibnamefont
  {Stevens}}, \bibinfo {author} {\bibfnamefont {E.}~\bibnamefont {van~der
  Poel}}, \bibinfo {author} {\bibfnamefont {S.}~\bibnamefont {Grossmann}},\
  and\ \bibinfo {author} {\bibfnamefont {D.}~\bibnamefont {Lohse}},\ }\href
  {https://doi.org/10.1017/jfm.2013.298} {\bibfield  {journal} {\bibinfo
  {journal} {Journal of Fluid Mechanics}\ }\textbf {\bibinfo {volume} {730}},\
  \bibinfo {pages} {295} (\bibinfo {year} {2013})}\BibitemShut {NoStop}%
\bibitem [{\citenamefont {Ecke}\ \emph {et~al.}(1986)\citenamefont {Ecke},
  \citenamefont {Haucke}, \citenamefont {Maeno},\ and\ \citenamefont
  {Wheatley}}]{Ecke86}%
  \BibitemOpen
  \bibfield  {author} {\bibinfo {author} {\bibfnamefont {R.~E.}\ \bibnamefont
  {Ecke}}, \bibinfo {author} {\bibfnamefont {H.}~\bibnamefont {Haucke}},
  \bibinfo {author} {\bibfnamefont {Y.}~\bibnamefont {Maeno}},\ and\ \bibinfo
  {author} {\bibfnamefont {J.~C.}\ \bibnamefont {Wheatley}},\ }\href
  {https://doi.org/10.1103/PhysRevA.33.1870} {\bibfield  {journal} {\bibinfo
  {journal} {Physical Review A}\ }\textbf {\bibinfo {volume} {33}},\ \bibinfo
  {pages} {1870} (\bibinfo {year} {1986})}\BibitemShut {NoStop}%
\bibitem [{\citenamefont {Heslot}\ \emph {et~al.}(1987)\citenamefont {Heslot},
  \citenamefont {Castaing},\ and\ \citenamefont {Libchaber}}]{Heslot87}%
  \BibitemOpen
  \bibfield  {author} {\bibinfo {author} {\bibfnamefont {F.}~\bibnamefont
  {Heslot}}, \bibinfo {author} {\bibfnamefont {B.}~\bibnamefont {Castaing}},\
  and\ \bibinfo {author} {\bibfnamefont {A.}~\bibnamefont {Libchaber}},\ }\href
  {https://doi.org/10.1103/PhysRevA.36.5870} {\bibfield  {journal} {\bibinfo
  {journal} {Physical Review A}\ }\textbf {\bibinfo {volume} {36}},\ \bibinfo
  {pages} {5870} (\bibinfo {year} {1987})}\BibitemShut {NoStop}%
\bibitem [{\citenamefont {Deane}\ and\ \citenamefont {Stokes}(2002)}]{Deane02}%
  \BibitemOpen
  \bibfield  {author} {\bibinfo {author} {\bibfnamefont {G.~B.}\ \bibnamefont
  {Deane}}\ and\ \bibinfo {author} {\bibfnamefont {M.~D.}\ \bibnamefont
  {Stokes}},\ }\href {https://doi.org/10.1038/nature00967} {\bibfield
  {journal} {\bibinfo  {journal} {Nature}\ }\textbf {\bibinfo {volume} {418}},\
  \bibinfo {pages} {839} (\bibinfo {year} {2002})}\BibitemShut {NoStop}%
\bibitem [{\citenamefont {Mukherjee}\ \emph {et~al.}(2019)\citenamefont
  {Mukherjee}, \citenamefont {Safdari}, \citenamefont {Shardt}, \citenamefont
  {Kenjere{\v{s}}},\ and\ \citenamefont {Van~den Akker}}]{Mukherjee19}%
  \BibitemOpen
  \bibfield  {author} {\bibinfo {author} {\bibfnamefont {S.}~\bibnamefont
  {Mukherjee}}, \bibinfo {author} {\bibfnamefont {A.}~\bibnamefont {Safdari}},
  \bibinfo {author} {\bibfnamefont {O.}~\bibnamefont {Shardt}}, \bibinfo
  {author} {\bibfnamefont {S.}~\bibnamefont {Kenjere{\v{s}}}},\ and\ \bibinfo
  {author} {\bibfnamefont {H.~E.}\ \bibnamefont {Van~den Akker}},\ }\href
  {https://doi.org/10.1017/jfm.2019.654} {\bibfield  {journal} {\bibinfo
  {journal} {Journal of Fluid Mechanics}\ }\textbf {\bibinfo {volume} {878}},\
  \bibinfo {pages} {221} (\bibinfo {year} {2019})}\BibitemShut {NoStop}%
\bibitem [{\citenamefont {Crialesi-Esposito}\ \emph {et~al.}(2023)\citenamefont
  {Crialesi-Esposito}, \citenamefont {Chibbaro},\ and\ \citenamefont
  {Brandt}}]{CrialesiEsposito23interaction}%
  \BibitemOpen
  \bibfield  {author} {\bibinfo {author} {\bibfnamefont {M.}~\bibnamefont
  {Crialesi-Esposito}}, \bibinfo {author} {\bibfnamefont {S.}~\bibnamefont
  {Chibbaro}},\ and\ \bibinfo {author} {\bibfnamefont {L.}~\bibnamefont
  {Brandt}},\ }\href {https://doi.org/10.1038/s42005-022-01122-8} {\bibfield
  {journal} {\bibinfo  {journal} {Communications Physics}\ }\textbf {\bibinfo
  {volume} {6}},\ \bibinfo {pages} {5} (\bibinfo {year} {2023})}\BibitemShut
  {NoStop}%
\bibitem [{\citenamefont {Roccon}\ \emph {et~al.}(2023)\citenamefont {Roccon},
  \citenamefont {Zonta},\ and\ \citenamefont {Soldati}}]{Roccon23}%
  \BibitemOpen
  \bibfield  {author} {\bibinfo {author} {\bibfnamefont {A.}~\bibnamefont
  {Roccon}}, \bibinfo {author} {\bibfnamefont {F.}~\bibnamefont {Zonta}},\ and\
  \bibinfo {author} {\bibfnamefont {A.}~\bibnamefont {Soldati}},\ }\href
  {https://doi.org/10.1103/PhysRevFluids.8.090501} {\bibfield  {journal}
  {\bibinfo  {journal} {Physical Review Fluids}\ }\textbf {\bibinfo {volume}
  {8}},\ \bibinfo {pages} {090501} (\bibinfo {year} {2023})}\BibitemShut
  {NoStop}%
\bibitem [{\citenamefont {Garrett}\ \emph {et~al.}(2000)\citenamefont
  {Garrett}, \citenamefont {Li},\ and\ \citenamefont {Farmer}}]{Garrett2000}%
  \BibitemOpen
  \bibfield  {author} {\bibinfo {author} {\bibfnamefont {C.}~\bibnamefont
  {Garrett}}, \bibinfo {author} {\bibfnamefont {M.}~\bibnamefont {Li}},\ and\
  \bibinfo {author} {\bibfnamefont {D.}~\bibnamefont {Farmer}},\ }\href
  {https://doi.org/10.1175/1520-0485(2000)030<2163:TCBBSS>2.0.CO;2} {\bibfield
  {journal} {\bibinfo  {journal} {Journal of Physical Oceanography}\ }\textbf
  {\bibinfo {volume} {30}},\ \bibinfo {pages} {2163} (\bibinfo {year}
  {2000})}\BibitemShut {NoStop}%
\bibitem [{\citenamefont {Benzi}\ \emph
  {et~al.}(2021{\natexlab{a}})\citenamefont {Benzi}, \citenamefont {Divoux},
  \citenamefont {Barentin}, \citenamefont {Manneville}, \citenamefont
  {Sbragaglia},\ and\ \citenamefont {Toschi}}]{benzi2021continuum}%
  \BibitemOpen
  \bibfield  {author} {\bibinfo {author} {\bibfnamefont {R.}~\bibnamefont
  {Benzi}}, \bibinfo {author} {\bibfnamefont {T.}~\bibnamefont {Divoux}},
  \bibinfo {author} {\bibfnamefont {C.}~\bibnamefont {Barentin}}, \bibinfo
  {author} {\bibfnamefont {S.}~\bibnamefont {Manneville}}, \bibinfo {author}
  {\bibfnamefont {M.}~\bibnamefont {Sbragaglia}},\ and\ \bibinfo {author}
  {\bibfnamefont {F.}~\bibnamefont {Toschi}},\ }\href
  {https://doi.org/10.1103/PhysRevE.104.034612} {\bibfield  {journal} {\bibinfo
   {journal} {Physical Review E}\ }\textbf {\bibinfo {volume} {104}},\ \bibinfo
  {pages} {034612} (\bibinfo {year} {2021}{\natexlab{a}})}\BibitemShut
  {NoStop}%
\bibitem [{\citenamefont {Benzi}\ \emph
  {et~al.}(2021{\natexlab{b}})\citenamefont {Benzi}, \citenamefont {Divoux},
  \citenamefont {Barentin}, \citenamefont {Manneville}, \citenamefont
  {Sbragaglia},\ and\ \citenamefont {Toschi}}]{benzi2021stress}%
  \BibitemOpen
  \bibfield  {author} {\bibinfo {author} {\bibfnamefont {R.}~\bibnamefont
  {Benzi}}, \bibinfo {author} {\bibfnamefont {T.}~\bibnamefont {Divoux}},
  \bibinfo {author} {\bibfnamefont {C.}~\bibnamefont {Barentin}}, \bibinfo
  {author} {\bibfnamefont {S.}~\bibnamefont {Manneville}}, \bibinfo {author}
  {\bibfnamefont {M.}~\bibnamefont {Sbragaglia}},\ and\ \bibinfo {author}
  {\bibfnamefont {F.}~\bibnamefont {Toschi}},\ }\href
  {https://doi.org/10.1103/PhysRevLett.127.148003} {\bibfield  {journal}
  {\bibinfo  {journal} {Physical Review Letters}\ }\textbf {\bibinfo {volume}
  {127}},\ \bibinfo {pages} {148003} (\bibinfo {year}
  {2021}{\natexlab{b}})}\BibitemShut {NoStop}%
\end{thebibliography}%

\end{document}